\documentclass[a4paper,12pt,notitlepage]{article}

\usepackage{monash}
\usepackage{NMSSM}

\xnewcommand{\multinest}{\code{MultiNest}}
\xnewcommand{\multinestv}{\code{MultiNest\allowbreak{}v3.10}}
\xnewcommand{\softsusy}{\code{SOFTSUSY}}
\xnewcommand{\softsusyv}{\code{SOFTSUSY-3.6.2}}
\xnewcommand{\superplot}{\code{SuperPlot}}

\newcommand{\figwidth}{0.3\textwidth}

\begin{document}

\title{Bayesian analysis and naturalness \\ of (Next-to-)Minimal Supersymmetric Models}

\author[1]{Peter Athron}
\author[1,2]{Csaba Balazs}
\author[3,4]{Benjamin Farmer}
\author[1]{Andrew Fowlie}
\author[5,6]{Dylan Harries}
\author[7]{Doyoun Kim}
\affiliation[1]{ARC Centre of Excellence for Particle Physics at the Tera-scale, Monash University, Melbourne, Victoria 3800, Australia}
\affiliation[2]{Monash Centre for Astrophysics, School of Physics and Astronomy, Monash University, Melbourne, Victoria 3800 Australia}
\affiliation[3]{Department of Physics, Stockholm University, AlbaNova, SE-106 91 Stockholm, Sweden}
\affiliation[4]{The Oskar Klein Centre for Cosmoparticle Physics, AlbaNova, SE-106 91 Stockholm, Sweden}
\affiliation[5]{ARC Centre of Excellence for Particle Physics at the
Tera-scale, Department of Physics, The University of Adelaide,
Adelaide, South Australia 5005, Australia}
\affiliation[6]{Institute of Particle and Nuclear Physics, 
Faculty of Mathematics and Physics, Charles University in Prague, 
V Hole\v{s}ovi\v{c}k\'{a}ch 2, 180 00 Praha 8, Czech Republic}
\affiliation[7]{Center for Theoretical Physics of the Universe, Institute for Basic Science (IBS), 

Daejeon, 34051, Korea}
\preprint{CTPU-17-33}

\date{\today} 

\maketitle

\begin{abstract}
The Higgs boson discovery stirred interest in next-to-minimal supersymmetric models, due to the apparent fine-tuning required to accommodate it in minimal theories.  To assess their naturalness, we compare fine-tuning in a \Z{3} conserving semi-constrained Next-to-Minimal Supersymmetric Standard Model (NMSSM) to the constrained MSSM (CMSSM). 
We contrast popular fine-tuning measures with naturalness priors, which automatically appear in statistical measures of the plausibility that a given model reproduces the weak scale. Our comparison shows that naturalness priors provide valuable insight into the hierarchy problem and rigorously ground naturalness in Bayesian statistics. For the \CMSSM and semi-constrained \NMSSM we demonstrate qualitative agreement between naturalness priors and 
popular fine tuning measures. Thus, we give a clear plausibility argument that favours relatively light superpartners.
\end{abstract}

\section{Introduction}

The absence of supersymmetry (SUSY) at the LHC \see{Agashe:2014kda}
and the discovery of a $125\gev$ Standard Model-like Higgs
boson\cite{Chatrchyan:2012xdj,Aad:2012tfa} raise the spectre of
fine-tuning in supersymmetric 
models\cite{Barbieri:1987fn,Ellis:1986yg}. This appears to undermine
the \textit{raison d'etre} for weak-scale supersymmetry: eliminating
fine-tuning in the Standard Model (SM) by cancelling quadratic
divergences\cite{Witten:1981nf}, thus solving the infamous hierarchy 
problem\cite{Weinberg:1975gm,Weinberg:1979bn,Susskind:1978ms,
Gildener:1976ai}. A $125\gev$ Higgs is particularly problematic for
minimal supersymmetric models \see{Draper:2011aa,Akula:2011aa,
Fowlie:2012im,Kadastik:2011aa,Buchmueller:2011ab,Cao:2011sn} because it
can only be achieved by large quantum corrections from massive
sparticles\cite{Draper:2013oza,Bagnaschi:2014rsa,Vega:2015fna,Bahl:2016brp,Athron:2016fuq,Bahl:2017aev,Harlander:2017kuc}. 

 In singlet extensions of minimal supersymmetry\cite{Fayet:1974pd,Fayet:1976et,Fayet:1977yc,Nilles:1982dy,Frere:1983ag,Derendinger:1983bz,Veselov:1985gd,Ellis:1988er,Drees:1988fc} the tree-level Higgs mass  can be raised beyond that of the $Z$-boson. The simplest singlet extension is the Next-to-Minimal Supersymmetric SM (\NMSSM), reviewed in
\refcite{Maniatis:2009re,Ellwanger:2009dp}. It is argued that the 
\NMSSM is more natural, that is less fine-tuned, than minimal
supersymmetric models\cite{BasteroGil:2000bw,Dermisek:2005ar,
Ellwanger:2011mu,King:2012is,Kang:2012sy,Gunion:2012zd,Cao:2012fz,Ellwanger:2012ke,King:2012tr,Gherghetta:2012gb}. Furthermore, there have been many
supersymmetric models, built in light of LHC results, that are
claimed to be more natural because they raise
the Higgs mass at tree-level \see{Li:2011ab,Moroi:2011aa,Kyae:2012ea,Boudjema:2012cq,
Basak:2012bd,Athron:2012sq,Benakli:2012cy,An:2012vp,Kyae:2012rv,Bae:2012ir,
Bhattacharyya:2012qj,Craig:2012bs,Basso:2013vla,Maru:2013ooa,Galloway:2013dma,Bandyopadhyay:2013lca,Bharucha:2013ela,
Chang:2014zva,Bertuzzo:2014bwa,Dimopoulos:2014aua,Bertuzzo:2014sma,Ding:2015wma,Belanger:2015cra,Kim:2015yha, Capdevilla:2015qwa,Nakai:2015swg,Okada:2016wlm,Hebbar:2017fit,Badziak:2017kjk}.

Checking such claims by calculating fine-tuning in various
supersymmetric models, however, is somewhat futile, as the results
would completely depend upon the definition of fine-tuning itself.
This subjectivity is a common criticism of naturalness arguments.
Rather than abandoning naturalness or relying on heuristic
judgments, we instead advocate for an approach that is based on
Bayesian statistics. In this approach, one has a well-defined means
of quantifying both how plausible a particular parameter space point
is in the context of a given model and which model in a given set is
the most plausible in light of experimental data.  Apart from these
being the most germane questions to pose, we argue that they also
capture the essence of ordinary naturalness arguments whilst evading
arbitrary aspects of naturalness by utilizing a unique logical
framework in Bayesian inference \see{Gregory,Jaynes:2003,Earman}.

Such calculations automatically incorporate so-called naturalness
priors that contain factors strongly resembling some traditional
measures of fine-tuning, but which have a rigorous probabilistic
interpretation.  In addition to being a well-founded
fine-tuning measure, the appearance of these
naturalness priors also leads to posterior probability densities
that tend to favor regions of parameter space that would be
considered as having low fine-tuning according to the na\"{\i}ve
tuning measures, as we show below. Thus Bayesian plausibility 
analyses automatically take into account fine-tuning in a
model and the effects of new experimental data on this tuning.
Moreover, through comparing the Bayesian evidence for different
models it is possible to make statistically meaningful comparisons 
between models.  The role of the naturalness priors in these comparisons 
is to ensure that the outcome of such a comparison is  reflective of 
whether one model is more natural
than another for a given set of experimental data. 


The Bayesian interpretation of naturalness was advocated numerous
times over the last decade\cite{Allanach:2007qk,Cabrera:2008tj,
Fowlie:2014xha,Fichet:2012sn,Cabrera:2010dh,Fowlie:2014faa,
Fowlie:2015uga,Clarke:2016jzm,Fundira:2017vip}.  However, since it
remains much less common than traditional fine-tuning measures, we
recapitulate the essential points in \refsec{Sec:Bayes}. We
illustrate this  methodology with a warm-up example of the hierarchy
problem in the SM in \refsec{Sec:SM}, define our semi-constrained
NMSSM and the \CMSSM models in \refsec{Sec:SUSY}, and describe
results from our fully-fledged Bayesian analysis in
\refsec{Sec:Results}. This completes our previous
study\cite{Kim:2013uxa} and complements previous Bayesian analyses
of the semi-constrained NMSSM\cite{Fowlie:2014faa,
LopezFogliani:2009np,Kowalska:2012gs} and
\CMSSM\cite{Akula:2012kk,Williams:2015bfa,Diamanti:2015kma,Catalan:2015cna,
Casas:2014eca,Roszkowski:2014wqa,Strege:2014ija,Fowlie:2014awa,
AbdusSalam:2013qba,Cabrera:2013jya,
Arina:2013zca,deAustri:2013saa,Fowlie:2013oua,Antusch:2013kna,
Cabrera:2012vu,Strege:2012bt,Balazs:2012bx,Fowlie:2012im,
Balazs:2013qva,Roszkowski:2012uf,Strege:2011pk,Fowlie:2011mb,
Bertone:2011pq,Cabrera:2011ds,Allanach:2011ya,Bertone:2011nj,
Fowlie:2011vf,Allanach:2011ut,Feroz:2011bj,Ripken:2010ja,
Cabrera:2010xx,Akrami:2010cz,Cabrera:2009dm,Akrami:2009hp,
Roszkowski:2009ye,Roszkowski:2009sm,Trotta:2008bp,Feroz:2008wr,
Allanach:2008iq,Allanach:2008tu,Allanach:2008zn,Roszkowski:2007fd,
Roszkowski:2006mi}. We close by summarizing our findings in
\refsec{Sec:Conclusions}. 

\section{Bayesian inference}\label{Sec:Bayes}

Bayesian statistics is a framework for quantifying the plausibility of a hypothesis, such as a scientific theory \see{Gregory}. The central equation for our analysis is Bayes' theorem for continuous variables,
\begin{equation}\label{Eq:Bayes}
\pg{\params}{\model, \data} \propto \pg{\data}{\model,\params} \cdot \pg{\params}{\model}.
\end{equation}
The theorem expresses that the prior probability density \pg{\params}{\model} for parameters \params in a model \model is updated by experimental data \data, resulting in the posterior \pg{\params}{\model, \data}. The updating factor \pg{\data}{\model,\params} is known as a likelihood function when interpreted as a function of \params. The posterior is often insensitive to the diffuseness of the prior, \ie whether one permits broad or narrow ranges for the parameters, but may be sensitive to the shape of the prior, though this sensitivity may be counterbalanced by sufficient data.

We may find the probability density for a subset of parameters by \emph{marginalization}, \ie integration. For example, the marginal posterior density for $x$ would be
\begin{equation}\label{Eq:Marginalization}
\pg{x}{\model, \data} = \int
\pg{\params}{\model, \data} \dif y \dif \dots .
\end{equation}
As simple as it seems, marginalization captures the traditional idea in physics that fine-tuned parameters are relatively implausible. We may rewrite \refeq{Eq:Marginalization} as
\begin{equation}
\pg{x}{\model, \data} = \int
\pg{x}{y, \dots, \model, \data} \cdot 
\pg{y, \dots}{\model, \data} 
\dif y \dif \dots ,
\end{equation}
which states that the posterior density for $x$ is the average conditional density \pg{x}{y, \dots, \model, \data}. For a given $x$, it may be possible to fine-tune the value of $y, \dots$ such that the conditional density is substantial. The average conditional density and thus the posterior, though, may be negligible. As we shall see in \refsec{Sec:SM}, in this way marginalization automatically penalizes fine-tuning related to the hierarchy problem. 

The second equation for our statistical analysis is Bayes' theorem for a discrete hypothesis,
\begin{equation}\label{Eq:BayesModel}
\Probg{\model}{\data} \propto \pg{\data}{\model} \cdot \Prob{\model}.
\end{equation}
We see that the plausibility of a model is updated by a factor known as the evidence, which may be expressed as
\begin{equation}
\pg{\data}{\model} = \int \pg{\data}{\params, \model} \cdot \pg{\params}{\model}\dif x \dif y \dif \dots \, .
\end{equation}
The evidence is a functional of the priors for the model's parameters. Model selection by evidences is somewhat controversial, partly since evidences may be sensitive to the diffuseness of prior densities and this sensitivity cannot be compensated by sufficient data. For that reason, we focus upon posterior distributions, though briefly compare models with evidences, which are a byproduct of our analysis. 

Computationally, the evidence is the average likelihood. As such, it penalizes fine-tuning automatically, since if, for a particular model, agreement with data is found in only a small region of the prior volume, the average likelihood will be small relative to a model in which agreement is found everywhere or more readily.



\section{Fine-tuning in the Standard Model}\label{Sec:SM}
We now consider fine-tuning of the weak scale in the Standard Model (SM) interpreted as an effective field theory with quadratic corrections from new physics.  Our toy-model of the effective SM is defined by a cut-off $\Lambda^2$ and parameters $\mu^2$ and $\lambda$ in the Higgs potential,
\begin{equation}
V = \mu^2 h^2 + \lambda h^4 .
\end{equation}
This toy model predicts that
\begin{equation}
\mz^2 = \frac{\bar{g}^2 v^2}{4} =
-\frac{\bar{g}^2}{8\lambda}\left(\mu^2 + \Lambda^2\right),
\end{equation}
where $\bar{g}^2 = g^2 + {g^\prime}^2$, $g$ and $g^\prime$ being the
$SU(2)_L$ and (non-GUT normalized) $U(1)_Y$ gauge couplings,
respectively. We assume, as happens in many specific cases, that new physics at the cut-off scale
results in quadratic corrections to $\mu^2$.  To keep the toy model
as simple as possible, we do not consider any coefficient from a
loop-factor in front of the quadratic correction and neglect the new
physics corrections to $\lambda$. 

The most common measure of fine-tuning  in particle physics is the
Barbieri-Giudice-Ellis-Nanopoulos (BGEN)
measure\cite{Barbieri:1987fn,Ellis:1986yg}, which is based upon
measuring the sensitivity of some observable quantity to variations
in the underlying, assumed to be fundamental, model parameters. In
discussions of the hierarchy problem, the measure is conventionally
formulated in terms of the predicted $Z$-boson mass, leading to
the tuning sensitivities defined by
\begin{equation} \label{Eq:DelBG}
\Delta_p  \equiv \left|\fp{\ln \mzpole^2}{\ln p}\right|,
\end{equation}
for each model parameter $p$.
This traditional measure leaves many questions. Are fine-tuned theories implausible? And if so, why? How much fine-tuning is too much and why? 
How should we adjust our conclusion in light of new experimental evidence?  There are no answers to these questions because the measure is only intuitively connected to physics and lacks rigorous mathematical roots.  In contrast, it is well known that in Bayesian statistics fine-tuning is intimately connected to model plausibility by a fine-tuning penalty automatically incorporated in the evidence\cite{Allanach:2008iq, Cabrera:2008tj, Cabrera:2010dh}.
%
%
%

Applying the traditional BGEN measure to the cut-off $\Lambda^2$ in
our toy model of the SM we find\footnote{The sensitivities $\Delta_p$ are
more commonly calculated with respect to the Lagrangian parameters
in a model. In a realistic model, one might consider applying
the measure to a heavy mass parameter characterizing the scale
of new physics; we use the generic cut-off here simply to illustrate
the effects of these (unspecified) parameters.}
\begin{equation}\label{eq:toyBG}
\Delta_{\Lambda^2} =
\frac{\bar{g}^2}{8\lambda}\frac{\Lambda^2}{\mz^2}, 
\end{equation}
which indicates that fine-tuning mounts as the cut-off exceeds the weak scale, that is if $\Lambda \gg \mz$. This is the SM hierarchy problem.


To illustrate that Bayesian statistics captures essential aspects of the hierarchy problem and fine-tuning, we consider the posterior for the SM cut-off, conditioned upon the measured $Z$-boson and Higgs mass, $\mz^\text{exp}$ and $\mh^\text{exp}$. If Bayesian statistics quantifies the hierarchy problem, the posterior should favor an SM cut-off close to the weak scale.
%
%
%
%
We begin by applying Bayes' theorem to calculate the posterior for $\Lambda^2$ given our toy version of the SM and the experimental measurement of the mass of the $Z$ boson, $\mz^\text{exp} = 91.1876 \pm 0.0021\gev$\cite{Olive:2016xmw}, 
\begin{align}
\pg{\Lambda^2}{\mz^\text{exp}, \SM} &= \int \pg{\Lambda^2, \mu^2, \lambda}{\mz^\text{exp}, \SM} 
                              \dif \mu^2 \dif \lambda \\
&= \frac{1}{\ev} \int \pg{\mz^\text{exp}}{\Lambda^2, \mu^2, \lambda, \SM} \cdot 
   \pg{\Lambda^2, \mu^2, \lambda}{\SM} \dif \mu^2 \dif \lambda,
\end{align}
where $\ev \equiv \pg{\mz^\text{exp}}{\SM}$ is the evidence. We continue by replacing the likelihood function $\pg{\mz^\text{exp}}{\Lambda^2, \mu^2,
\lambda, \SM}$ with a Dirac $\delta$-function, because \mz is measured
with such high precision,\footnote{We approximate the likelihood
function by a Dirac $\delta$-function under integration, \ie
\begin{equation}\label{eq:dirac_approx}
\int \pg{\mz^\text{exp}}{\Lambda^2, \mu^2, \lambda, \SM} \cdot \pg{\Lambda^2, \mu^2, \lambda}{\SM}
\dif \mu^2 \dif \lambda
\approx
\int \delta(\mz - \mz^\text{exp}) \cdot \pg{\Lambda^2, \mu^2, \lambda}{\SM} 
\dif \mu^2 \dif \lambda.
\end{equation}
} 
and change the variable in the Dirac $\delta$-function from \mz to $\mu^2$,
\begin{equation}\label{eq:before_mu2_integration}
= \frac{1}{\ev}\int \frac{\delta(\mu^2 - \mu^2_Z)}{\left|\dpartial{\mz}{\mu^2}\right|} 
  \cdot \pg{\Lambda^2, \mu^2, \lambda}{\text{SM}} \dif \mu^2 \dif \lambda,
\end{equation}
where $\mu_Z^2$ reproduces the measured \mz,
\begin{equation}
\mu^2_Z = - \frac{8\lambda}{\bar{g}^2}(\mz^\text{exp})^2 - \Lambda^2.
\end{equation}
We identify the integral over $\mu^2$ as an effective prior for the SM quartic and cut-off scale,
\begin{align}\label{eq:sm-effective-prior}
p_\text{eff.}(\Lambda^2, \lambda) \equiv \pg{\Lambda^2, \lambda}{\SM, \mz}
&= 
\frac{1}{\ev} \int \frac{\delta(\mu^2 - \mu^2_Z)}{\left|\dpartial{\mz}{\mu^2}\right|} 
\cdot \pg{\Lambda^2, \mu^2, \lambda}{\text{SM}} \dif \mu^2\\
&=
\frac{1}{\ev} \left|\frac{1}{\dpartial{\mz}{\mu^2}}\right|_{\mu^2_Z} 
\!\!\!\cdot\, 
\pg{\Lambda^2, \mu_Z^2, \lambda}{\text{SM}} .
\end{align}
In \refsec{Sec:EffectivePrior} we identify similar effective priors
in supersymmetric models. The effective prior is conditioned upon
measurement of \mz. By using an effective prior with one Lagrangian parameter fixed such that the measured \mz is obtained, one obtains a prior which is logically identical to the case in which no parameters are fixed and \mz is simply input as a constraint in the likelihood. However, the effective prior allows for vastly more efficient scanning, since one can scan only the hypersurface of parameter space in which the correct \mz is predicted.
In the SM, the fixed parameter was the
Higgs Lagrangian mass, $\mu^2$, and the specific form of the
effective prior obtained contains the same derivative that 
would appear when the traditional fine-tuning measure,
\refeq{Eq:DelBG}, is applied to the parameter $\mu^2$.

Performing the $\mu^2$ integration to obtain the marginal density for $\Lambda^2$ in \refeq{eq:before_mu2_integration} we find
\begin{equation}
\pg{\Lambda^2}{\mz^\text{exp}, \SM} = \frac{1}{\ev} \int
\left|\frac{16\mz^\text{exp}\lambda}{\bar{g}^2}\right| \cdot 
  \pg{\Lambda^2, \mu^2_Z, \lambda}{\text{SM}} \dif \lambda.
\end{equation}
We pick logarithmic priors for the SM parameters, such that 
\begin{equation}
\pg{\Lambda^2, \mu^2_Z, \lambda}{\text{SM}} = 
\begin{cases} 
\frac{N}{\Lambda^2 |\mu^2_Z| \lambda} & \text{inside prior ranges $\mathcal{R}$,}\\
0 & \text{otherwise.}\\
\end{cases} 
\end{equation}
The prior for \eg the SM cut-off favors no particular scale ---
logarithmic priors equally weight every order of magnitude, i.e.  
$\pg{\ln \Lambda}{\SM} = \text{const.}$ The normalization factor
$N$ is defined such that the integral of the prior over the chosen
prior ranges is unity. We take the prior ranges to
be $10^{-4} < \lambda < 10$ and $10^{-10}\gev^2 < |\mu^2| <
10^{40}\gev^2$. The prior range for the cut-off affects only the
overall normalization of the posterior and the ranges outside of
which it is zero. 

Thus with our priors the posterior is,
\begin{align}\label{eq:post}
\pg{\Lambda^2}{\mz^\text{exp}, \SM} &= \frac{N}{\ev} \int_\mathcal{R}
\left|\frac{16\mz^\text{exp}\lambda}{\bar{g}^2}\right| \cdot 
   \frac{1}{\Lambda^2|\mu^2_Z|\lambda}\dif \lambda\\
%
%
&= \frac{16N}{\ev} \frac{\mz^\text{exp}}{\Lambda^2} \int_\mathcal{R} 
   \frac{1}{|8\lambda (\mz^\text{exp})^2 + \bar{g}^2 \Lambda^2|}\dif \lambda .
\end{align}
The prior distribution is substantially updated by the data because we have taken $\mu$ to have a logarithmic prior instead of fixing it at the outset to reproduce the measured \mz and then treating the latter as a nuisance parameter.  As a result, after the $\mu$ integration a factor of $\frac{1}{|\mu_Z^2|}$ appears in the remaining integrand which is approximately $(\bar{g}^2 \Lambda^2)^{-1}$ when $\Lambda \gg \mz^\text{exp}$.  The impact of this is to update the prior distribution such that large values of
$\Lambda$ are strongly disfavored.    

\begin{figure}[t]
\centering
\includegraphics[width=0.5\textwidth]{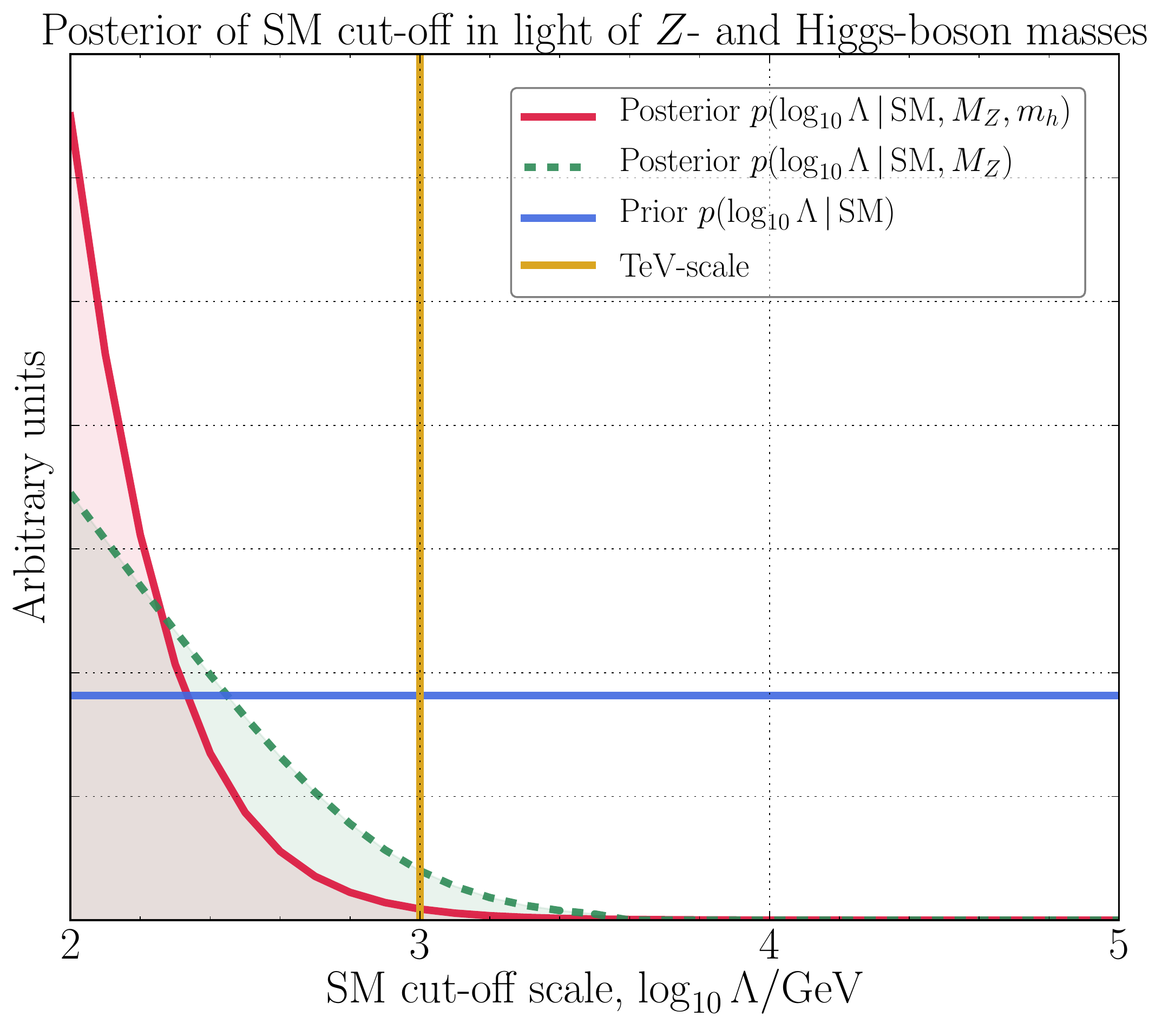}
\caption{Distribution of $\log_{10} \Lambda$, the log of the SM cut-off. The
prior distribution (blue line) was flat in $\log\Lambda$. Once
conditioned upon the weak scale ($\mz^\text{exp}$), the posterior distribution
(green dashed line, filled) favors a small SM cut-off. This is 
the gist of  the SM hierarchy problem caused by quadratic
corrections, $\Lambda^2$. Further conditioning upon $\mh$ makes
little difference (red line, filled). A cut-off of $\Lambda = 1\tev$
is shown for reference by the vertical (yellow) line.} 
\label{fig:SM_Lambda}
\end{figure}

This is illustrated in \reffig{fig:SM_Lambda} where this posterior distribution and a similar one from a calculation that includes the Higgs boson mass $\mh^\text{exp} \simeq 125\gev$ in the likelihood are plotted as functions of $\log\Lambda$. We find, as expected, that the application of Bayes' 
theorem captures the gist of the hierarchy problem: quadratic corrections 
in the SM Higgs mass mean that we ought to expect new physics close to the measured weak scale. 
The prior distribution for the magnitude of the cut-off was flat, but once conditioned upon
the weak scale (\ie measurements of the $Z$-boson mass and the Higgs mass), a sub-TeV SM cut-off was favored. 

Finally, we calculate $\ev$, the evidence of the SM in light of the
measurement of \mz, by 
integrating \refeq{eq:post} with respect to the SM cut-off and rearranging to find the evidence, $\ev$. We find that if the lower
limit on the prior for $\Lambda$ is $\Lambda_\text{SM}$ and
$\Lambda_\text{SM} \gg \mz^\text{exp}$, then
\begin{equation}
\pg{\log\mz^\text{exp}}{\SM} \approx
\pg{\log\mz^\text{exp},\log\mh^\text{exp}}{\SM}
\approx 
c \left(\frac{\mz^\text{exp}}{\Lambda_\text{SM}}\right)^2,
\end{equation}
where $c$ is a coefficient determined by factors
from integration and priors, which we calculated to be $\mathcal{O}(1)$ for our choices. For comparison with dimensionless fine-tuning ratios, 
we expressed the evidences as \eg $\pg{\log\mz^\text{exp}}{\SM} = \mz^\text{exp} \pg{\mz^\text{exp}}{\SM}$.

This tells us that if the cut-off is of the order of the Planck
scale, \mpl, then the evidence is very small, 
${\mathcal O}(10^{-34})$.  But if the model allows the cut-off to be
of order \mz then the evidence is $\mathcal{O}(1)$.  Therefore the
evidence strongly prefers an SM effective theory that is valid only
up to the electroweak or TeV scale (with new physics such as
supersymmetry appearing at that scale) to an SM effective theory
with no new physics below \mpl. This is the essence of the
well-known hierarchy problem, but expressed in a statistically
rigorous manner.

Besides its coherency and connection to statistics, an advantage of this formulation over ad-hoc fine-tuning measures is that the evidence calculation can be repeated in any new extension of the SM, and consistently compared to the evidence computed in other models. If there is no cancellation of the quadratic divergences within that model then one should obtain a similar result as obtained in the toy example.  In supersymmetry the quadratic divergences are cancelled; however, soft-breaking introduces corrections of order $\msusy^2$, which may result in fine-tuning if $\msusy \gg \mz$.

In supersymmetry the quadratic divergences are cancelled; however,
soft-breaking introduces corrections of the order of the squared
soft masses, which may result in fine-tuning if these soft masses
are required to be substantially larger than $\mz$. For this reason
one should expect that in supersymmetric models a similar result
approximately holds, \ie
\begin{equation}\label{Eq:Evidence}
\pg{\log\mz^\text{exp}}{\SUSY} \approx
\pg{\log\mz^\text{exp},\log\mh^\text{exp}}{\SUSY}
\approx 
c \left(\frac{\mz^\text{exp}}{\msusy}\right)^2,
\end{equation}
where in this expression $\msusy$ characterizes the minimal size of
the soft masses consistent with the likelihood and chosen priors. We
now explicitly repeat our calculations in two supersymmetric models
to see whether this is the case.

\section{Supersymmetric models}\label{Sec:SUSY}

\subsection{Models}
We consider two models: a semi-constrained \NMSSM and the 
constrained \MSSM (\CMSSM). The models are tractable examples of a
minimal supersymmetric model and a singlet extension that we 
investigate with Bayesian statistics.

\subsubsection{Semi-constrained NMSSM}
The \NMSSM solves the $\mu$-problem\cite{Kim:1983dt} of the \MSSM
by replacing the \MSSM superpotential term $\mu \hat{H}_d \cdot
\hat{H}_u$ by one of the form $\lambda \hat{S} \hat{H}_d \cdot
\hat{H}_u$, where $\hat{S}$ is a new gauge singlet
superfield.\footnote{We use the notation $\hat{A} \cdot \hat{B}
\equiv \epsilon_{\alpha\beta} \hat{A}^\alpha \hat{B}^\beta = \hat{A}^2
\hat{B}^1 - \hat{A}^1 \hat{B}^2$ to denote a contraction between
$SU(2)_L$ doublets.} An effective $\mu$-term, given by
\begin{equation} \label{eq:nmssm-mu-eff}
\mueff = \lambda \vev{S},
\end{equation}
is then generated when the scalar component $S$ of this singlet
superfield develops a vacuum expectation value (VEV), $\vev{S}$. 
The most general renormalizable superpotential of the \NMSSM should
also contain additional terms beyond those found in the \MSSM
involving the singlet $\hat{S}$. Here we restrict our attention to
the \Z{3}-conserving \NMSSM \see{Maniatis:2009re,Ellwanger:2009dp},
for which the full superpotential is
\begin{align}
\W{\NMSSM} &= y^u_{ij} \hat{u}^c_i \hat{H}_u \cdot \hat{Q}_j 
  + y^d_{ij} \hat{d}^c_i \hat{Q}_j \cdot \hat{H}_d 
  + y^e_{ij} \hat{e}^c_i \hat{L}_j \cdot \hat{H}_d 
  +\lambda \hat{S} \hat{H}_d \cdot \hat{H}_u
  + \frac{1}{3} \kappa \hat{S}^3 
  \label{eq:z3-nmssm-superpotential} \\
&= \W{\MSSM}|_{\mu=0}  + \lambda \hat{S} \hat{H}_d \cdot 
  \hat{H}_u + \frac{1}{3} \kappa \hat{S}^3 . \nonumber
\end{align}
Here the notation $\W{\MSSM}|_{\mu = 0}$ refers to the usual \MSSM
superpotential, \ie
\begin{equation} \label{eq:mssm-superpotential}
\W{\MSSM} =  y^u_{ij} \hat{u}^c_i \hat{H}_u \cdot \hat{Q}_j 
  + y^d_{ij} \hat{d}^c_i \hat{Q}_j \cdot \hat{H}_d 
  + y^e_{ij} \hat{e}^c_i \hat{L}_j \cdot \hat{H}_d
  + \mu \hat{H}_d \cdot \hat{H}_u ,
\end{equation}
evaluated with $\mu = 0$. The cubic singlet coupling $\kappa$ is
required to explicitly break a global $U(1)$ Peccei-Quinn symmetry,
which would otherwise give rise to a massless axion when it is
spontaneously broken by the scalar field $S$ acquiring a VEV.

As usual in phenomenological SUSY models, in the \NMSSM SUSY is softly broken by a set of explicit soft terms,
\begin{equation} \label{eq:general-soft-breaking}
\L{soft}^{\text{\NMSSM}} = \L{soft-scalar} + \L{soft-gaugino} +
  \L{soft-trilinear} ,
\end{equation}
where the soft scalar masses, gaugino masses and soft trilinear
terms are taken to be
\begin{align}
-\L{soft-scalar} &= m^2_{S} |S|^2 + m_{H_u}^2 |H_u|^2 
  + m_{H_d}^2 |H_d|^2 + m_{Q_{ij}}^2 \tilde{Q}^\dagger_i \tilde{Q}_j
  \nonumber \\
& \quad {} + m_{u^c_{ij}}^2 \tilde{u}^{c\, \dagger}_i  \tilde{u}^c_j
  + m_{d^c_{ij}}^2 \tilde{d}^{c \, \dagger}_i \tilde{d}^c_j
  + m_{L_{ij}}^2 \tilde{L}^\dagger_i \tilde{L}_j 
  + m_{e^c_{ij}}^2 \tilde{e}^{c \, \dagger}_i \tilde{e}^c_j ,
  \label{eq:z3-nmssm-soft-scalar-masses} \\
-\L{soft-gaugino} &= \frac{1}{2} ( M_1 \tilde{B} \tilde{B} 
 + M_2 \tilde{W} \tilde{W} + M_3 \tilde{g}\tilde{g} + \hc ) ,
 \label{eq:z3-nmssm-soft-gaugino-masses} \\
-\L{soft-trilinear} &= a^u_{ij} \tilde{u}^c_i H_{u} \cdot
  \tilde{Q}_j + a^d_{ij}\tilde{d}^c_i \tilde{Q}_j \cdot H_d 
  + a^e_{ij} \tilde{e}^c_i \tilde{L}_j \cdot H_d \nonumber \\
& \quad {} + a_\lambda S H_d \cdot H_u + \frac{1}{3} a_\kappa S^3 
+ \hc ,
  \label{eq:z3-nmssm-soft-trilinears}
\end{align} 
respectively. To construct the semi-constrained NMSSM that we consider
here, the above soft parameters are assumed to satisfy a set of
relationships at the grand unification (GUT) scale \mgut motivated by
those  found in minimal supergravity 
(mSUGRA)\cite{Chamseddine:1982jx,Arnowitt:1992aq}. These GUT scale
boundary conditions are as follows:
\begin{itemize}
\item The soft-breaking trilinears are parameterized by
\begin{equation} 
\begin{gathered}
a^u_{ij}(\mgut) \equiv y^u_{ij}(\mgut) A^u_{ij} , \, 
a^d_{ij}(\mgut) \equiv y^d_{ij}(\mgut) A^d_{ij} , \,  
a^e_{ij}(\mgut) \equiv y^e_{ij}(\mgut) A^e_{ij} , \\  
a_\lambda(\mgut) \equiv \lambda(\mgut) A_\lambda , \text{ and } 
a_\kappa(\mgut) \equiv \kappa(\mgut) A_\kappa ,
\end{gathered} \label{eq:gut-scale-trilinears}
\end{equation}
where the reduced trilinear couplings are partially unified
at the GUT scale, that is,
\begin{equation} \label{eq:azero-definition}
A^u_{ij} = A^d_{ij} = A^e_{ij} \equiv \azero \delta_{ij},
\end{equation}
while $A_\lambda$ and $A_\kappa$ are allowed to vary separately.
\item The soft-breaking scalar masses are partially unified at $\mgut$,
\begin{equation}
\begin{gathered}
m^2_{Q_{ij}}(\mgut) = m^2_{L_{ij}}(\mgut) = m^2_{u^c_{ij}}(\mgut) = m^2_{d^c_{ij}}(\mgut) = m^2_{e^c_{ij}}(\mgut) \equiv \mzero^2 
\delta_{ij} , \\
m_{H_d}^2(\mgut) = m_{H_u}^2(\mgut) \equiv \mzero^2 .
\end{gathered} \label{eq:mzero-definition}
\end{equation}
The exception is the soft-breaking scalar mass for the singlet,
$m_S^2(\mgut) \equiv m_{S_0}^2$, which is taken to be free.
\item The soft-breaking gaugino masses are unified at the GUT scale,
\begin{equation} \label{eq:mhalf-definition}
M_1(\mgut) = M_2(\mgut) = M_3(\mgut) \equiv \mhalf .
\end{equation}
\end{itemize}
In addition to the GUT scale values of the soft parameters,
the values of the Yukawa couplings $\lambda$ and $\kappa$ at the
GUT scale, $\lambda(\mgut) \equiv \lambda_0$ and $\kappa(\mgut)
\equiv \kappa_0$, must also be specified. This semi-constrained 
model is therefore described by the nine GUT scale parameters
\begin{equation}
\{\azero, A_\lambda, A_\kappa, \mzero, m_{S_0}^2, \mhalf, \lambda_0, \kappa_0, \text{sign}\,\mueff\} .
\end{equation}
In the MSSM, the effects of $A$ terms are absorbed into the RG
evolution of the soft terms such that the VEVs have no explicit
dependence on them. On the other hand, the VEVs depend on $A$ terms
directly in the NMSSM. It is, therefore, important to have flexible
constraints on $A$ terms in the semi-constrained NMSSM. 
 
Note that, depending on the literature, the semi-constrained NMSSM is
defined by slightly different assumptions. A more strict convention
allows only the singlet specific parameters to be unconstrained such
that $\alambda=\azero$ is implied at the GUT
scale\cite{Ellwanger:2011mu}, while the more flexible version lets
non-universal Higgs masses be free parameters as well as
\alambda\cite{AbdusSalam:2011fc}.

Hereafter, NMSSM is used to simply denote the semi-constrained NMSSM,
if there is no special reason to distinguish it from the general
NMSSM.

\subsubsection{\CMSSM}
For comparison purposes we use the
\CMSSM\cite{Chamseddine:1982jx,Arnowitt:1992aq,Kane:1993td}, 
one of the most-studied supersymmetric models. In the
parameterization that we consider, the model can be characterized
by five parameters at the GUT scale. These are a common scalar mass,
\mzero, a common gaugino mass, \mhalf, a common trilinear, \azero,
the GUT scale value of the $\mu$ parameter appearing in
\refeq{eq:mssm-superpotential}, $\mu_0 \equiv \mu(\mgut)$, and the
GUT scale value of the corresponding soft-breaking bilinear
coupling, $B_0 \mu_0 \equiv B\mu(\mgut)$, where 
$\L{soft}^{\text{\MSSM}} \supset B\mu (H_d \cdot H_u + \hc)$. The
unified soft parameters \mzero, \mhalf, and \azero have analogous
definitions to those used in the semi-constrained \NMSSM; that is, 
in the \CMSSM the boundary conditions \refeq{eq:azero-definition}, 
\refeq{eq:mzero-definition}, and \refeq{eq:mhalf-definition}
are assumed to hold.

\subsection{Likelihood and priors}

We include Particle Data Group (PDG)
world-averages\cite{Agashe:2014kda} of measurements of the Higgs and
$Z$-boson masses in \reftable{Table:Like} in our likelihood function.
Under integration, we approximate the Gaussian likelihood function for
the $Z$-boson mass by a Dirac $\delta$-function, as in
\refeq{eq:dirac_approx}. We added in quadrature a $1\gev$ theoretical
uncertainty in the calculation of the SM-like Higgs boson mass by
\softsusy\cite{Allanach:2001kg,Allanach:2013kza}.

\begin{table}[htbp]
\centering
\begin{tabular}{cccc}
\toprule
Parameter & PDG & Theory error & Distribution\\
\midrule
\mz & $91.1876\pm0.0021\gev$\cite{Agashe:2014kda} & & Dirac\\
\mh & $125.09\pm0.24\gev$\cite{Agashe:2014kda} & $1\gev$ & Gaussian\\
\bottomrule
\end{tabular}
\caption{Likelihoods in our analysis for the Higgs and 
$Z$-boson masses. We added in quadrature a $1\gev$ theory error upon
the \softsusy calculation of \mh.}
\label{Table:Like}
\end{table}

Our chosen priors for the parameters of the \CMSSM and the 
semi-constrained \NMSSM are shown in \reftable{Table:Priors}. 
Because we are ignorant of the  soft-breaking mass scale, we pick
logarithmic priors, where possible, that equally weight every order
of magnitude.  Note that the trilinear couplings in both models,
$\azero$, $A_\lambda$, and $A_\kappa$, are allowed to take both
signs, and we use the piecewise prior
\begin{equation} \label{eq:piecewise-trilinear-prior}
p(A) \propto 
\begin{cases} 
    1 & |A| \leq 100 \gev ,\\
    \frac{100\gev}{|A|} & 100 \gev < |A| \leq 20 \tev , \\
    0 & |A| > 20\tev .\\
\end{cases}
\end{equation}
This choice corresponds to a logarithmic prior with special
treatment at $|A| \simeq 0$ such that the prior remains proper.

In addition to the relevant GUT scale parameters, the models share
nuisance parameters that are not of particular interest in
this analysis, but which could impact our results.
The most important nuisance parameters are the top quark mass, \mt, and the strong coupling,
\alphas. We pick Gaussian priors for them, with means and standard deviations determined by PDG
world-averages of experimental measurements\cite{Agashe:2014kda}, as shown in
\reftable{Table:Priors}. We fix other SM nuisance parameters, including the bottom mass
and weak coupling, to their measured values.



\mxnewcommand{\dash}{\text{, }}

\begin{table}[htbp]
\centering
\begin{tabular}{cc}
\toprule
Parameter & Distribution\\
\midrule
\CMSSM\\
\midrule
\mzero & Log, $1\gev\dash20\tev$\\
\mhalf & Log, $1\gev\dash15\tev$\\
\azero & Log for $100 \gev < |A| \le 20\tev$, Flat for $|A| \le 100 \gev$\\
$|\mu_0|$ & Log, $100\gev\dash20\tev$\\
$B_0\mu_0$ & Log, $(100\gev)^2\dash(20\tev)^2$\\
\sgnmu & $\pm1$ with equal probability\\
\midrule
\NMSSM\\
\midrule
$\lambda_0$ & Log, $10^{-6}\dash1$\\
$\kappa_0$ & Log for $10^{-10}< |\kappa| < 1$\\
$m_{S_0}$ & Same as $\mzero$\\
$A_\lambda$ & Same as \azero\\
$A_\kappa$ & Same as \azero\\
\midrule
SM\\
\midrule
\alphas   & Gaussian, $0.1185\pm0.0006$\cite{Agashe:2014kda}\\
\mt       & Gaussian, $173.34\pm0.76\gev$\cite{Agashe:2014kda}\\
\bottomrule
\end{tabular}
\caption{Priors for the \CMSSM and semi-constrained \NMSSM model
parameters.  In the \CMSSM, $|\mu_0|$ is marginalized in accordance
with $\mzpole$, while in the \NMSSM $\msinglet$ is marginalized as
$|\mu_0|$ via $\vev{S}$, as described in \refsec{Sec:EffectivePrior}.
The full set of parameters in our scan of the \CMSSM includes the SM
parameters, and the same priors are used in the \NMSSM for those
parameters that are shared with the \CMSSM. In the case of the \NMSSM,
the parameters $|\mu_0|$ and $B_0\mu_0$ are absent and instead we
specify priors for $\lambda_0$ and $\kappa_0$.}
\label{Table:Priors}
\end{table}

\subsubsection{Effective naturalness priors}
\label{Sec:EffectivePrior}

As in the SM in \refeq{eq:sm-effective-prior}, from these initial
priors we find effective priors in the \CMSSM and \NMSSM in which
one of the GUT scale parameters is fixed to reproduce the observed
value of $\mzpole^2$. This corresponds quite closely
to the approach taken in spectrum generators for the \MSSM and \NMSSM,
such as \softsusy, where some
of the presumed fundamental parameters are traded for
phenomenological parameters at the weak scale. The models can then 
be parameterized in terms of the remaining GUT scale parameters and
a set of precisely known electroweak (EW) parameters. It should be
noted, however, that this is equivalent to working directly in terms
of the GUT scale parameters and marginalizing with the chosen EW
observables using a $\delta$-function likelihood. This provides an
economic way to survey  the entirety of parameter space, discarding
points that lead to hardly justifiable low-energy spectra.

In the \MSSM, the effective priors arise from making the conventional
trade
\begin{equation} \label{eq:mssm-change-of-variables}
\{ |\mu_0| , \, B_0 \mu_0 , \, \sgnmu ,\dots \}
\to \{ \mzpole^2, \, \tanb, \, \sgnmu , \dots \} ,
\end{equation}
where as usual $\tanb \equiv v_2 / v_1$ is defined as the ratio of
the two Higgs VEVs,
\begin{equation} \label{eq:higgs-vev-convention}
\vev{H_d^0} = \frac{v_1}{\sqrt{2}} , \quad 
\vev{H_u^0} = \frac{v_2}{\sqrt{2}} .
\end{equation}
In practice, we achieve this trade in two steps. First, the GUT
scale parameters are evolved to $\msusy \equiv \sqrt{ m_{\tilde t_1}
m_{\tilde t_2}}$, the scale of EW symmetry breaking (EWSB), 
using two-loop renormalization group equations (RGEs). The
EWSB conditions,
\begin{align}
\frac{1}{2} \mzpole^2 &= -\mu^2 + \frac{\bar{m}^2_{H_d} 
  - \bar{m}^2_{H_u} \tan^2{\beta}}{\tan^2{\beta} - 1} 
  - \frac{1}{2} \re \Pi_{ZZ}^T , \label{eq:mssm-mz-prediction} \\
\sin 2\beta &=  \frac{2 B\mu}{\bar{m}_{H_u}^2 + \bar{m}_{H_d}^2 
  + 2 \mu^2} , \label{eq:mssm-beta-prediction}
\end{align}
can then be used to exchange the low-energy values of $\mu$ and
$B\mu$ for $\mzpole^2$ and $\tanb$. In these expressions, the one-
and two-loop corrections from the  Coleman-Weinberg 
potential\cite{Coleman:1973jx} have been absorbed
into the quantities $\bar{m}_{H_{d,u}}^2$, and $\re \Pi_{ZZ}^T$ is
the transverse part of the $Z$-boson self-energy. Since the EWSB
conditions cannot fix the phase of the $\mu$-parameter, \sgnmu is 
an additional parameter. This trade is convenient, since we may now
input the measured $Z$-boson and fermion masses, the latter being
related to their Yukawa couplings via \tanb.

The priors for the two choices of parameter sets are related by the
Jacobian, $\jac^{\text{\text{\CMSSM}}}$, associated with this change
of variables,
\begin{align}  
\pg{|\mu_0|, \sgnmu, B_0\mu_0, \dots}{\text{\text{\CMSSM}}} &\equiv
\pg{\mu_0, B_0\mu_0, \dots}{\text{\text{\CMSSM}}} \nonumber \\
&= \jac^{\text{\text{\CMSSM}}} \cdot \pg{\mzpole^2, \tanb,
\dots}{\text{\text{\CMSSM}}} ,
\label{eq:cmssm-priors-change-of-variables}
\end{align}
where $\jac^{\text{\text{\CMSSM}}}$ is given by
$\jac^{\text{\text{\CMSSM}}} = |\det J^{\text{\text{\CMSSM}}}|$,
$J^{\text{\text{\CMSSM}}}$ being the appropriate Jacobian matrix.
Here we treat the RG evolution from $\mgut$ to $\msusy$ and the
subsequent solution of the EWSB conditions as two consecutive
changes of variables, so that $\jac^{\text{\text{\CMSSM}}}$  may be
written as a product of the Jacobian determinant associated with
each, \ie
\begin{equation} \label{eq:generic-jacobian-structure}
\jac^M = \jac^M_{\msusy} \jac^M_{\mgut} ,
\end{equation}
where $M = \text{\text{\CMSSM}}, \text{\CNMSSM}$ denotes the
particular model under consideration. The forms of $\jac^M_{\msusy}$
and $\jac^M_{\mgut}$ are given in \refapp{App:Jacobian}.

The effective prior results from conditioning on the measurement
of $\mzpole$ and then marginalizing over $\mzpole$. This yields
\begin{align}
p_{\text{eff.}}(\tanb, \dots) &= \int \pg{\mzpole^2, \tanb, \dots}
{\text{\text{\CMSSM}}, \mzpole^\text{exp}} \dif \mzpole^2
\label{eq:cmssm-effective-prior-defn} \\
&\equiv \frac{1}{\ev} \int \delta((\mzpole^\text{exp})^2 - \mzpole^2) \cdot
\pg{\mzpole^2, \tanb, \dots}{\text{\text{\CMSSM}}} \dif \mzpole^2 \nonumber \\
&= \frac{1}{\ev} \pg{(\mzpole^\text{exp})^2, \tanb, \dots}{\text{\text{\CMSSM}}} \nonumber \\
&= \frac{1}{\ev} \left . \frac{1}{\jac^{\text{\text{\CMSSM}}}} \right
|_{\mzpole^\text{exp}} \pg{\mu_Z, B_Z\mu_Z, \dots}{\text{\text{\CMSSM}}} ,
\end{align}
where $\mu_Z = \mu_0(\mzpole = \mzpole^\text{exp})$ and $B_Z \mu_Z =
B_0\mu_0(\mzpole = \mzpole^\text{exp})$ are the values of the high-scale
parameters that result for $\mzpole = \mzpole^\text{exp}$, for the given
value of $\tanb$ and all other model parameters. The form of 
the effective prior is identical to that in the SM in
\refeq{eq:sm-effective-prior}. It is worth noting that we do not
develop any nontrivial, or misleading, behavior in the effective
prior according to our choice of EW parameters, $\{\mzpole^2, 
\tanb\}$. This choice is not unique; for instance, one could choose
the VEVs $\{v_1, v_2\}$ instead. In this case, the effective
prior would differ only by the additional non-singular Jacobian 
factor\footnote{In general, additional terms involving derivatives
of the $Z$-boson self-energy are also present, but these are
numerically small and may be neglected here.}
\begin{equation} \label{eq:vevs-jacobian}
\left | \det \frac{\partial (v_1, v_2)}{\partial (\mzpole^2,
\tanb)} \right | \approx \frac{v_1^2}{2 \mzrun^2} ,
\end{equation}
where $\mzrun$ is the tree-level $Z$-boson mass.

In the \CNMSSM, the imposed \Z{3} symmetry forbids an explicit
superpotential bilinear term for $\hat{H}_d$ and $\hat{H}_u$, 
along with the corresponding soft-breaking parameter. We instead
make the trade
\begin{equation} \label{eq:snmssm-change-of-variables}
\{ \lambda_0, \kappa_0, m_{S_0}^2 \} \to
\{ \lambda, \mzpole^2, \tanb \} . 
\end{equation}
After the intermediate step of exchanging GUT scale parameters
for their low-energy counterparts through RG running, we can make
use of the three \NMSSM EWSB conditions to obtain the new set of
input parameters. Note that exchanging $\lambda_0$ for $\lambda$
is achieved solely by integrating the RGEs, so that at the EWSB
scale we need only trade $\{\kappa, m_S^2\} \to \{ \mzpole^2,
\tanb\}$.

To do so, we first use the \MSSM-like EWSB condition
\begin{equation} \label{eq:nmssm-mz-prediction}
\frac{1}{2} \mzpole^2 = - \mueff^2 
+ \frac{\bar{m}^2_{H_d} - \bar{m}^2_{H_u} \tan^2{\beta}}
{\tan^2{\beta} - 1} - \frac{1}{2} \re \Pi_{ZZ}^T ,
\end{equation}
where the effective $\mu$-parameter, $\mueff$, is
defined in \refeq{eq:nmssm-mu-eff},
to express the effective $\mu$-parameter
in terms of $\mzpole^2$ and $\tanb$.
Since in this approach we retain
$\lambda$ as a free input parameter, this has the effect of
determining the singlet VEV, $\vev{S} \equiv s / \sqrt{2}$, as a function of $\mzpole^2$ and
$\tanb$.

Second, we trade $s$ for $m_S^2$ via the EWSB condition,
\begin{equation} \label{eq:nmssm-ms2-equation}
\bar{m}_S^2 = -\kappa^2 s^2 - \frac{1}{2}\lambda^2 v^2
- \frac{a_\kappa s}{\sqrt{2}} + \frac{v^2}{2 s}\sin 2\beta
\left ( \frac{a_\lambda}{\sqrt{2}} + \lambda \kappa s \right ),
\end{equation}
where we make the usual definition $v^2 = v_1^2 + v_2^2$, and have
absorbed the loop-corrections from the Coleman-Weinberg potential
into $\bar{m}_S^2$. Finally, we make \tanb an input parameter by
trading $\kappa$ for \tanb via the second \MSSM-like EWSB condition,
\begin{equation} \label{eq:nmssm-beta-prediction}
\sin 2\beta = \frac{2 B_\text{eff.}\mueff}
{\bar{m}_{H_u}^2 + \bar{m}_{H_d}^2 + 2 \mueff^2
+ \frac{\lambda^2 v^2}{2}},
\end{equation}
where we define an effective soft-breaking bilinear
\begin{equation}
B_\text{eff.} \mueff \equiv \frac{s}{\sqrt{2}}
\left ( a_\lambda + \frac{\lambda \kappa s}{\sqrt{2}} \right ) .
\end{equation}
Thus, ultimately, in our analysis $m_S^2$ plays the role of $\mu^2$
via an effective $\mu$-term and $\kappa$ plays the role of $B\mu$
via an effective $B\mu$-term. The final effective prior is defined
in the same way as in the \CMSSM, \ie it has the form
\begin{equation}
p_{\text{eff.}}(\tanb, \lambda, \dots) = \frac{1}{\ev} \left . \frac{1}{\jac^{\text{\CNMSSM}}} \right
|_{\mzpole^\text{exp}} \pg{\lambda_Z, \kappa_Z, m_{S_Z}^2 \dots}{\text{\CNMSSM}} .
\end{equation}

The effective priors automatically disfavor fine-tuned regions of
parameter space. Indeed, from their explicit forms in
\refapp{App:Jacobian}, we see that the effective priors favor RG
evolution that results in weak-scale parameters similar in magnitude
to the weak scale. The region of parameter space in which this
occurs is known as the ``focus point''\cite{Chan:1997bi,
Feng:1999hg,Feng:1999mn}. In these regions of parameter space,
the RG evolution of the soft masses is such that, at the SUSY scale,
$\mhu^2 \sim \mzpole^2$ almost independently of the initial
value of $\mhu^2(\mgut) = \mzero^2$. In the \CMSSM, the dependence
of $\mhu^2(\msusy)$ on the universal soft-breaking masses can be
quantified using semi-analytic solutions to the RGEs, which take
the form
\begin{align}
m^2(\msusy) &= c^{m^2}_{\mzero^2}(\msusy) \mzero^2
+ c^{m^2}_{\mhalf^2}(\msusy) \mhalf^2 
+ c^{m^2}_{\mhalf \azero}(\msusy) \mhalf \azero \nonumber \\
& \quad {} + c^{m^2}_{\azero^2}(\msusy) \azero^2 ,
\label{eq:cmssm-semi-analytic-soft-mass}
\end{align}
for $m^2 = \mhu^2, \mhd^2$ and  where the coefficients $c^i_j(Q)$
depend only on the gauge and Yukawa couplings. In the 
semi-constrained \NMSSM the semi-analytic solutions instead take the
form
\begin{align}
m^2(\msusy) &= c^{m^2}_{\mzero^2}(\msusy) \mzero^2
+ c^{m^2}_{m_{S_0}^2}(\msusy) m_{S_0}^2
+ c^{m^2}_{\mhalf^2}(\msusy) \mhalf^2 \nonumber \\
& \quad {} + c^{m^2}_{\mhalf \azero}(\msusy) \mhalf \azero
+ c^{m^2}_{\mhalf A_\lambda}(\msusy) \mhalf A_\lambda \nonumber \\
& \quad {} + c^{m^2}_{\mhalf A_\kappa}(\msusy) \mhalf A_\kappa
+ c^{m^2}_{\azero A_\lambda}(\msusy) \azero A_\lambda
+ c^{m^2}_{\azero A_\kappa}(\msusy) \azero A_\kappa \nonumber \\
& \quad {} + c^{m^2}_{A_\lambda A_\kappa}(\msusy) A_\lambda A_\kappa
+ c^{m^2}_{\azero^2}(\msusy) \azero^2
+ c^{m^2}_{A_\lambda^2}(\msusy) A_\lambda^2 \nonumber \\
& \quad {} + c^{m^2}_{A_\kappa^2}(\msusy) A_\kappa^2
\label{eq:snmssm-semi-analytic-soft-mass}
\end{align}
for $m^2 = \mhu^2, \mhd^2, \msinglet^2$.

\subsection{Comparison to other fine-tuning measures}
\label{Subsec:Measures}
As discussed in \refsec{Sec:Bayes} and \refsec{Sec:SM}, Bayesian
methods automatically incorporate some of the common intuitions
relating to fine-tuning. It is therefore useful to compare the
results obtained in the Bayesian approach to other measures of
tuning. The traditional sensitivity measure defined in
\refeq{Eq:DelBG} is one example. In addition to the ambiguities
related to this measure discussed in \refsec{Sec:SM}, there is also
no agreement as to how a collection of sensitivities $\{\Delta_p\}$
should be calculated or combined to produce a tuning measure. For
instance, it is not necessarily clear whether the sensitivities
should be summed, profiled, added in quadrature, or combined in some
other way, nor is there agreement on the renormalization scale of
the parameters with respect to which we differentiate \mz. For our
purposes, we define
\begin{equation} \label{eq:bg-tuning-defn}
\Delta_\text{BG} \equiv \max_p \Delta_p \big|_{\mgut} ,
\end{equation}
where $\Delta_p$ is defined as in \refeq{Eq:DelBG} and the notation
$\Delta_p \big |_{Q}$ indicates that the parameters to differentiate
with respect to are those defined at the scale $Q$; here, this is
taken to be $Q = \mgut$, the scale of gauge coupling unification. In the \CMSSM, we pick the maximum from the measures for the parameters
$\{\azero, \mzero, \mhalf, \mu_0, B_0\mu_0\}$.
In the \CNMSSM, on the other hand, we consider
$\{\azero, A_\lambda, A_\kappa, \mzero, m_{S_0}^2, \mhalf, \lambda_0, \kappa_0\}$.

The BGEN measure defined in \refeq{eq:bg-tuning-defn} identifies
fine-tuning with sensitivity to small parameter variations.
Alternative naturalness measures have been proposed that instead seek
to quantify the size of any cancellations that must take place to
reproduce the observed EW scale. An example of this class of measures
is the so-called electroweak fine-tuning\cite{Baer:2012up}, which is
defined as
\begin{equation} \label{eq:ew-tuning-defn}
\Delta_{\text{EW}} \equiv \max_i \frac{|C_i|}{\mzrun^2 / 2} .
\end{equation}
The $C_i$ are the terms appearing in the expression for $\mzrun^2$ in
the model (see \refeq{eq:mssm-mz-prediction} and \refeq{eq:nmssm-mz-prediction}), evaluated at the renormalization scale $Q = \msusy$.
The expressions for the $C_i$ appropriate to each of the \CMSSM and
\CNMSSM are given in \refapp{App:ew-tuning}.

The measures in \refeq{eq:bg-tuning-defn} and
\refeq{eq:ew-tuning-defn} are pointwise measures that can be
compared with the (marginalized) posterior densities obtained in a
complete Bayesian analysis. Evaluating the latter in general
involves calculating non-trivial evidence integrals over the full
model parameter space. However, even without carrying out the full
computation, it can be seen that doing so nevertheless involves 
a simple pointwise tuning measure involving the Jacobian for the
change of variables from parameters to observables. Regions of
parameter space for which this quantity is large are penalized by a
factor of $1 / \jac$ in evidence integrals; that is, the effective
prior in such regions is suppressed. This motivates the definition of
the tuning measures\cite{Kim:2013uxa}
\begin{align}
\Delta_{\jac} \big |_{\mgut} &\equiv \left | \det \frac{\partial
\ln O_i}{\partial \ln p_j(\mgut)} \right | = \left | 
\frac{\prod_j p_j(\mgut)}{\prod_i O_i}\right | \jac^M ,
\label{eq:gut} \\
\Delta_{\jac} \big |_{\msusy} &\equiv \left | \det \frac{\partial
\ln O_i}{\partial \ln p_j(\msusy)} \right | = \left | 
\frac{\prod_j p_j(\msusy)}{\prod_i O_i}\right | \jac^M_{\msusy} .
\label{eq:susy}
\end{align}
The set $\{O_i\}$ contains the observables for which the
parameters $p_j$ are traded in each model, \ie $\{\mzpole^2, \tanb\}$
in the \CMSSM and $\{\lambda, \mzpole^2, \tanb\}$ in the \CNMSSM.
The measures in \refeq{eq:gut} and \refeq{eq:susy} differ in the
scales at which the parameters $p_j$ are defined. The first
involves the parameters $p_j$ defined at $\mgut$, namely $\{\mu_0,
B_0\mu_0\}$ in the \CMSSM and $\{\lambda_0, \kappa_0, m_{S_0}^2\}$
in the \CNMSSM, and includes the effect of running from the GUT scale
to low energies. \refeq{eq:susy} only involves the trade from
SUSY scale parameters to observables, so that the set of $p_j$
is $\{\mu, B\mu\}$ in the \CMSSM and $\{\lambda, \kappa, m_S^2\}$
in the \CNMSSM. It can be seen from the expressions in
\refapp{App:Jacobian} that the Jacobian factors $\jac^M$ and
$\jac^M_{\msusy}$ resemble traditional\cite{deCarlos:1993rbr,
deCarlos:1993ca,Chankowski:1997zh,Agashe:1997kn,Wright:1998mk,
Kane:1998im, BasteroGil:1999gu,Feng:1999zg, Allanach:2000ii,
Dermisek:2005ar,Barbieri:2005kf,Allanach:2006jc, Gripaios:2006nn,
Dermisek:2006py, Barbieri:2006dq,Kobayashi:2006fh, Perelstein:2012qg,
Antusch:2012gv,Cheng:2012pe,CahillRowley:2012rv, Ross:2012nr,
Basak:2012bd,Boehm:2013gst}  and alternative fine-tuning
measures\cite{Anderson:1994dz,Anderson:1994tr, Anderson:1995cp,
Anderson:1996ew,Ciafaloni:1996zh, Chan:1997bi, Barbieri:1998uv,
Giusti:1998gz,Casas:2003jx, Casas:2004uu, Casas:2004gh,
Casas:2006bd,Kitano:2005wc, Athron:2007ry, Baer:2012up}.


In the following, we will compare the results obtained using the 
framework of Bayesian statistics with the fine-tuning measures
defined above, and illustrate how the former can encapsulate
traditional notions of naturalness. To compare parameter inference
with fine-tuning measures and Bayesian statistics, in
\refsec{Sec:Results} we compare ``heat-maps'' of fine-tuning measures
with posterior densities. Because the fine-tuning measures are not
densities, we must compare them in a particular parameterization, and
bear in mind that densities 
are not invariant
under
reparameterizations (whereas the transformation of a fine-tuning
measure is ambiguous) and that densities are dimensionful (whereas
fine-tuning measures are dimensionless).

To compare model selection with fine-tuning measures and Bayesian
statistics, in \refsec{Sec:Results} we compare Bayes factors with
ratios of fine-tuning measures. The fine-tuning measure in
supersymmetric models is roughly
\begin{equation}
\Delta  \sim \frac{\msusy^2}{\mz^2}.
\end{equation}
By comparing with \refeq{Eq:Evidence} we see that the evidence for a
supersymmetric model (written in terms of $\log\mz$) may be crudely
written as
\begin{equation}
\pg{\log\mz}{\SUSY} \sim \frac{1}{\Delta}.
\end{equation}
The parametric behavior for $\msusy^2 \gg \mz^2$ is identical. Thus,
in this case, there is reason to expect that fine-tuning measures and Bayes factors may result in similar conclusions.




\subsection{Numerical methods} \label{Sec:NumericalMethods}

We computed statistical quantities --- posterior densities and
evidences --- with
\multinestv\cite{Feroz:2013hea,Feroz:2007kg,Feroz:2008xx} and 
plotted them with \superplot\cite{Fowlie:2016hew}. For the evidence integration, we modified the convergence criteria by defining the tolerance using the average likelihood of the live points, instead of the maximum. We performed two scans for each model: one with only \mz, and one with \mz and \mh in the likelihood.
We scanned 10 million and 100 million points for each scan of the \CMSSM and \CNMSSM, respectively. To calculate
the likelihoods and effective priors in each model, we computed
the mass spectrum and Jacobian factors for each parameter
point using a modified version of \softsusyv.
As described in detail in \refapp{App:Jacobian}, the required
Jacobian is written as the product of the Jacobian determinants for
the change of variables from the GUT scale parameters to the
low-energy Lagrangian parameters, and for the transformation
from these parameters to the derived parameters $\mzpole^2$ and
$\tanb$, so that $\jac$ may be expressed as in
\refeq{eq:generic-jacobian-structure}. The particular derivatives
required for the construction of $\jac^M_{\msusy}$ and
$\jac^M_{\mgut}$ are given in \refapp{App:Jacobian}. We implemented
subroutines to evaluate these derivatives numerically in \softsusy.
In the case of $\jac^M_{\mgut}$, this is achieved by varying the
high-scale model parameters at $\mgut$ and calculating the resulting
values of the low-energy Lagrangian parameters using the two-loop
RGEs. In a similar fashion, to determine the derivatives appearing in
$\jac^M_{\msusy}$, we vary the low-energy Lagrangian parameters and
recalculate the predicted values of $\mzpole^2$ and $\tanb$. The
underlying changes in the VEVs are found by numerically solving the
EWSB conditions for $v_1$, $v_2$ and $s$ after perturbing the model
parameters. Two-loop RG evolution of all the model parameters such as
the soft-breaking gaugino masses, scalar masses and 
trilinear terms is applied for the entire calculation. One-loop
threshold corrections for the gauge and Yukawa couplings are
included.

For each point in our scan, we also computed the
measures of fine-tuning given in \refeq{eq:bg-tuning-defn},
\refeq{eq:ew-tuning-defn}, \refeq{eq:gut} and \refeq{eq:susy}. In
the \MSSM, we make use of the existing implementation of the 
BGEN measure provided by \softsusy to find
$\Delta_\text{BG}$. As analogous
routines are not yet provided for the \NMSSM in \softsusy, we also
implemented the necessary numerical calculation of 
$\Delta_\text{BG}$ in the \NMSSM in our modified code. Derivatives
of $\mzpole^2$ are obtained numerically by perturbing the GUT
scale model parameters and calculating the predicted $Z$-boson
pole mass, after running to the SUSY scale and solving the EWSB
conditions for the VEVs at two-loop order.

\section{Results and discussion}\label{Sec:Results}

\begin{figure}
    \centering
    \begin{subfigure}[t]{\figwidth}
        \includegraphics[height=0.975\textwidth]{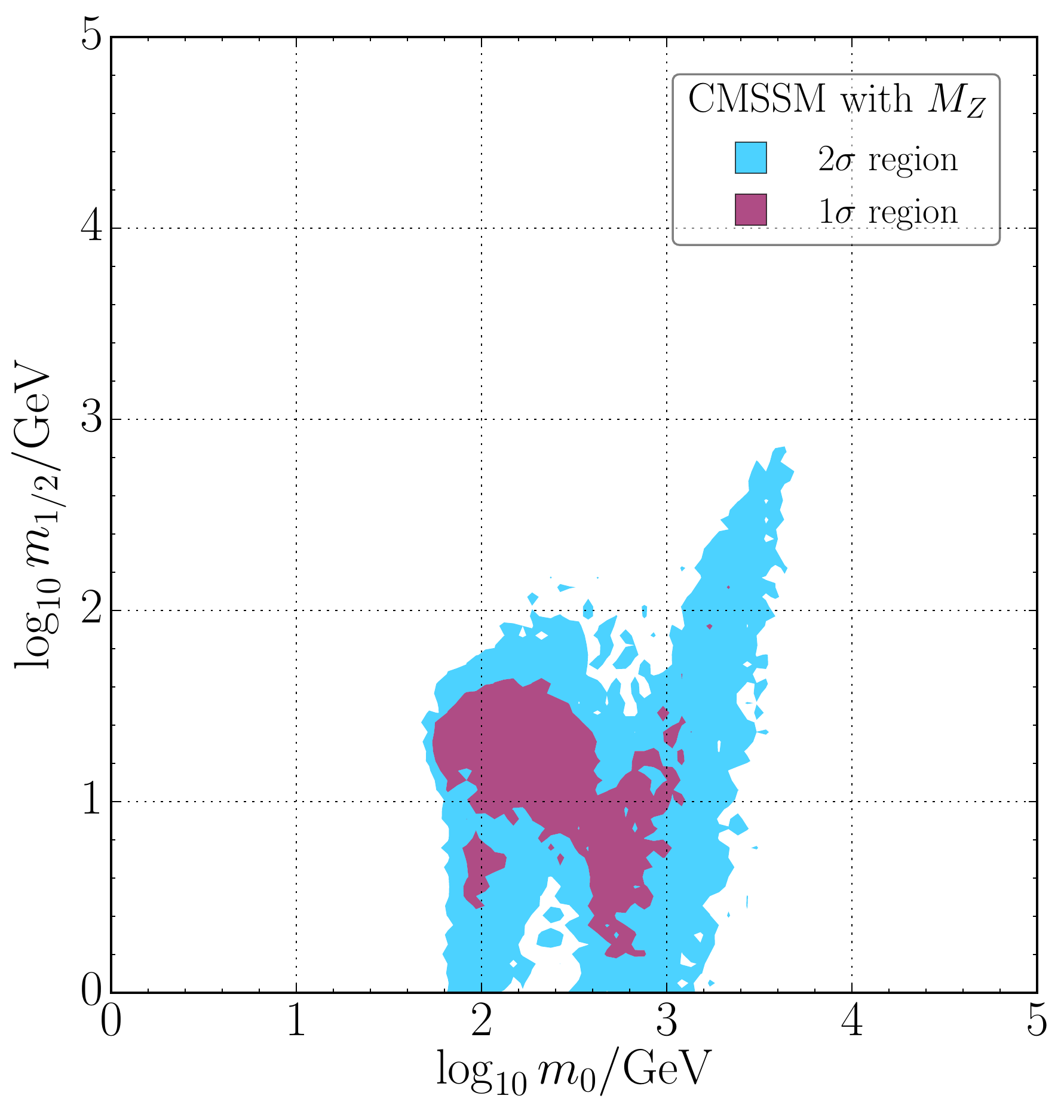}
        \caption{\CMSSM credible regions}
        \label{fig:CMSSM_pdf_mz_m0m12}
    \end{subfigure}
    \begin{subfigure}[t]{\figwidth}
        \includegraphics[height=0.975\textwidth]{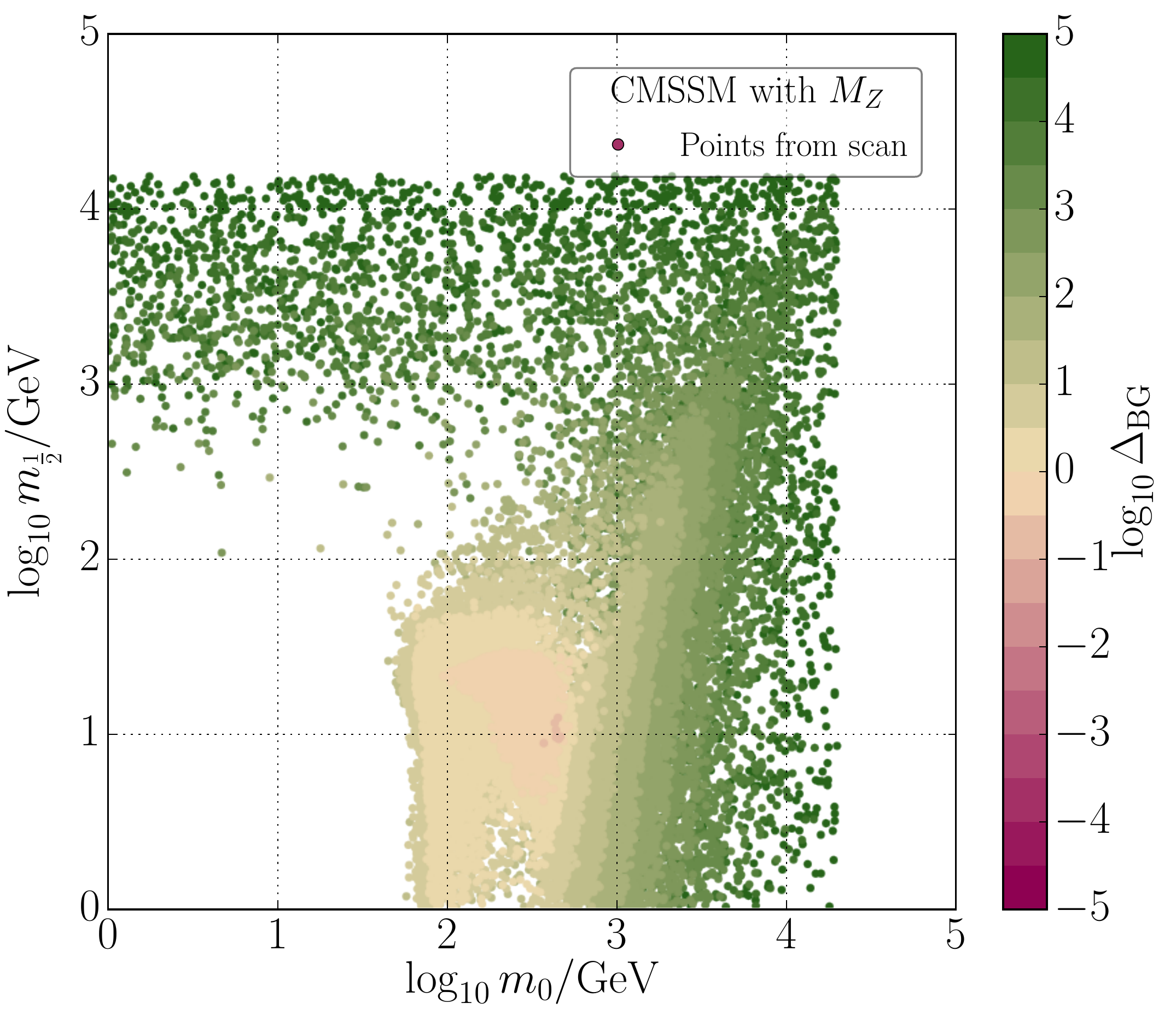}
        \caption{\CMSSM $\Delta_\text{BG}$}
        \label{fig:CMSSM_BG_m0m12}
    \end{subfigure}

    \begin{subfigure}[t]{\figwidth}
        \includegraphics[height=0.975\textwidth]{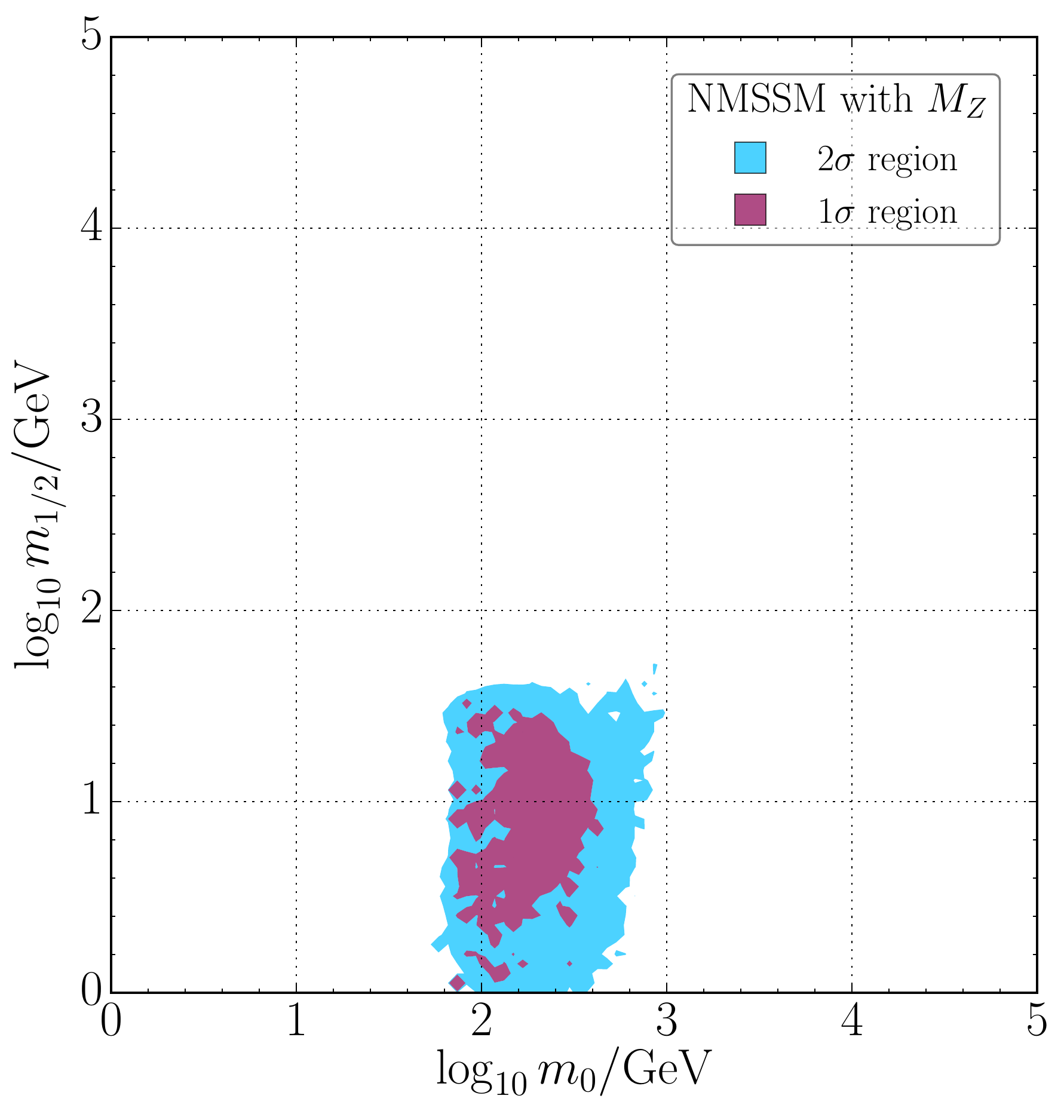}
        \caption{\CNMSSM credible regions}
        \label{fig:CNMSSM_pdf_mz_m0m12}
    \end{subfigure}
    \begin{subfigure}[t]{\figwidth}
        \includegraphics[height=0.975\textwidth]{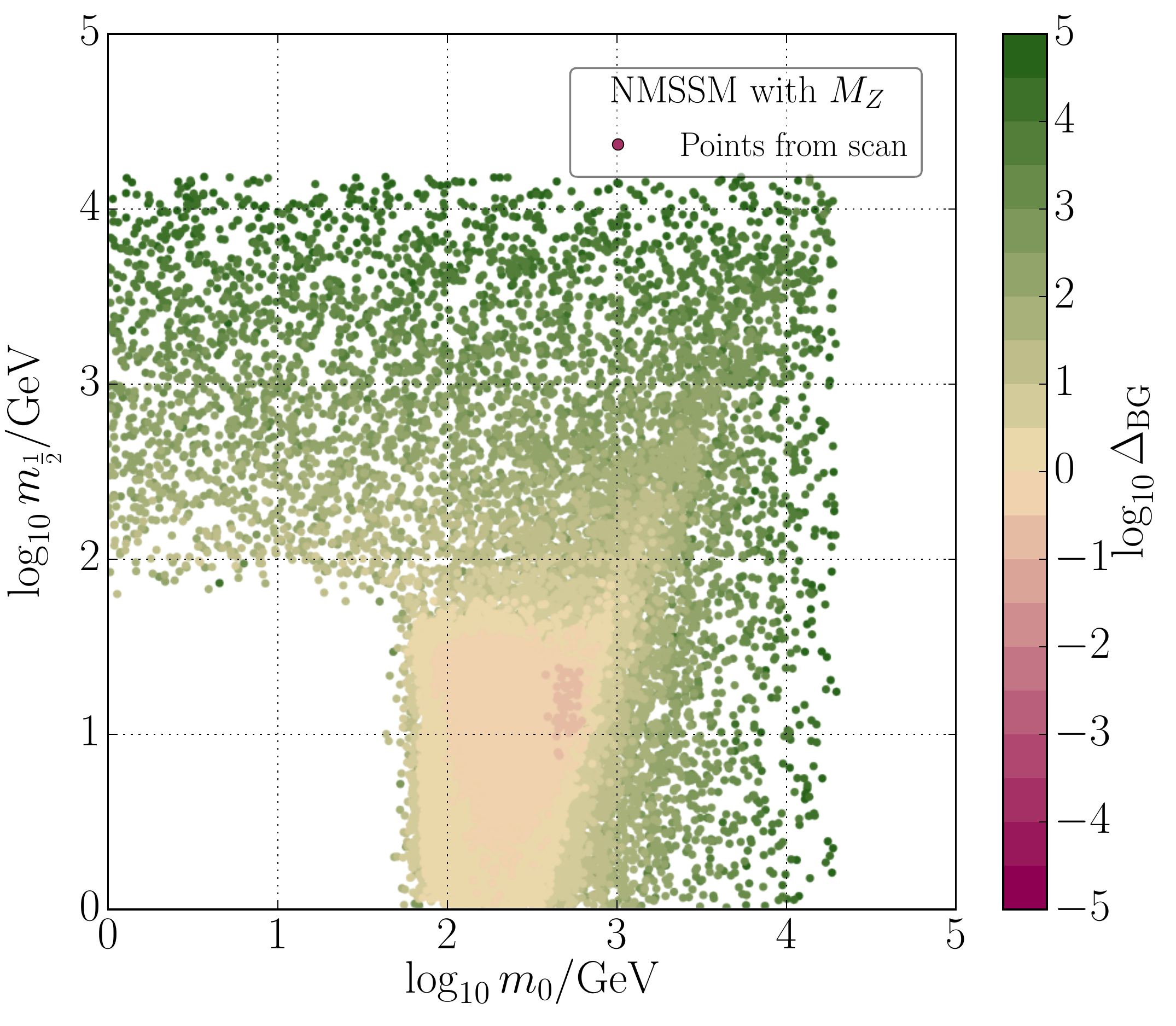}
        \caption{\CNMSSM $\Delta_\text{BG}$}
        \label{fig:CNMSSM_BG_m0m12}
    \end{subfigure}
    
    \caption{Credible regions of marginalized posterior probability density conditioned upon \mz (left) and profiled BGEN measure (right) for the \CMSSM (upper) and \CNMSSM (lower) on the (\mzero, \mhalf) plane.}
    \label{fig:mz_m0m12}
\end{figure}

In \reffig{fig:mz_m0m12}, we compare credible regions of the
marginalized posterior probability density conditioned upon \mz (left
frames) with the profiled BGEN fine-tuning measure (right frames) on
the (\mzero, \mhalf) planes of the \CMSSM (top frames) and \CNMSSM
(bottom frames).  On the posterior density plots we show the smallest $1\sigma$
(red) and $2\sigma$ (blue) credible regions, containing $68\%$ and
$95\%$ of the posterior mass respectively.  On the right frames different colors trace constant contours of the profiled BGEN measure.  According to the posterior plots, most probability density (that is, most of the low tuned area) lies in the weak scale valued \mzero and \mhalf region.  This not only confirms our qualitative expectations in \refeq{Eq:Evidence}, but also coincides with expectations for the scale of supersymmetry before the LHC operation.  As anticipated, the BGEN measure reflects the same expectations, agreeing fairly well with the trend shown by the posterior probability.  This is not a surprise considering that the dominant term in this measure appears in the posterior after trading the $\mu$ parameter to the $Z$ mass.  While most of the low tuned area lies in the bulk region, which was eliminated by the LHC, parts of the focus point also feature low tuning and are still experimentally feasible.  Low tuning in the focus point is prominently highlighted by the $2\sigma$ credible region of the posterior density, and supported by the BGEN measure.

The scatter plots in \reffig{fig:mz_m0m12}, and all other scatter plots, show points with appreciable posterior weight. The density of points results from the posterior density and the nested sampling algorithm. The \CMSSM and \CNMSSM (\mzero, \mhalf) planes feature a region with no points at $\mzero \lesssim 100 \gev$ and $\mhalf \lesssim 100 \gev$. The \CMSSM, furthermore, shows few points at $\mhalf \simeq 0$ and $\mzero \simeq 250\gev$. Such regions are disallowed as they fail to realize a physically sensible EWSB vacuum. This problem is particularly prevalent for large \azero and \tanb. Such regions were, moreover, ruled out prior to the LHC by LEP searches for supersymmetric particles and for the Higgs boson. 



\begin{figure}
    \centering
    \begin{subfigure}[t]{\figwidth}
        \includegraphics[width=\textwidth]{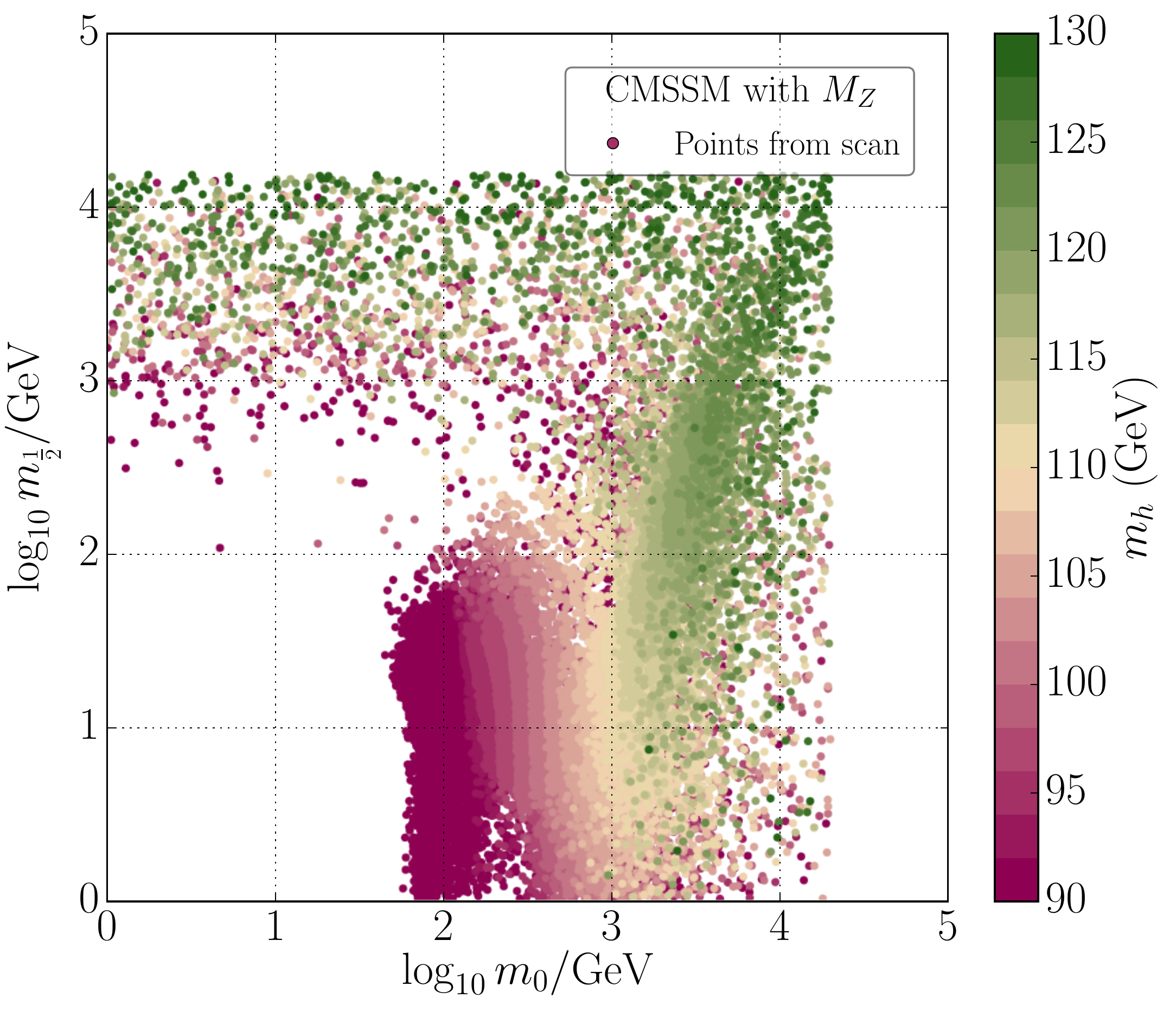}
        \caption{\CMSSM}
        \label{fig:CMSSM_mh}
    \end{subfigure}
    \begin{subfigure}[t]{\figwidth}
        \includegraphics[width=\textwidth]{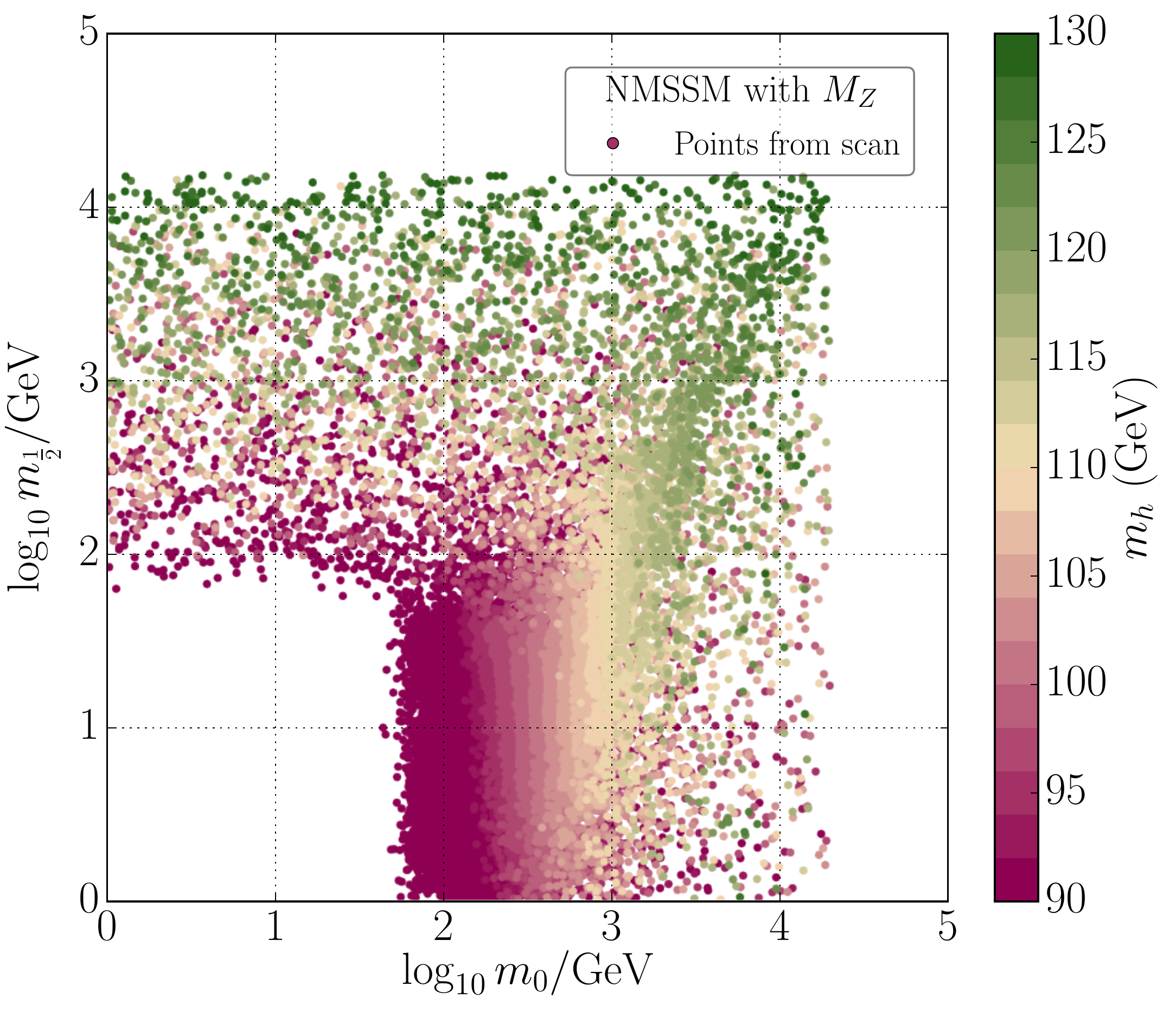}
        \caption{\CNMSSM}
        \label{fig:CNMSSM_mh}
    \end{subfigure}
    \caption{The lightest Higgs mass on the (\mzero, \mhalf) plane for the \CMSSM (left) and \CNMSSM (right).}
    \label{fig:mh_m0m12}
\end{figure}

Foreshadowing our inclusion of the Higgs mass in the likelihood, in \reffig{fig:mh_m0m12} we show the lightest Higgs boson mass on the (\mzero, \mhalf) plane for the \CMSSM (left) and \CNMSSM (right). The color scale indicates Higgs masses from $90 \gev$ (red) to $130\gev$ (green). We see that low fine-tuned regions and credible regions of the (\mzero, \mhalf) plane in \reffig{fig:mz_m0m12} correspond to $\mh \lesssim 100 \gev$. A Higgs mass of $\mh \approx 125 \gev$ requires large quantum corrections from massive sparticles and thus multi-TeV soft-breaking masses. Such points lie outside the credible regions of the posterior and are, by traditional measures, fine-tuned.


\begin{figure}
    \centering
    \begin{subfigure}[t]{\figwidth}
        \includegraphics[height=0.975\textwidth]{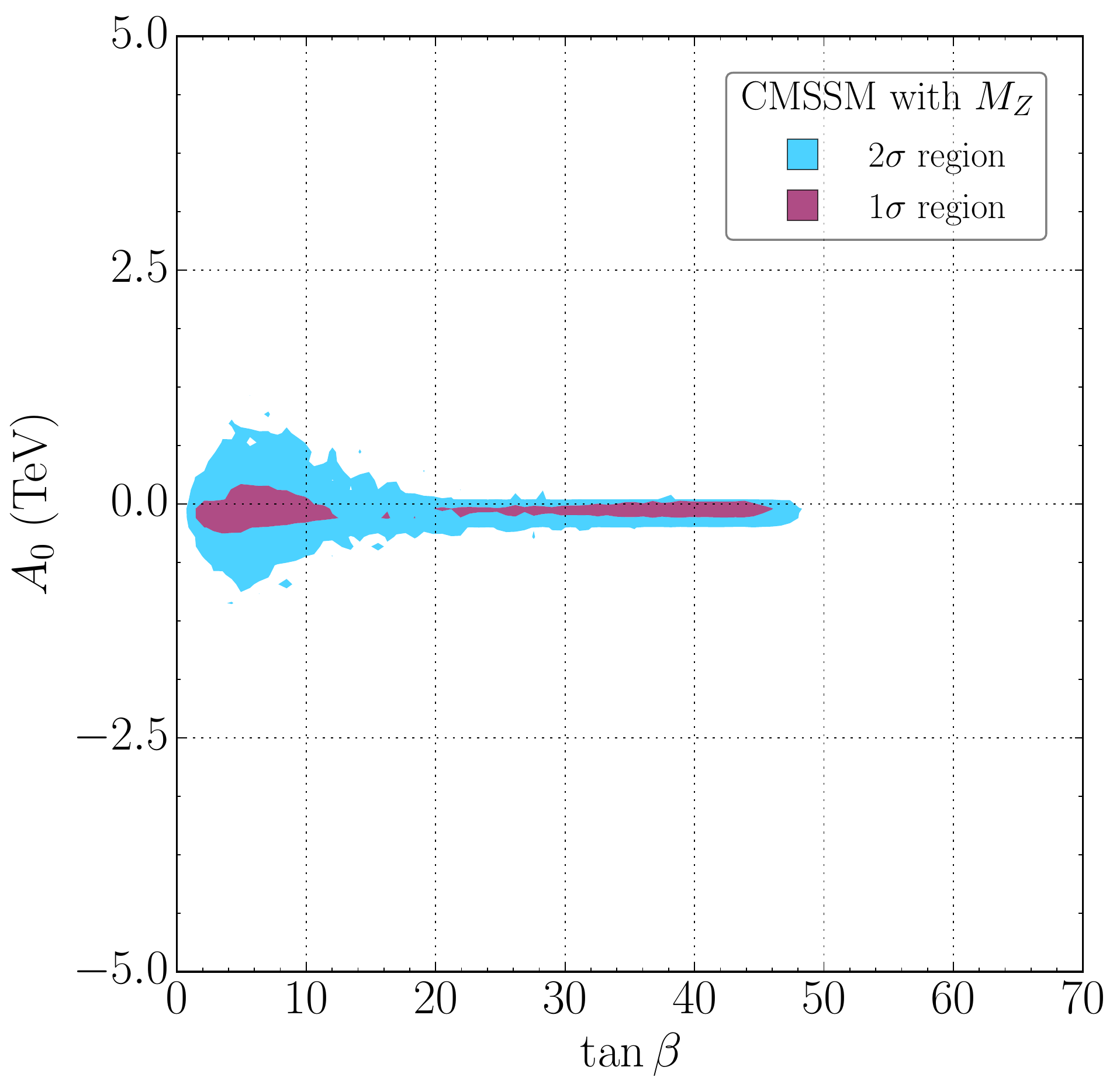}
        \caption{\CMSSM credible regions}
        \label{fig:CMSSM_pdf_mz_tanba0}
    \end{subfigure}
    \begin{subfigure}[t]{\figwidth}
        \includegraphics[height=0.975\textwidth]{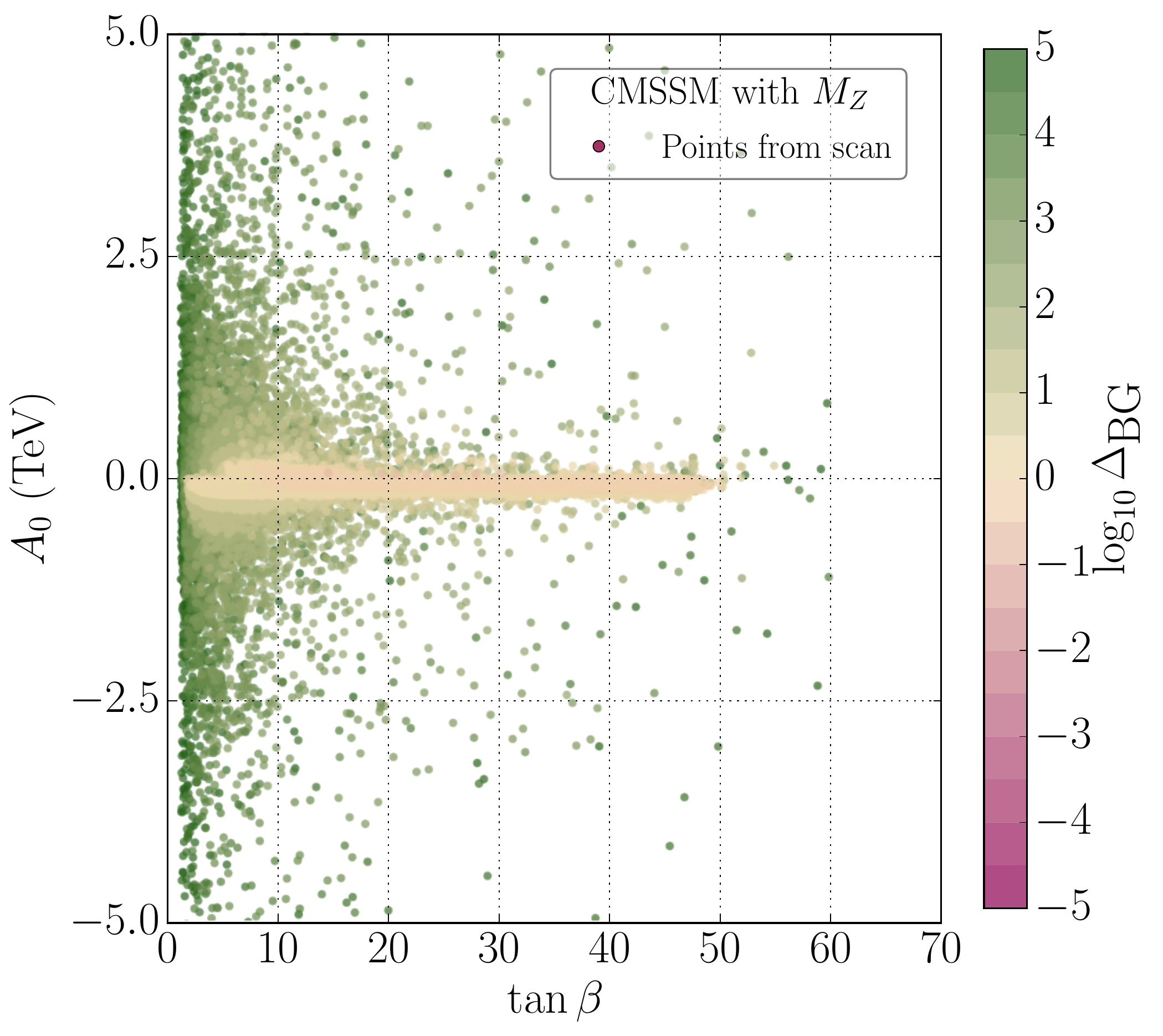}
        \caption{\CMSSM $\Delta_\text{BG}$}
        \label{fig:CMSSM_BG_tanba0}
    \end{subfigure}

    \begin{subfigure}[t]{\figwidth}
        \includegraphics[height=0.975\textwidth]{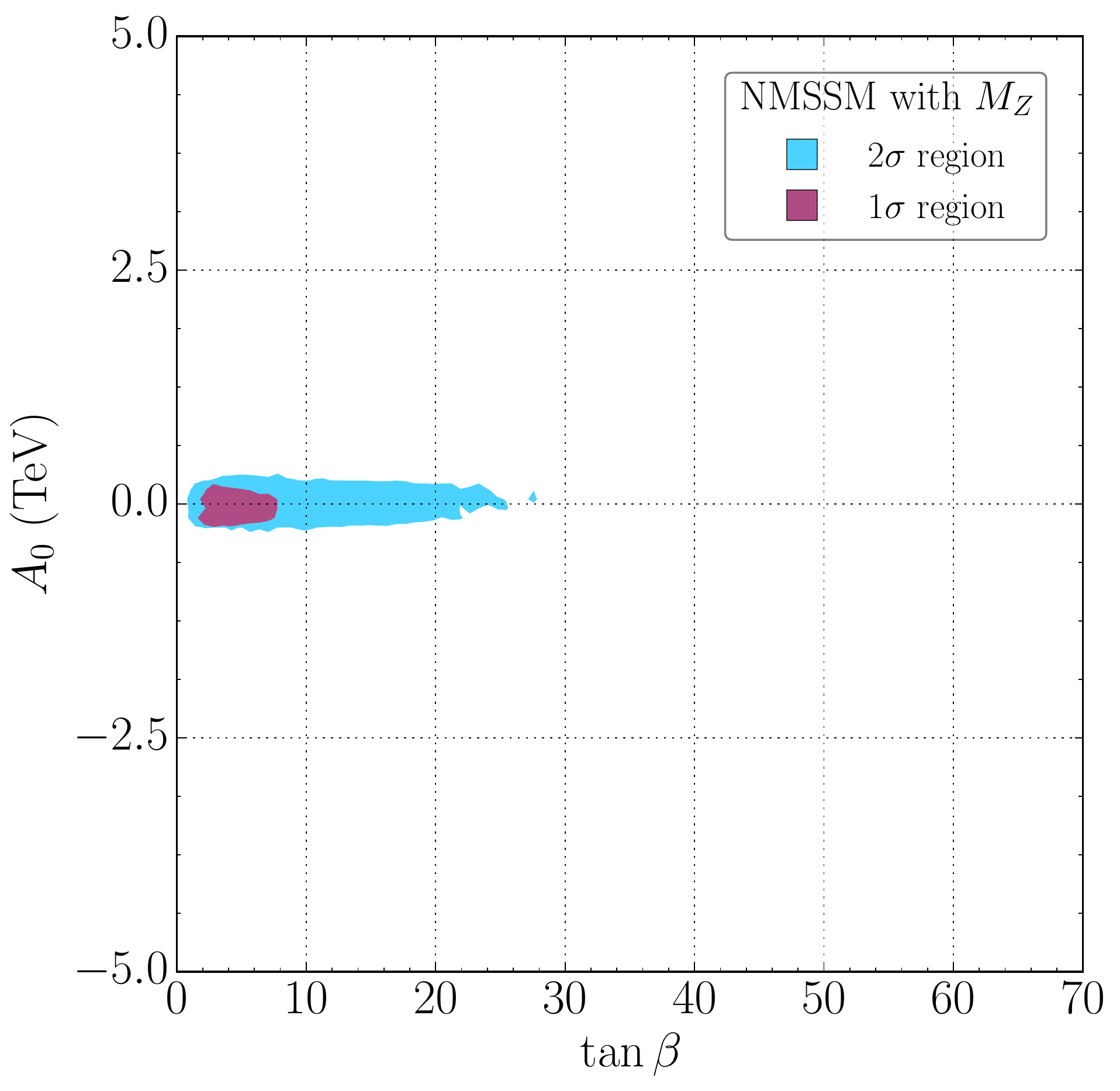}
        \caption{\CNMSSM credible regions}
        \label{fig:CNMSSM_pdf_mz_tanba0}
    \end{subfigure}
    \begin{subfigure}[t]{\figwidth}
        \includegraphics[height=0.975\textwidth]{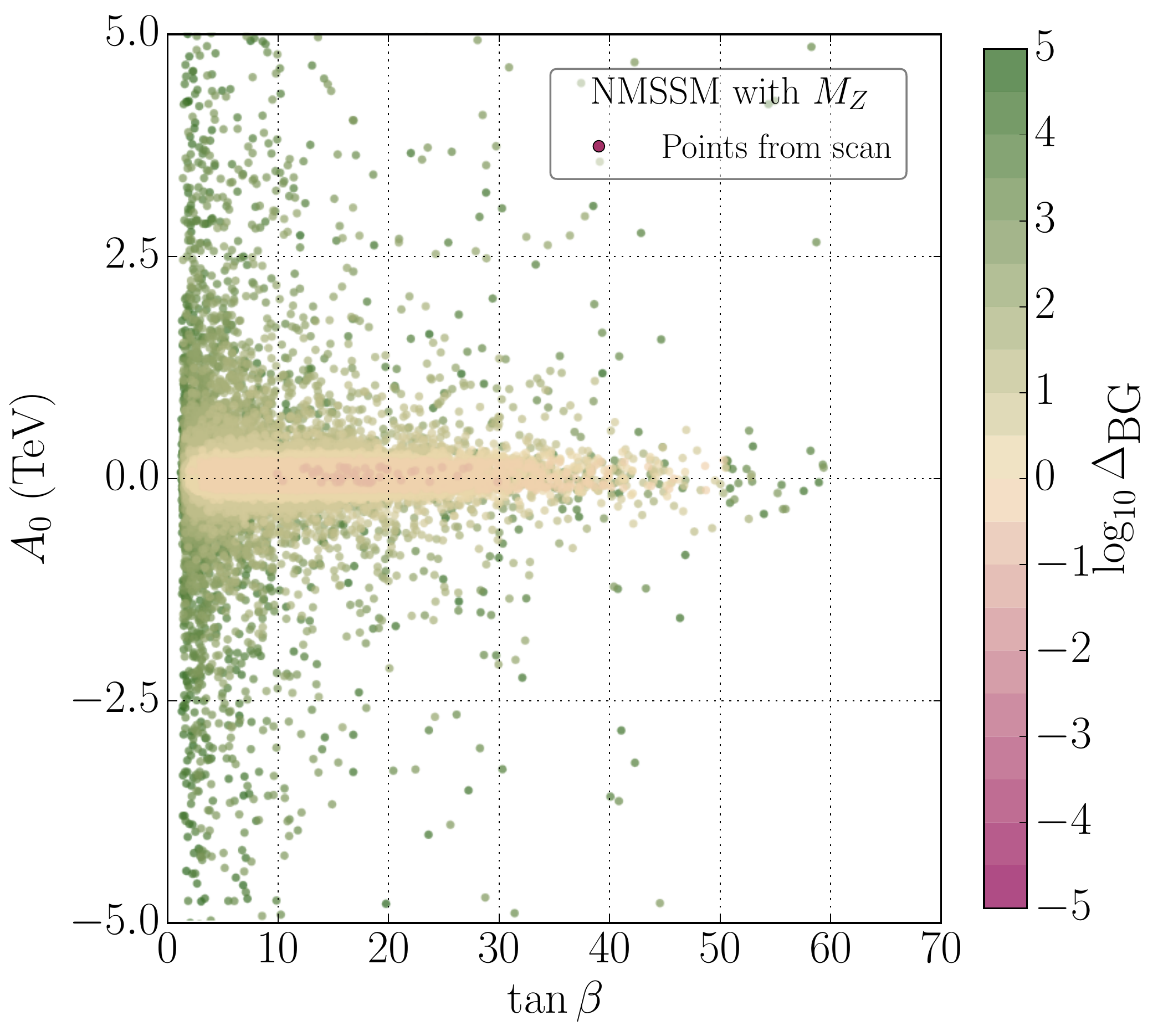}
        \caption{\CNMSSM $\Delta_\text{BG}$}
       \label{fig:CNMSSM_BG_tanba0}
    \end{subfigure}
    
    \caption{Credible regions of the marginalized posterior probability density conditioned upon \mz (left) and profiled BGEN measure (right) for the \CMSSM (upper) and \CNMSSM (lower) on the (\tanb, \azero) plane.}
    \label{fig:pdf_BG_tanba0}
\end{figure}

For completeness, we show the credible regions of the marginalized posterior density and the BGEN measure on the (\tanb, \azero) plane in \reffig{fig:pdf_BG_tanba0}. The posterior and BGEN measure agree with intuition before the LHC:  $\azero \approx 0$, for which loop corrections to the Higgs mass are small, is natural and most plausible.  As was also known in the absence of constraints on the Higgs mass, the credible regions suggest that naturalness issues are largely independent of \tanb in the \CMSSM, while they might slightly prefer low \tanb in the \CNMSSM. 

\begin{figure}
    \centering
    \begin{subfigure}[t]{\figwidth}
        \includegraphics[height=0.975\textwidth]{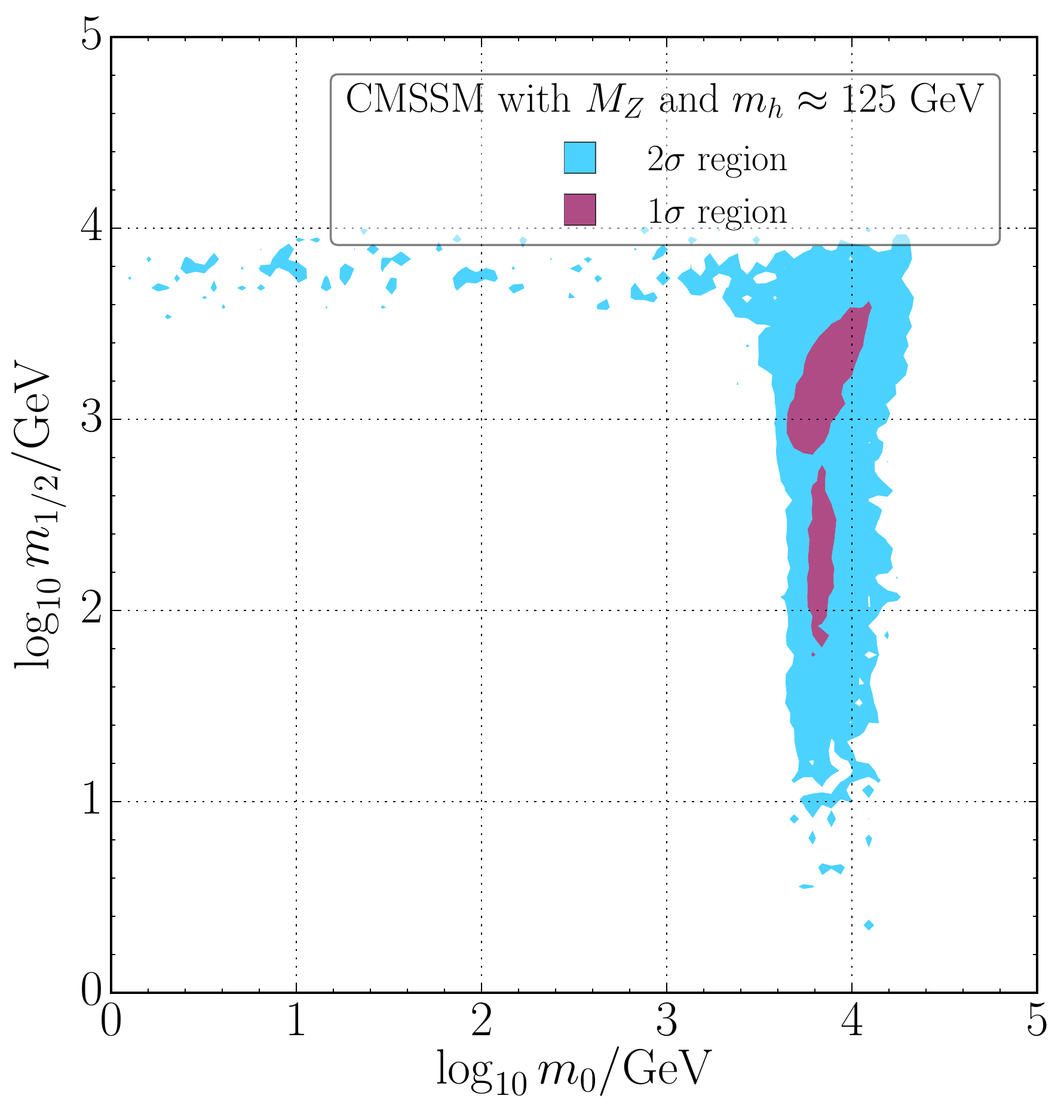}
        \caption{\CMSSM credible regions with \mz and \mh}
    \end{subfigure}
    \begin{subfigure}[t]{\figwidth}
        \includegraphics[height=0.975\textwidth]{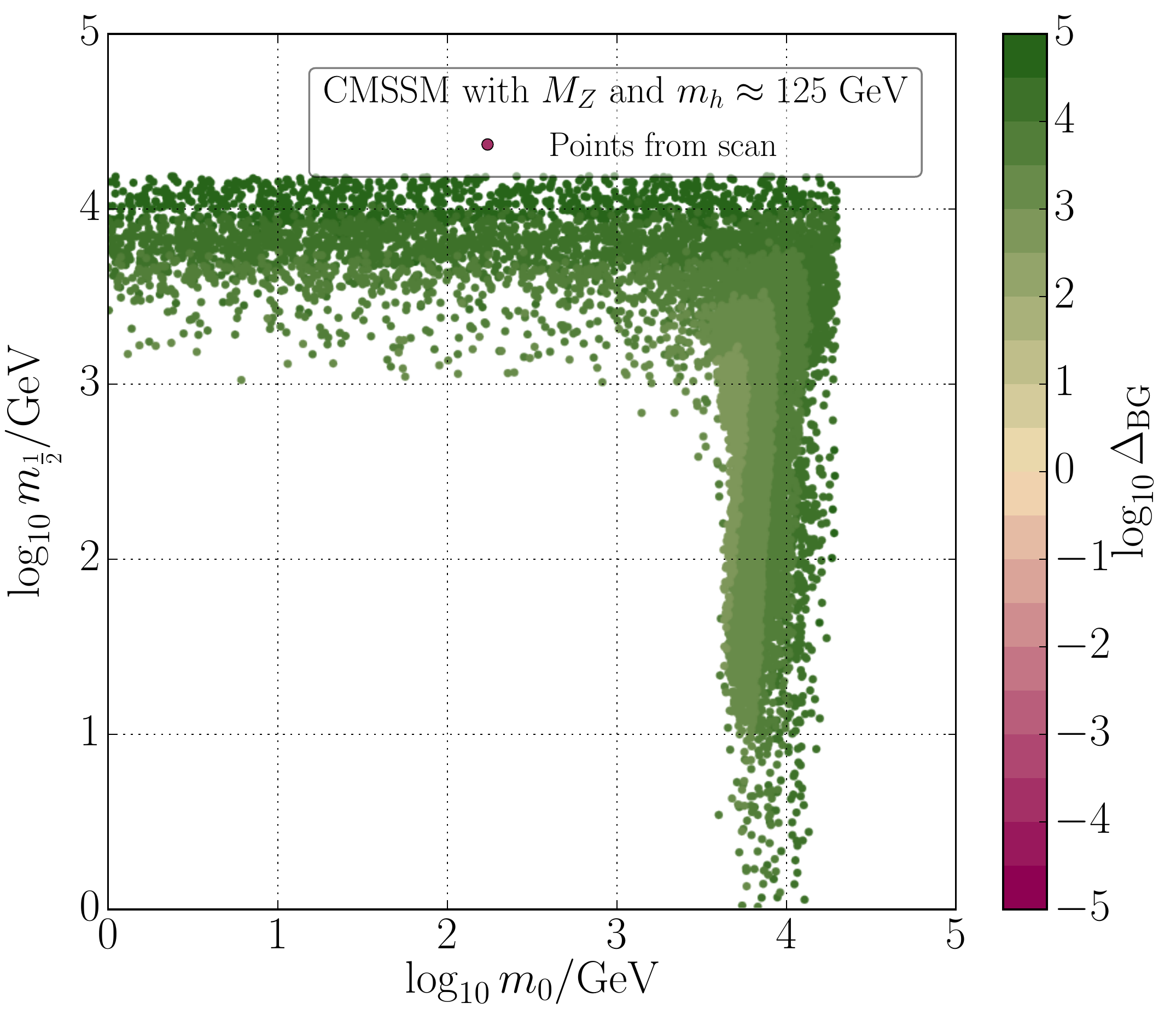}
        \caption{\CMSSM $\Delta_\text{BG}$}
        \label{fig:CMSSM_BG_mh_tanba0}
    \end{subfigure}

    \begin{subfigure}[t]{\figwidth}
        \includegraphics[height=0.975\textwidth]{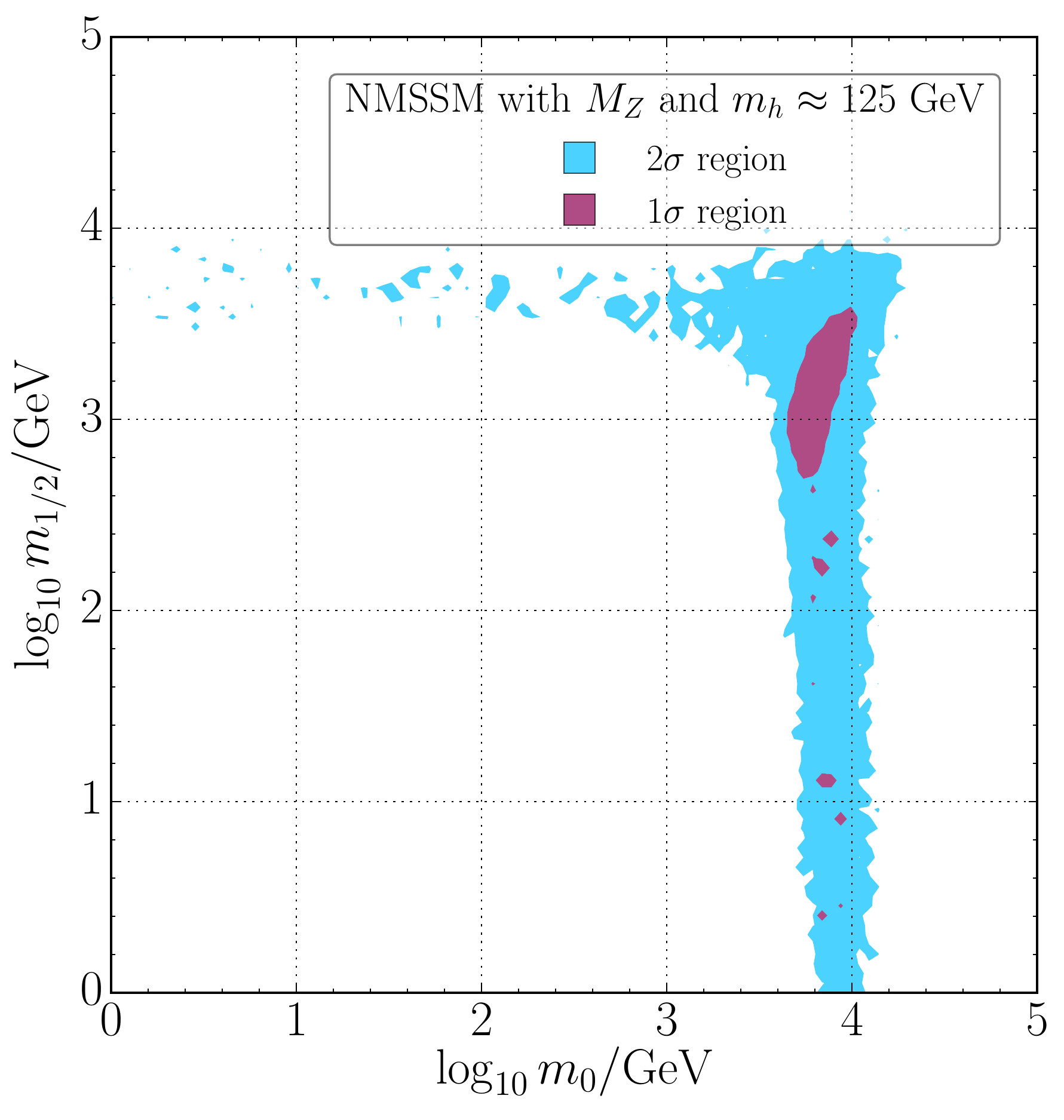}
        \caption{\CNMSSM credible regions with \mz and \mh}
    \end{subfigure}
    \begin{subfigure}[t]{\figwidth}
        \includegraphics[height=0.975\textwidth]{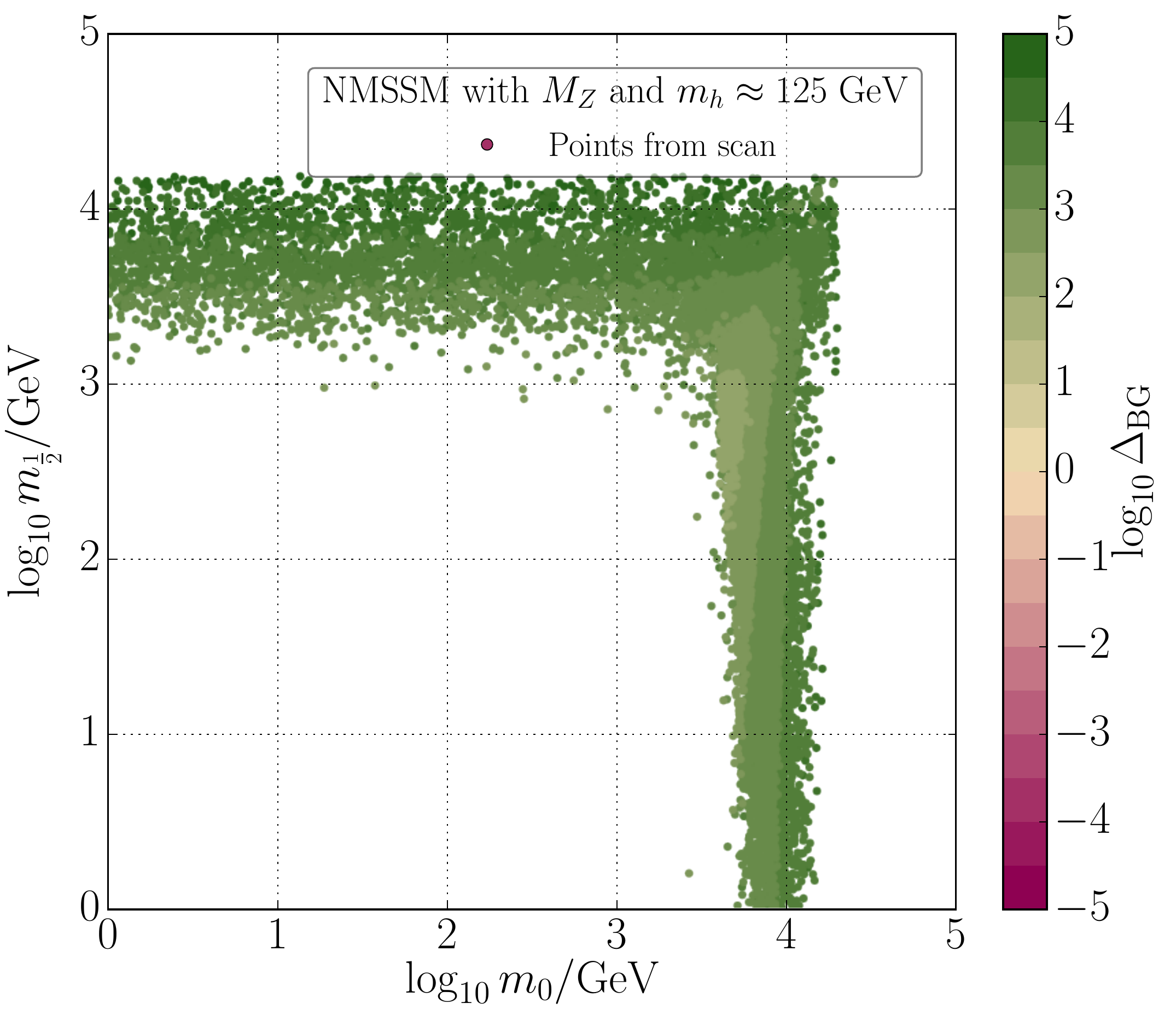}
        \caption{\CNMSSM $\Delta_\text{BG}$}
        \label{fig:CNMSSM_BG_mh_tanba0}
    \end{subfigure}
    
    \caption{Credible regions of the marginalized posterior probability density conditioned upon \mz and \mh (left), and profiled BGEN measure (right) for the \CMSSM (upper) and \CNMSSM (lower) on the (\mzero, \mhalf) plane.}
    \label{fig:mz_mh_m0m12}
\end{figure}

\begin{figure}
    \centering
    \begin{subfigure}[t]{\figwidth}
        \includegraphics[height=0.975\textwidth]{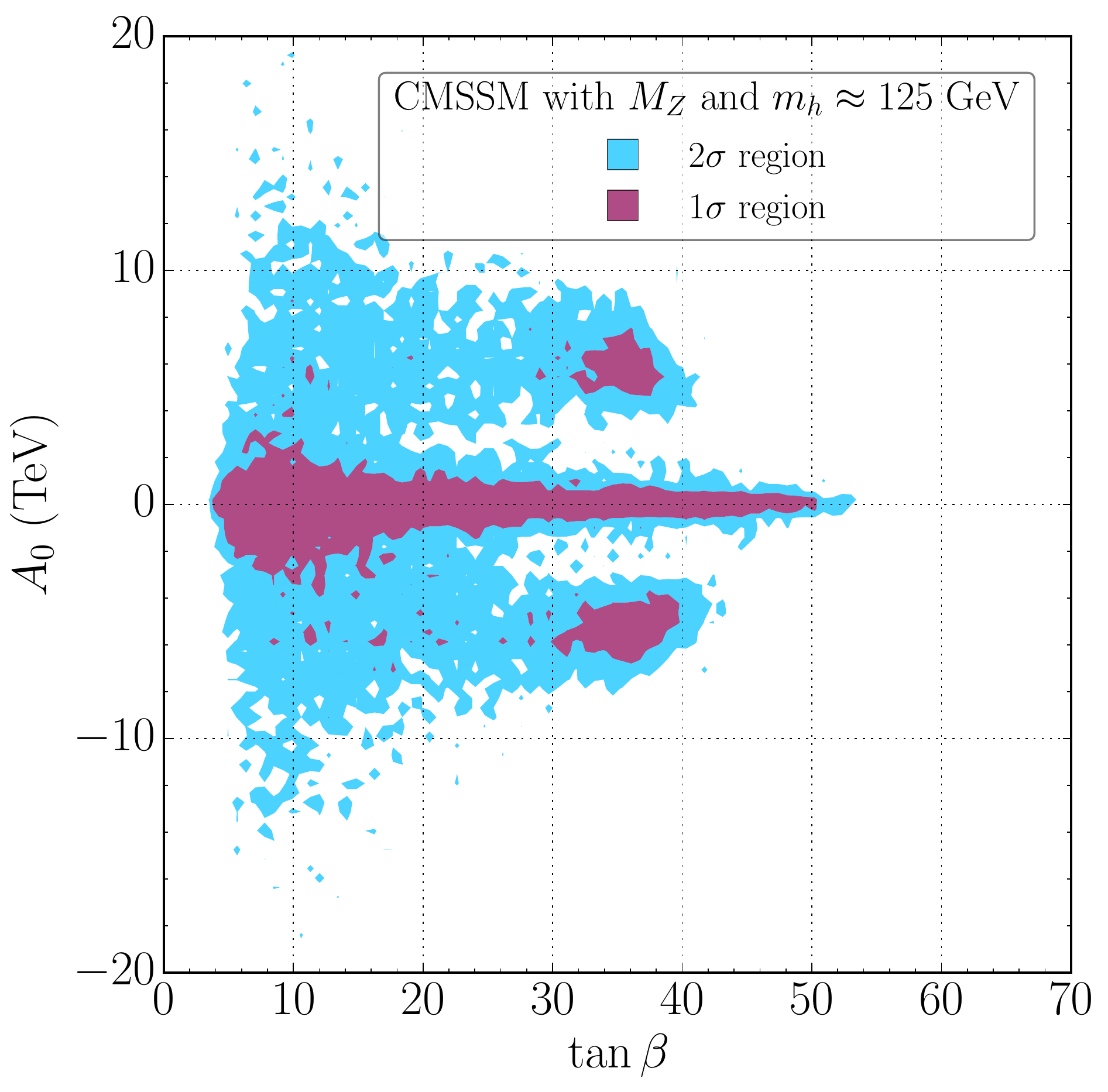}
        \caption{\CMSSM credible regions}
    \end{subfigure}
    \begin{subfigure}[t]{\figwidth}
        \includegraphics[height=0.975\textwidth]{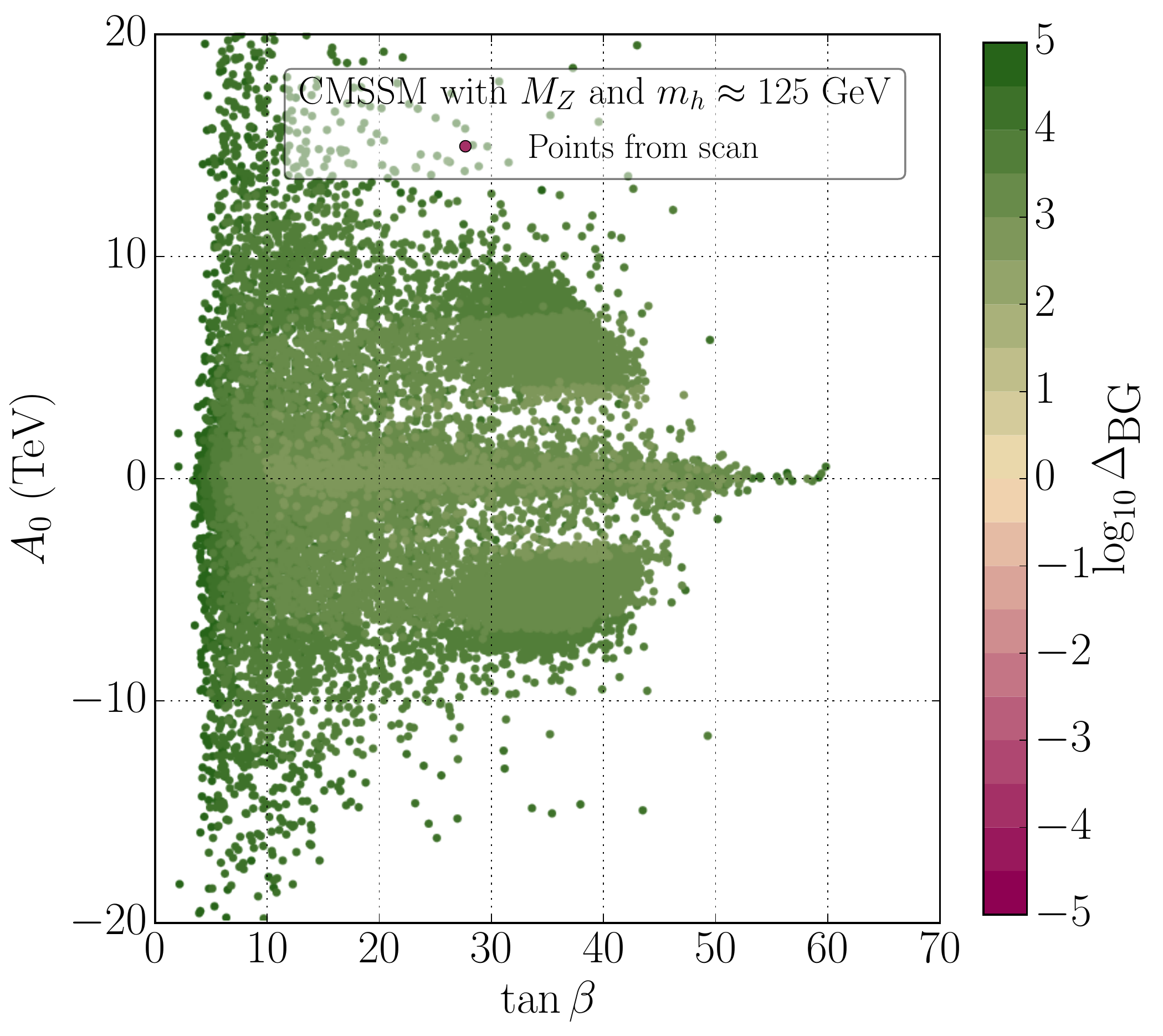}
        \caption{\CMSSM $\Delta_\text{BG}$}
    \end{subfigure}

    \begin{subfigure}[t]{\figwidth}
        \includegraphics[height=0.975\textwidth]{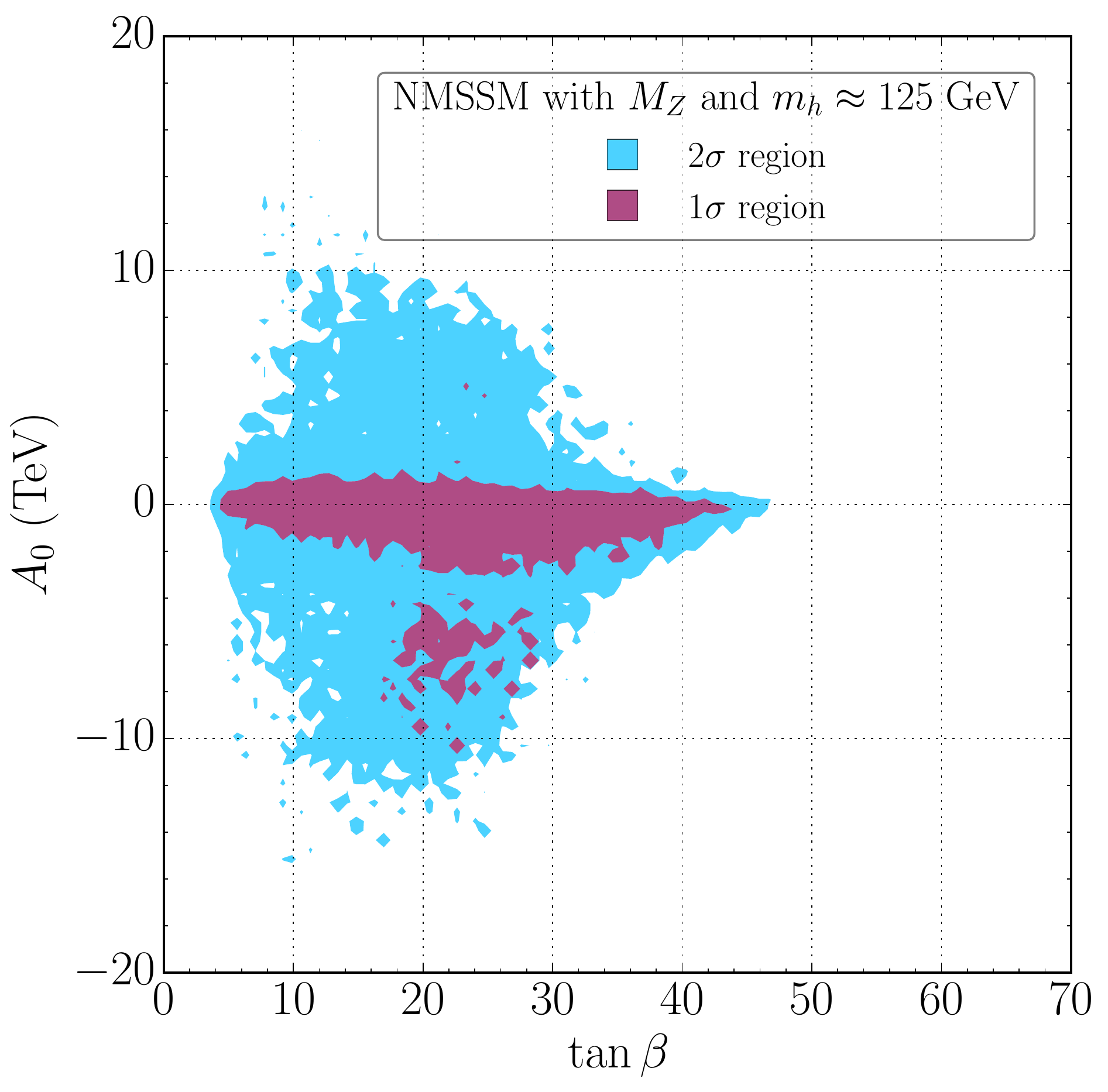}
        \caption{\CNMSSM credible regions}
    \end{subfigure}
    \begin{subfigure}[t]{\figwidth}
        \includegraphics[height=0.975\textwidth]{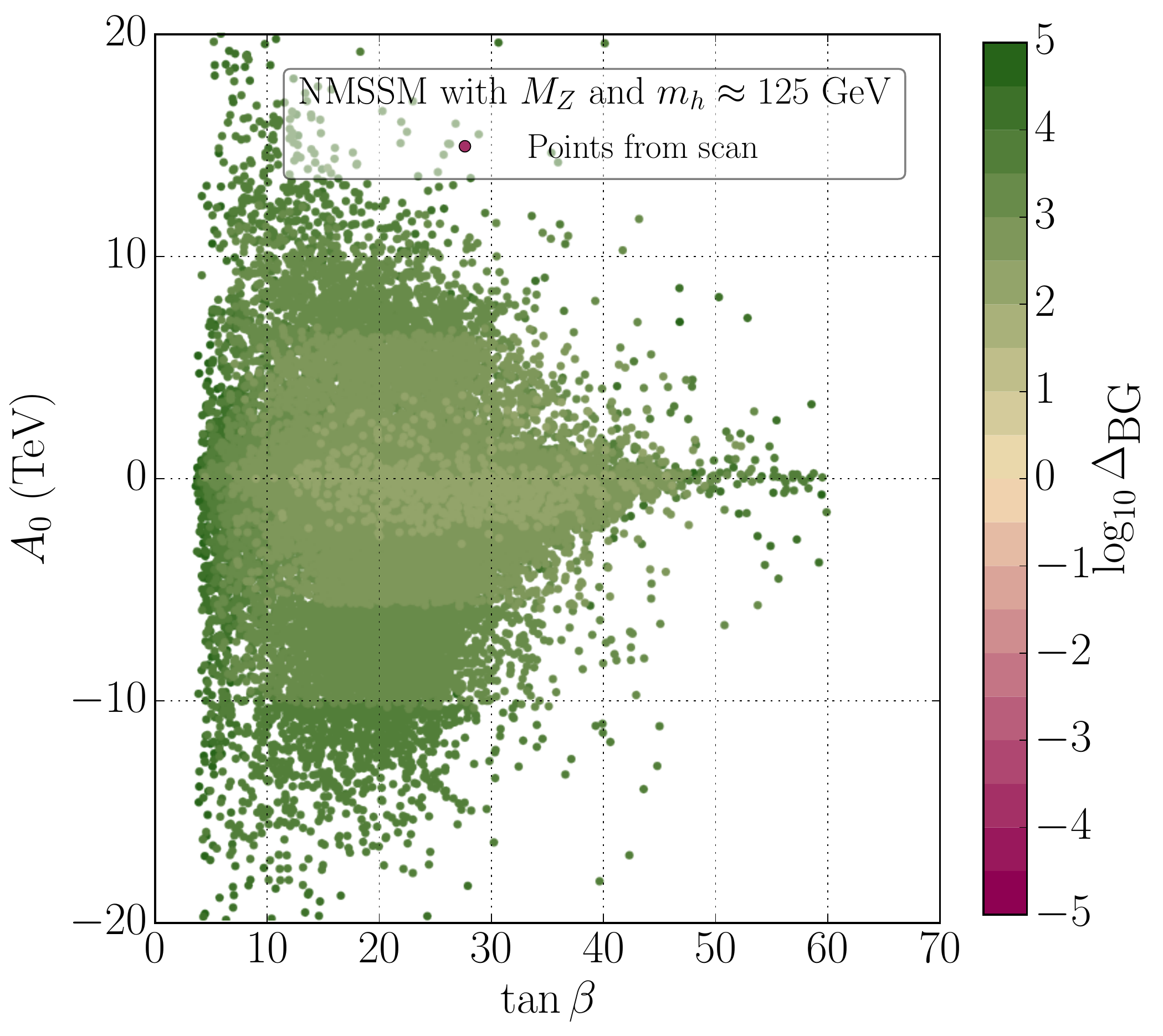}
        \caption{\CNMSSM $\Delta_\text{BG}$}
    \end{subfigure}
    
    \caption{Credible regions of the marginalized posterior probability density conditioned upon \mz and \mh (left) and profiled BGEN measure (right) for the \CMSSM (upper) and \CNMSSM (lower) on the (\tanb, \azero) plane.}
    \label{fig:pdf_BG_mh_tanba0}
\end{figure}

In light of our growing confidence in the posterior measuring 
fine-tuning, it is interesting to see how it fares against the
addition of the most relevant piece of new information from the LHC:
the lightest Higgs mass.  This is shown in \reffig{fig:mz_mh_m0m12}.
Our first observation is that the most plausible regions, indicated
by the $1\sigma$ and $2\sigma$ credible regions, are dramatically
shifted to about two orders of magnitude higher \mzero and \mhalf
values.  This is, of course, the well-known quantitative conclusion
from the LHC Run 1: supersymmetry is effectively eliminated, \ie
relatively implausible, below $1\tev$.  After folding in the
lightest Higgs mass the least fine-tuned regions lie in the focus
point, signalled by the slanted 1$\sigma$ region at large $m_0$ and $m_{1/2}$ for both the \CMSSM and the \CNMSSM.  In this region low fine
tuning is achieved with relatively small values of \azero (\reffig{fig:pdf_BG_mh_tanba0}).  In the
vertical region spanning between \mhalf $\sim$ 0.1-1 TeV \azero
increases with decreasing \mhalf.  This still allows for acceptable
fine-tuning in the \CMSSM. 

Just as in the case when only \mz was included in the likelihood, the BGEN measure confirms the picture painted by the posterior distribution.  The former signals the narrow vertical region at $\mzero \sim 10\tev$ and between $\mhalf \sim 0.1-1 \tev$ as the least fine-tuned.  This long vertical strip represents the focus point solution, thus confirming that Bayesian naturalness does find a naturalness benefit from focus point supersymmetry\cite{Chan:1997bi,
Feng:1999hg,Feng:1999mn}.


\begin{figure}
    \centering
    \begin{subfigure}[t]{\figwidth}
        \includegraphics[width=\textwidth]{./figs/CMSSM_BG_m0m12.pdf}
        \caption{$\Delta_\text{BG}$}
    \end{subfigure}
    \begin{subfigure}[t]{\figwidth}
        \includegraphics[width=\textwidth]{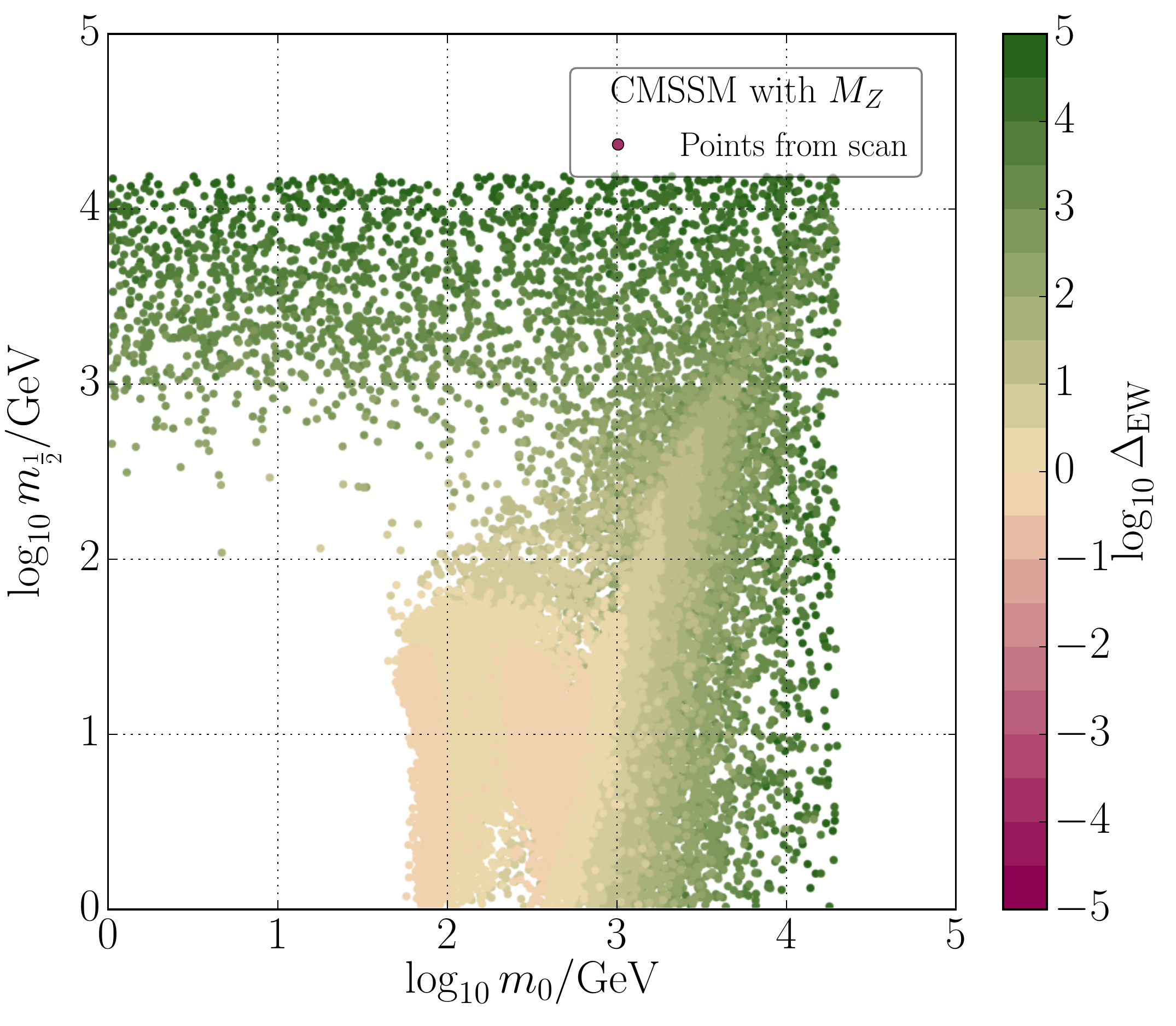}
        \caption{$\Delta_\text{EW}$}
    \end{subfigure}
    
    \begin{subfigure}[t]{\figwidth}
        \includegraphics[width=\textwidth]{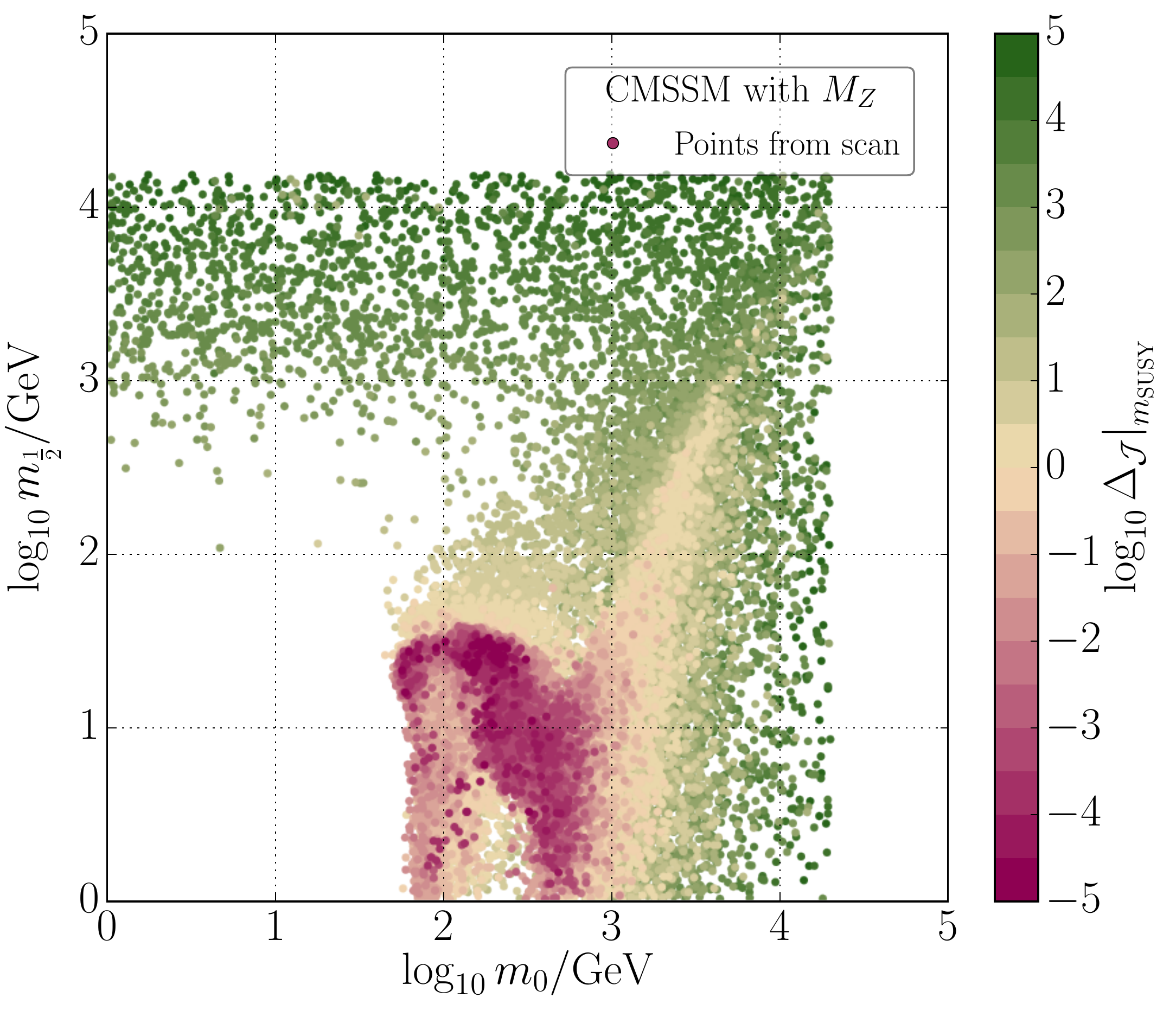}
        \caption{$\Delta_{\jac} \big |_{\msusy}$}
    \end{subfigure}
    \begin{subfigure}[t]{\figwidth}
        \includegraphics[width=\textwidth]{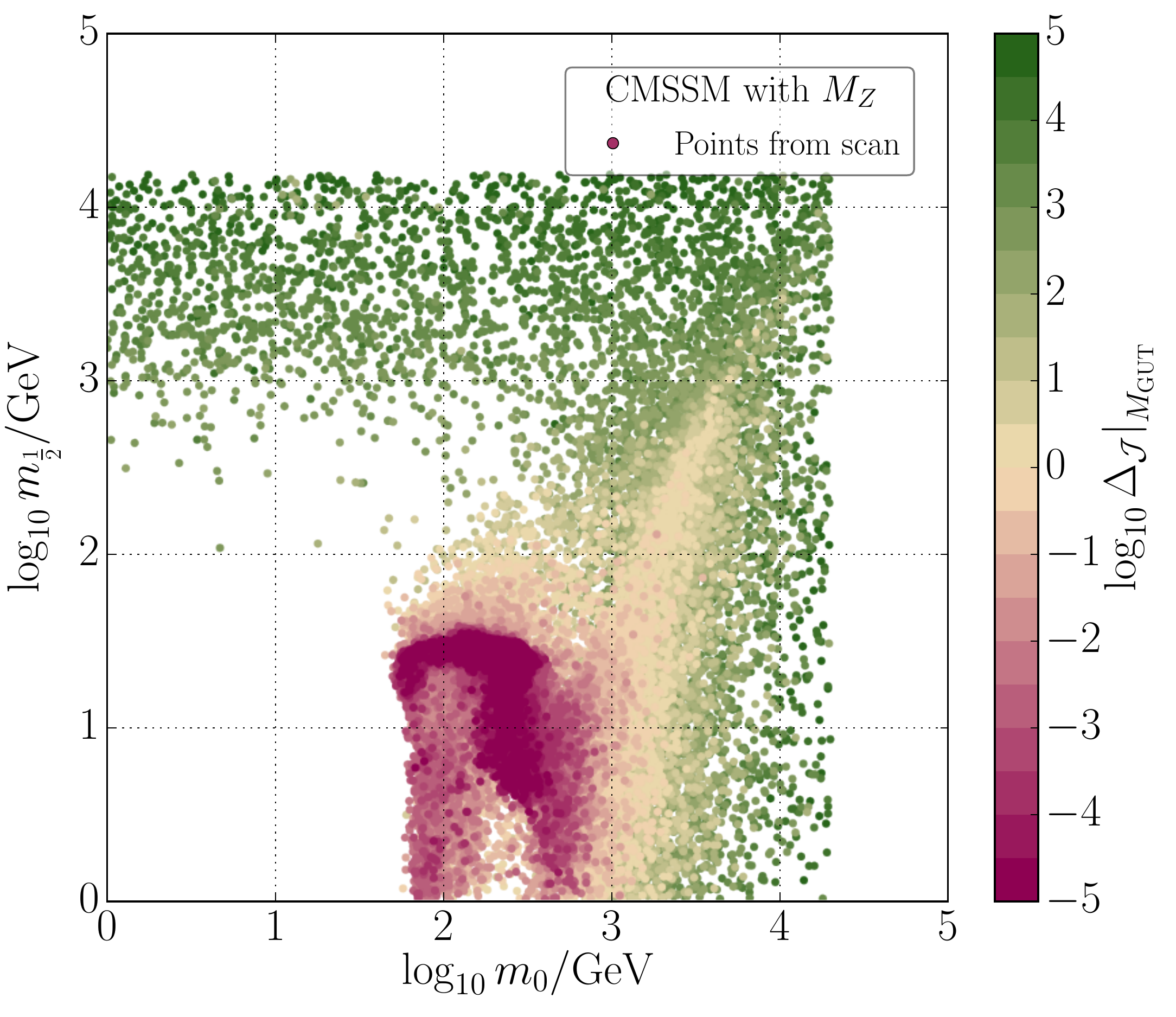}
        \caption{$\Delta_{\jac} \big |_{\mgut}$}
    \end{subfigure}  
    \caption{Comparison of fine-tuning measures in the \CMSSM on the (\mzero, \mhalf) plane. For their definitions, see \refsec{Subsec:Measures}.}
    \label{fig:cmssm_fine_tuning}
\end{figure}

\begin{figure}
    \centering
    \begin{subfigure}[t]{\figwidth}
        \includegraphics[width=\textwidth]{./figs/CNMSSM_BG_m0m12.pdf}
        \caption{$\Delta_\text{BG}$}
    \end{subfigure}
    \begin{subfigure}[t]{\figwidth}
        \includegraphics[width=\textwidth]{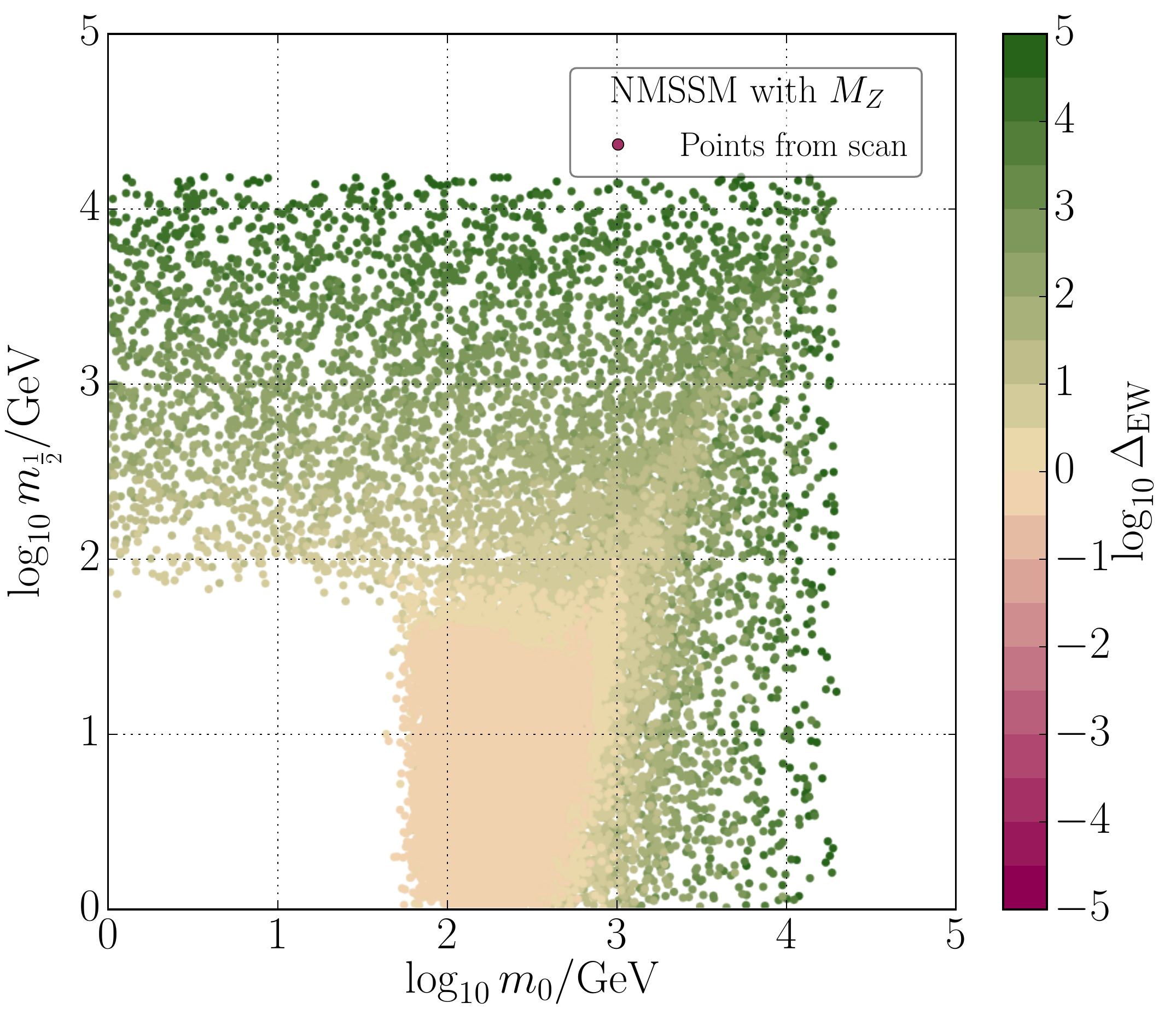}
        \caption{$\Delta_\text{EW}$}
    \end{subfigure}
    
    \begin{subfigure}[t]{\figwidth}
        \includegraphics[width=\textwidth]{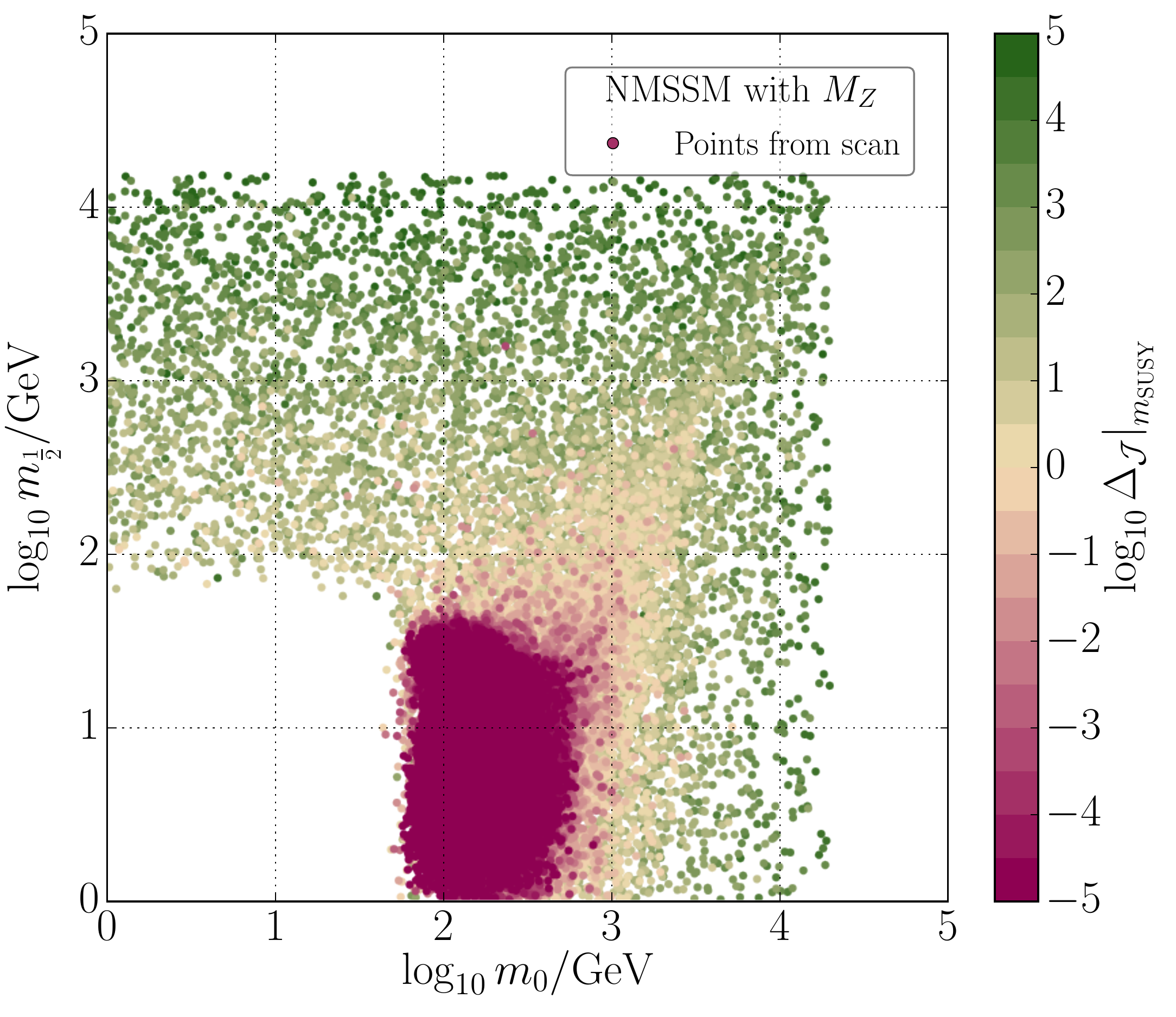}
        \caption{$\Delta_{\jac} \big |_{\msusy}$}
    \end{subfigure}
    \begin{subfigure}[t]{\figwidth}
        \includegraphics[width=\textwidth]{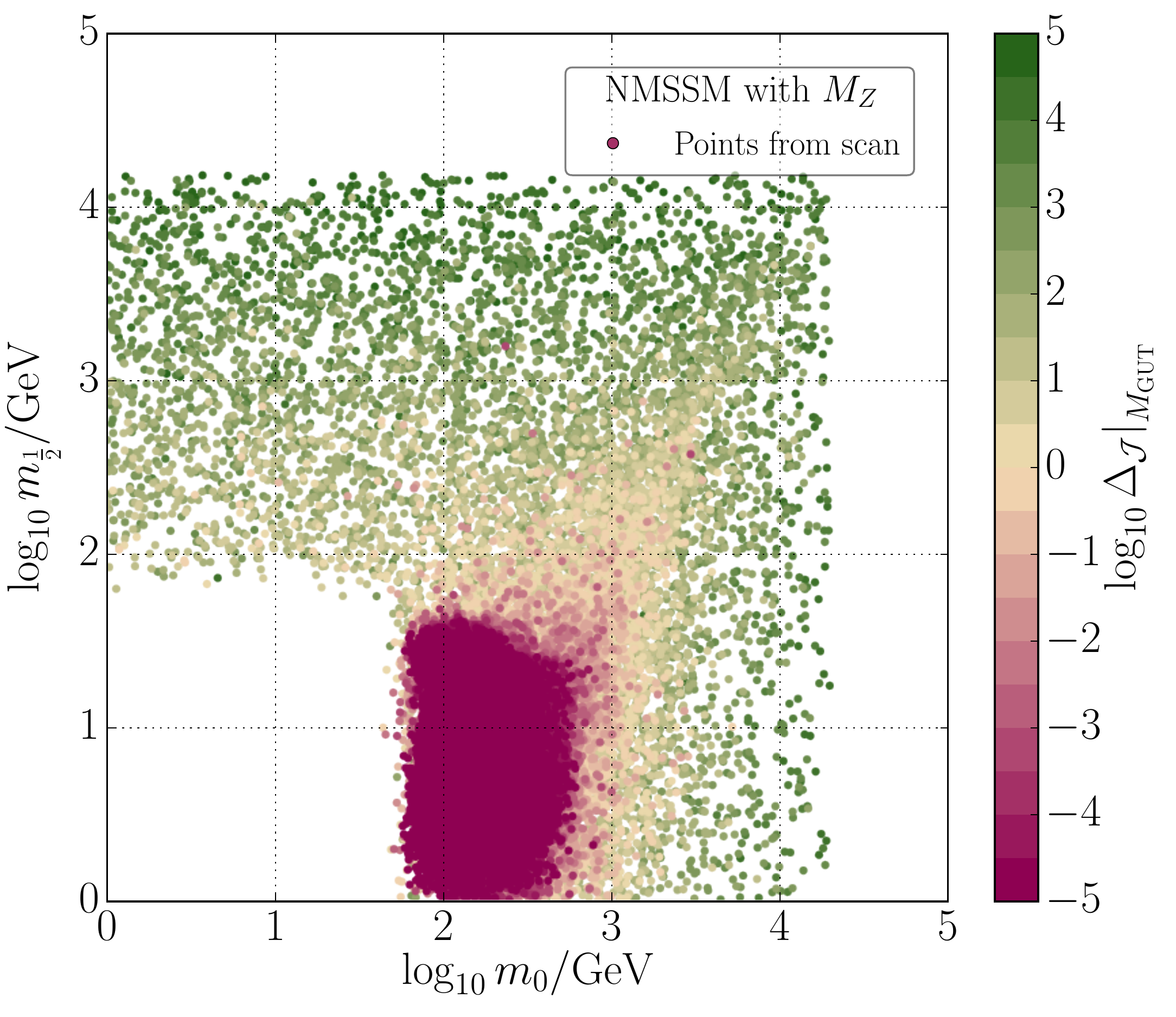}
        \caption{$\Delta_{\jac} \big |_{\mgut}$}
    \end{subfigure}  
    \caption{Comparison of fine-tuning measures in the \CNMSSM on the (\mzero, \mhalf) plane. For their definitions, see \refsec{Subsec:Measures}.}
    \label{fig:cnmssm_fine_tuning}
\end{figure}

To gauge their consistency with each other, we compare the fine-tuning measures defined in \refsec{Subsec:Measures} in the \CMSSM in \reffig{fig:cmssm_fine_tuning} and in the \CNMSSM in \reffig{fig:cnmssm_fine_tuning} on the (\mzero, \mhalf) planes. In each plot, parameters other than \mzero and \mhalf, such as \azero and \tanb, were chosen such that the fine-tuning measure was minimized. All fine-tuning measures are qualitatively similar, with a region of low fine-tuning at $\msusy \sim \mz$, and fine-tuning increases as \mzero and \mhalf are increased, as expected. The Jacobian-based fine-tuning measures, however, are substantially smaller than the traditional BGEN measure and EW measure. We should not, however, be mislead into a superficial comparison of the measures. The Jacobian based measures, $\Delta_\mathcal{J}$, are volumes of multidimensional hypercubes, \eg a two-dimensional volume in the \MSSM. The BGEN measure, $\Delta_\text{BG}$, on the other hand, corresponds to the length of a line element and $\Delta_\textrm{EW}$ measures the relative size of terms contributing to \mz.

\begin{figure}
    \centering
    \begin{subfigure}[t]{\figwidth}
        \includegraphics[width=\textwidth]{./figs/CMSSM_BG_mh_m0m12.pdf}
        \caption{$\Delta_\text{BG}$}
    \end{subfigure}
    \begin{subfigure}[t]{\figwidth}
        \includegraphics[width=\textwidth]{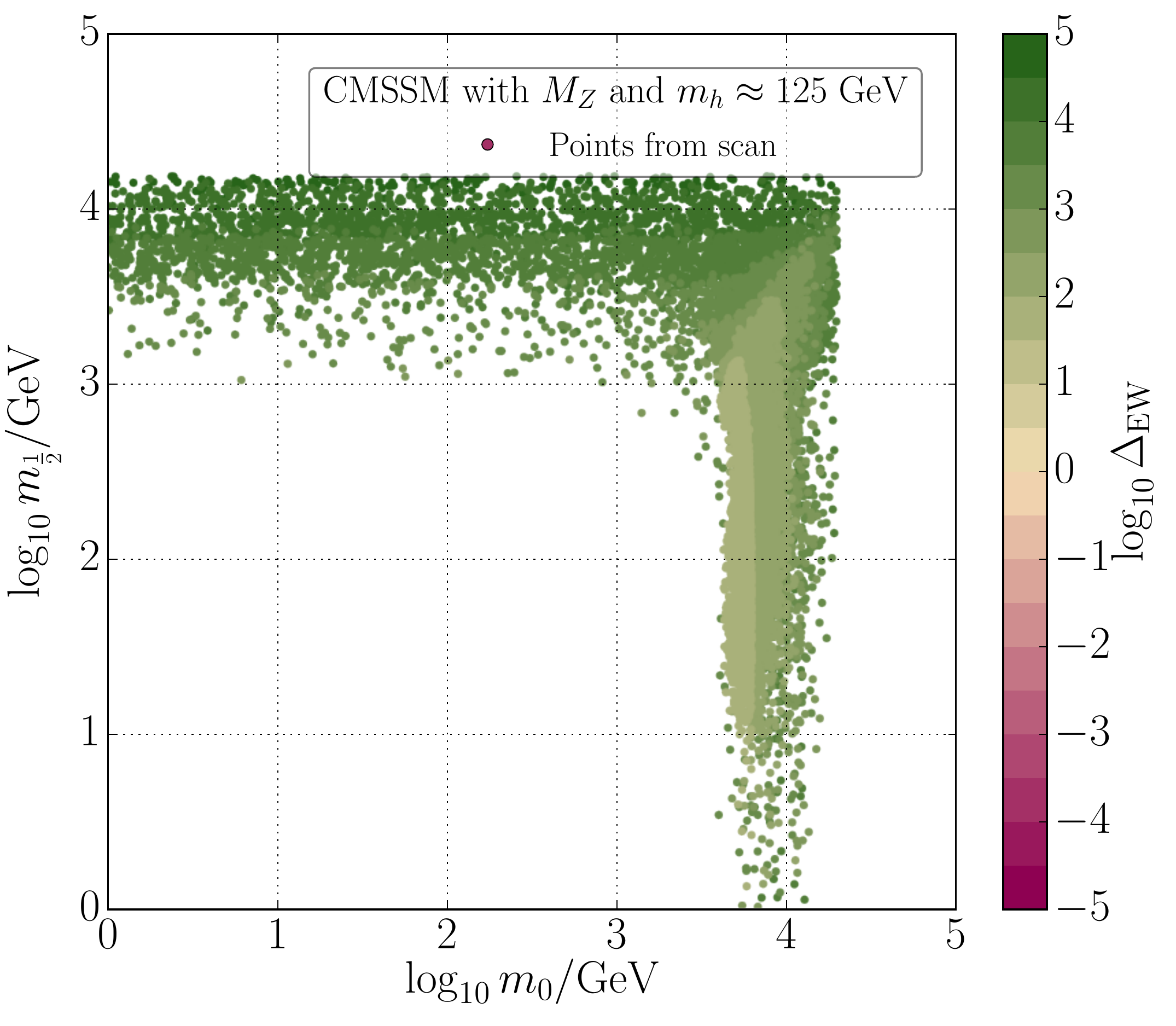}
        \caption{$\Delta_\text{EW}$}
    \end{subfigure}
    
    \begin{subfigure}[t]{\figwidth}
        \includegraphics[width=\textwidth]{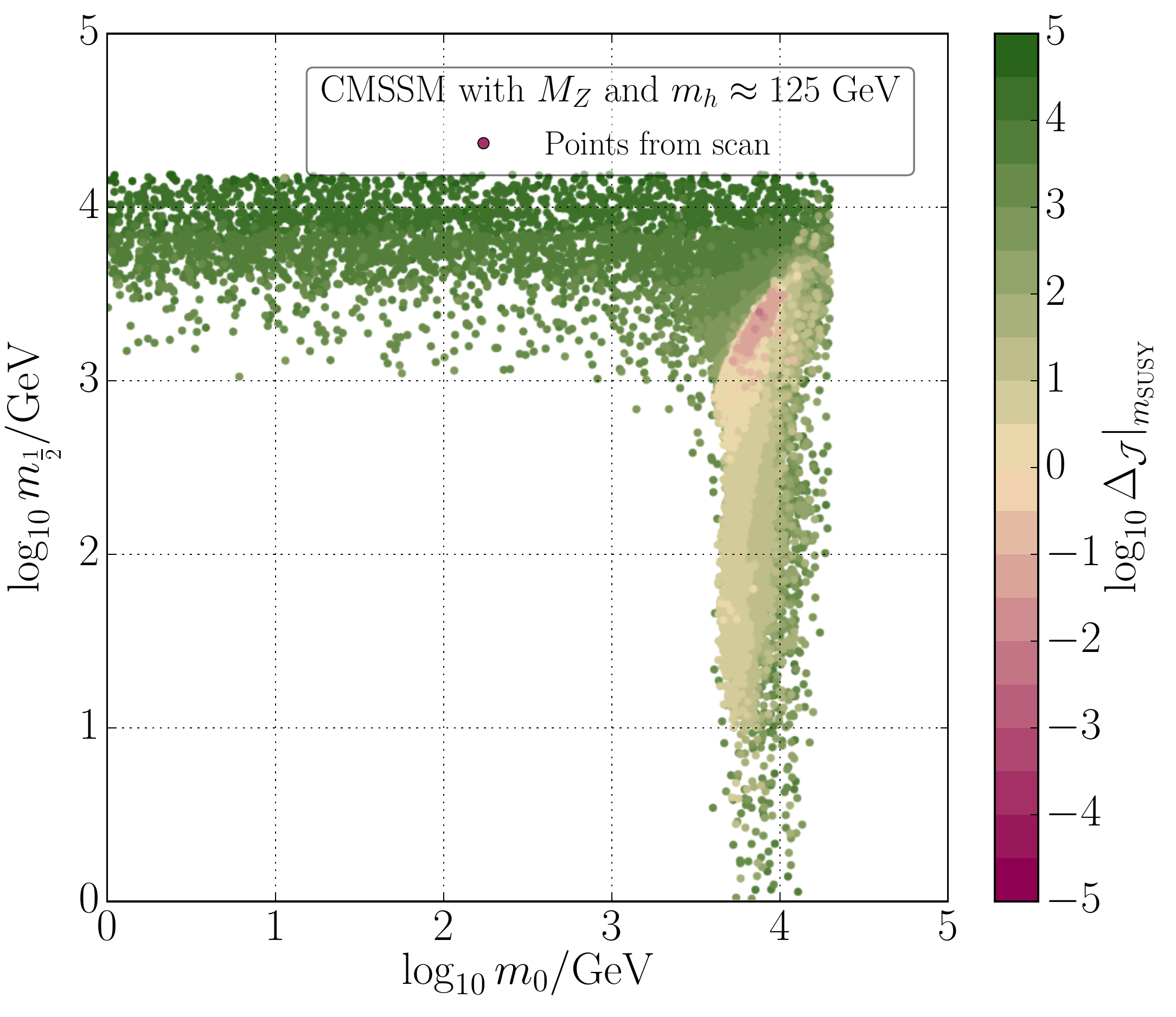}
        \caption{$\Delta_{\jac} \big |_{\msusy}$}
    \end{subfigure}
    \begin{subfigure}[t]{\figwidth}
        \includegraphics[width=\textwidth]{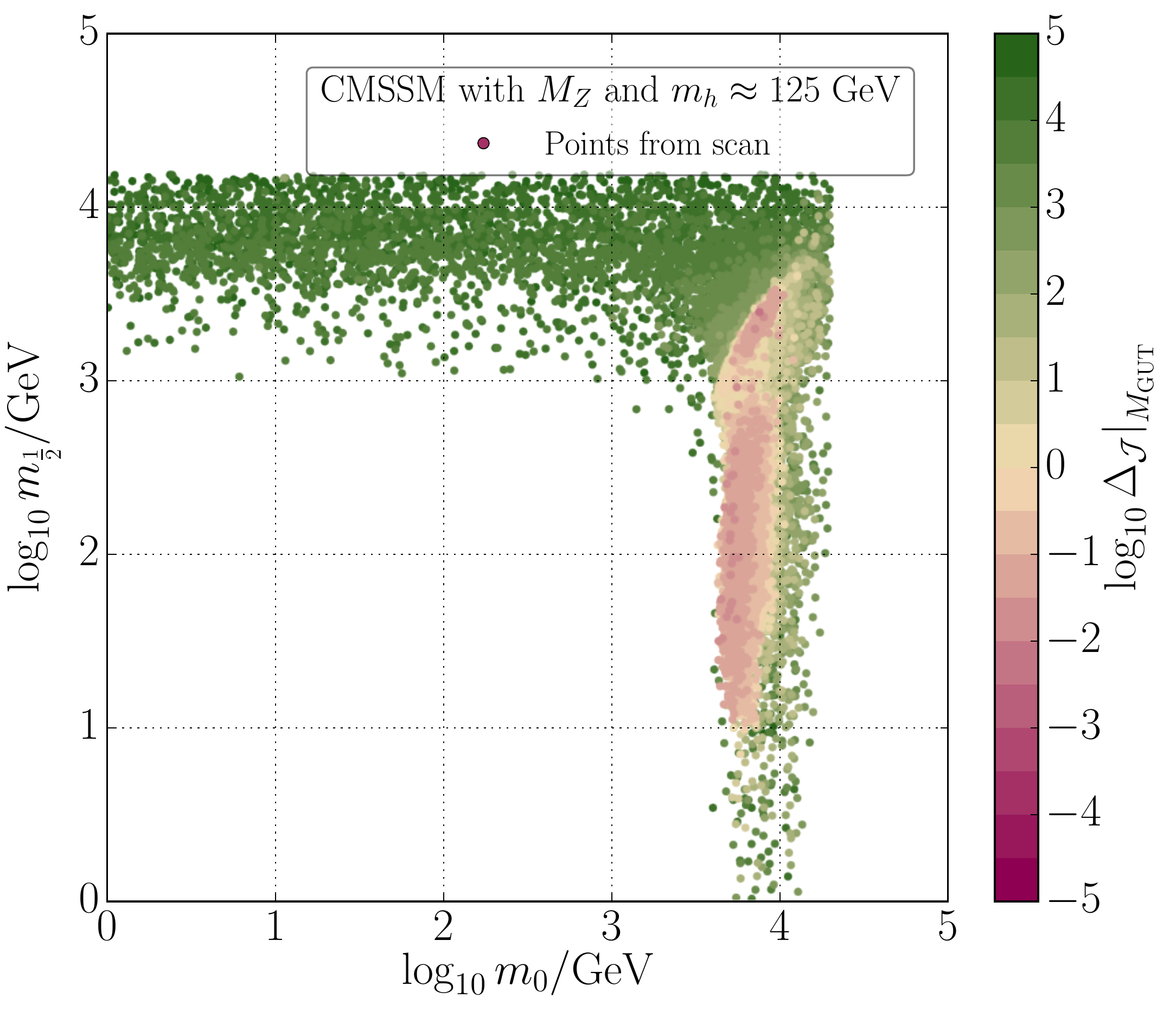}
        \caption{$\Delta_{\jac} \big |_{\mgut}$}
    \end{subfigure}  
    \caption{Comparison of fine-tuning measures in the \CMSSM with $\mh \approx 125\gev$ on the (\mzero, \mhalf) plane.  For their definitions, see \refsec{Subsec:Measures}.}
    \label{fig:cmssm_fine_tuning_mh}
\end{figure}

\begin{figure}
    \centering
    \begin{subfigure}[t]{\figwidth}
        \includegraphics[width=\textwidth]{./figs/CNMSSM_BG_mh_m0m12.pdf}
        \caption{$\Delta_\text{BG}$}
    \end{subfigure}
    \begin{subfigure}[t]{\figwidth}
        \includegraphics[width=\textwidth]{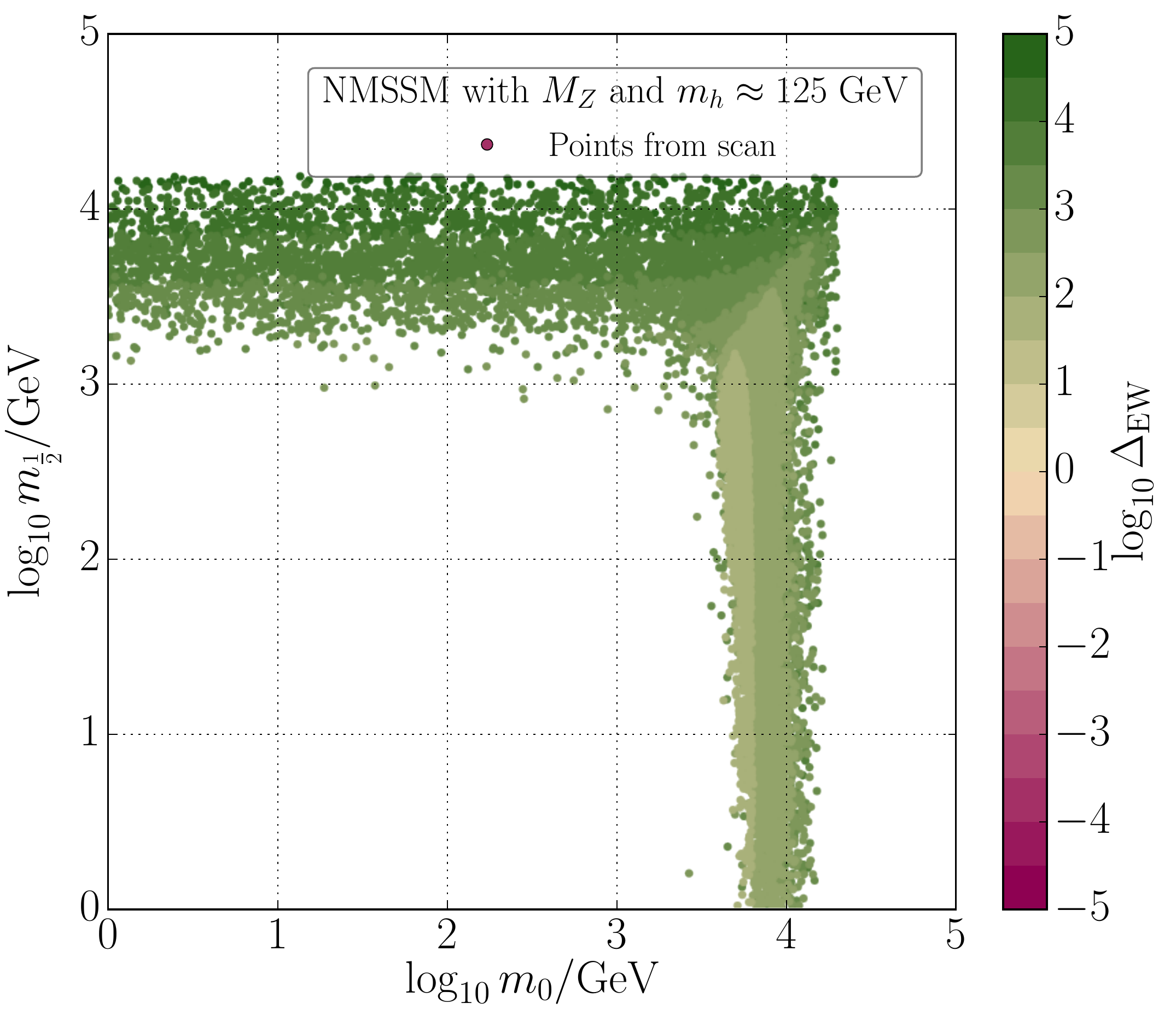}
        \caption{$\Delta_\text{EW}$}
    \end{subfigure}
    
    \begin{subfigure}[t]{\figwidth}
        \includegraphics[width=\textwidth]{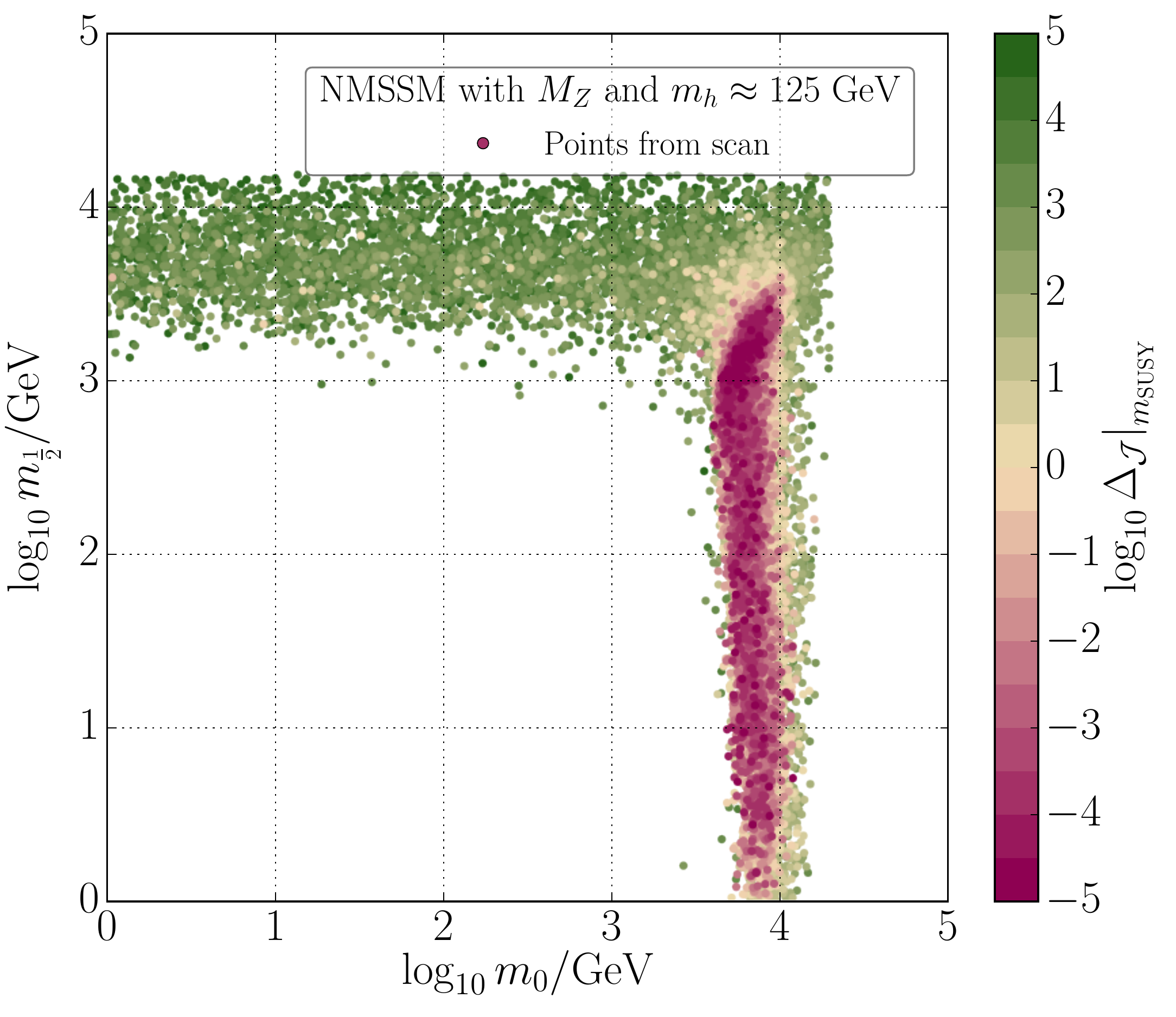}
        \caption{$\Delta_{\jac} \big |_{\msusy}$}
    \end{subfigure}
    \begin{subfigure}[t]{\figwidth}
        \includegraphics[width=\textwidth]{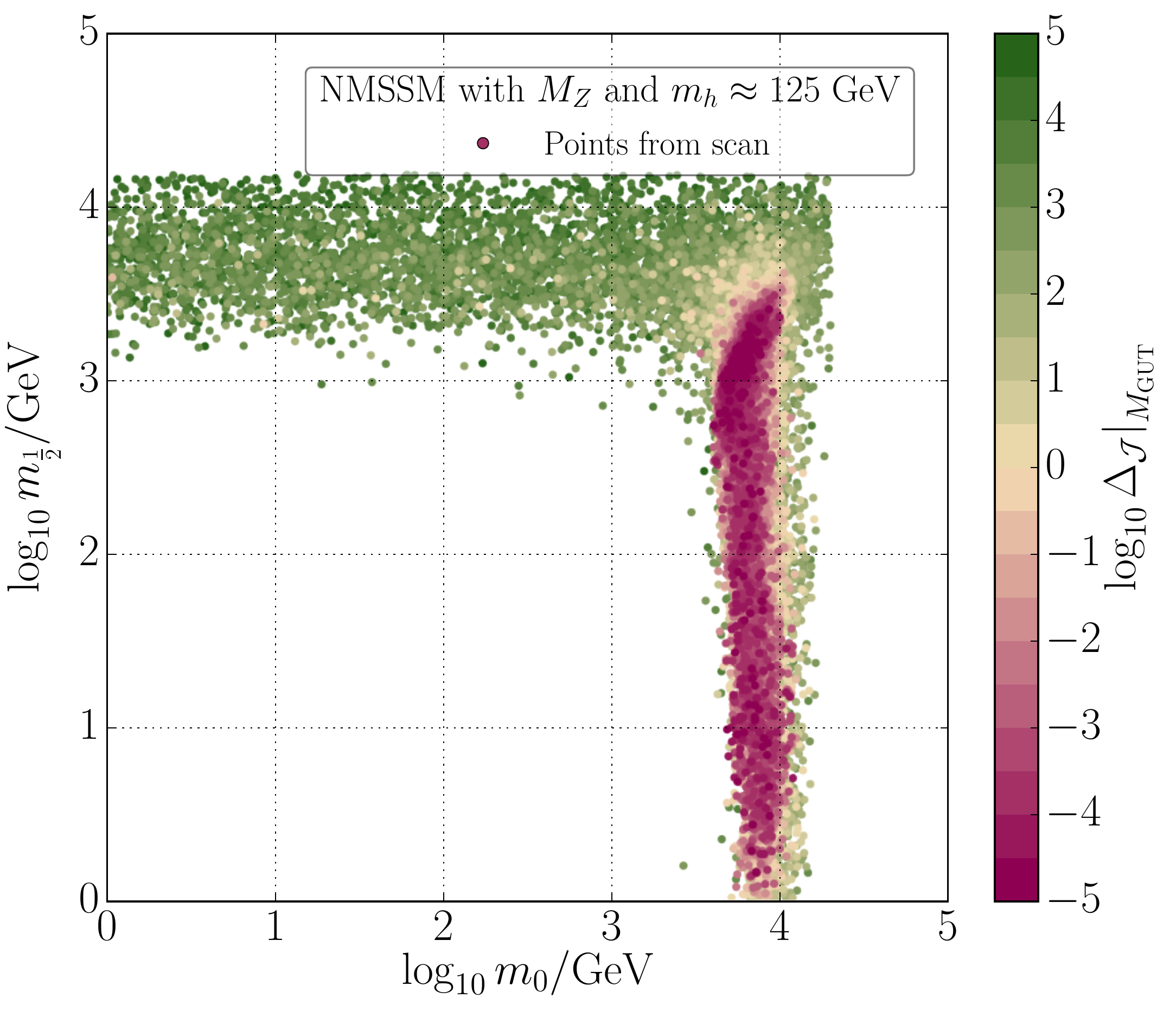}
        \caption{$\Delta_{\jac} \big |_{\mgut}$}
    \end{subfigure}  
    \caption{Comparison of fine-tuning measures in the \CNMSSM with $\mh \approx 125\gev$ on the (\mzero, \mhalf) plane. For their definitions, see \refsec{Subsec:Measures}.}
    \label{fig:cnmssm_fine_tuning_mh}
\end{figure}

Requiring that $\mh \simeq 125\gev$ increases the fine-tuning measures in the \CMSSM (\reffig{fig:cmssm_fine_tuning_mh}) and \CNMSSM (\reffig{fig:cnmssm_fine_tuning_mh}), and further structure is revealed. We find diagonal strips of low fine-tuning for Jacobian-based measures at about $\mzero \sim 10 \tev$ and $\mhalf \sim 1\tev$. The GUT scale Jacobian measure furthermore exhibits a vertical strip of low fine-tuning at $\mzero \sim 10\tev$. This indicates that the Jacobian based measure has a much sharper preference for the focus point region than $\Delta_{BG}$.  Note that this is the case even though we have not included the top mass or top Yukawa coupling in the set of parameters for which we take logarthmic derivatives for $\Delta_\text{BG}$. The Jacobian based measures in the \CNMSSM are also visibly smaller than those in the \CMSSM.

\begin{figure}
    \centering
    \begin{subfigure}[t]{\figwidth}
        \includegraphics[height=\textwidth]{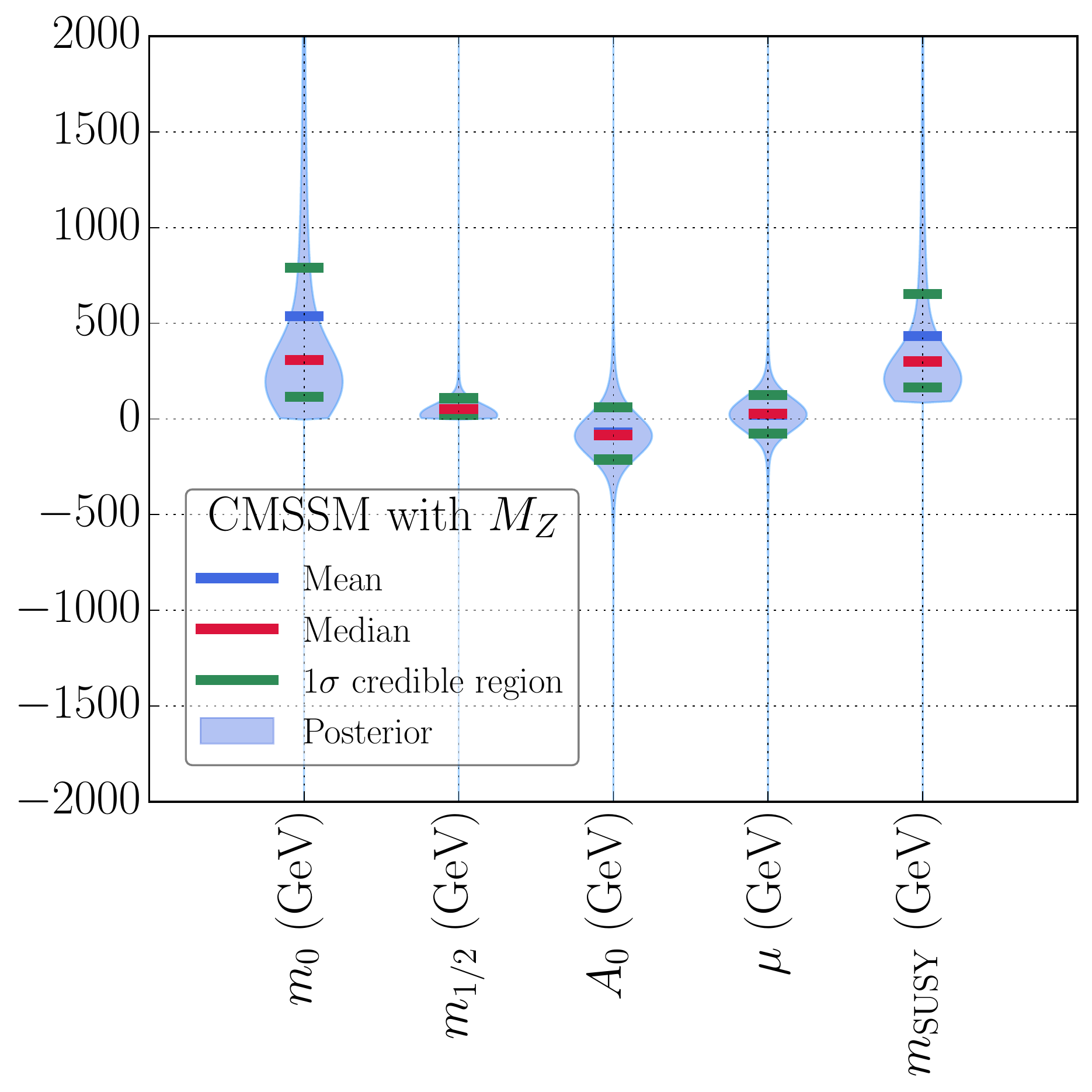}
        \caption{\CMSSM with \mz}
    \end{subfigure}
    \begin{subfigure}[t]{\figwidth}
        \includegraphics[height=0.985\textwidth]{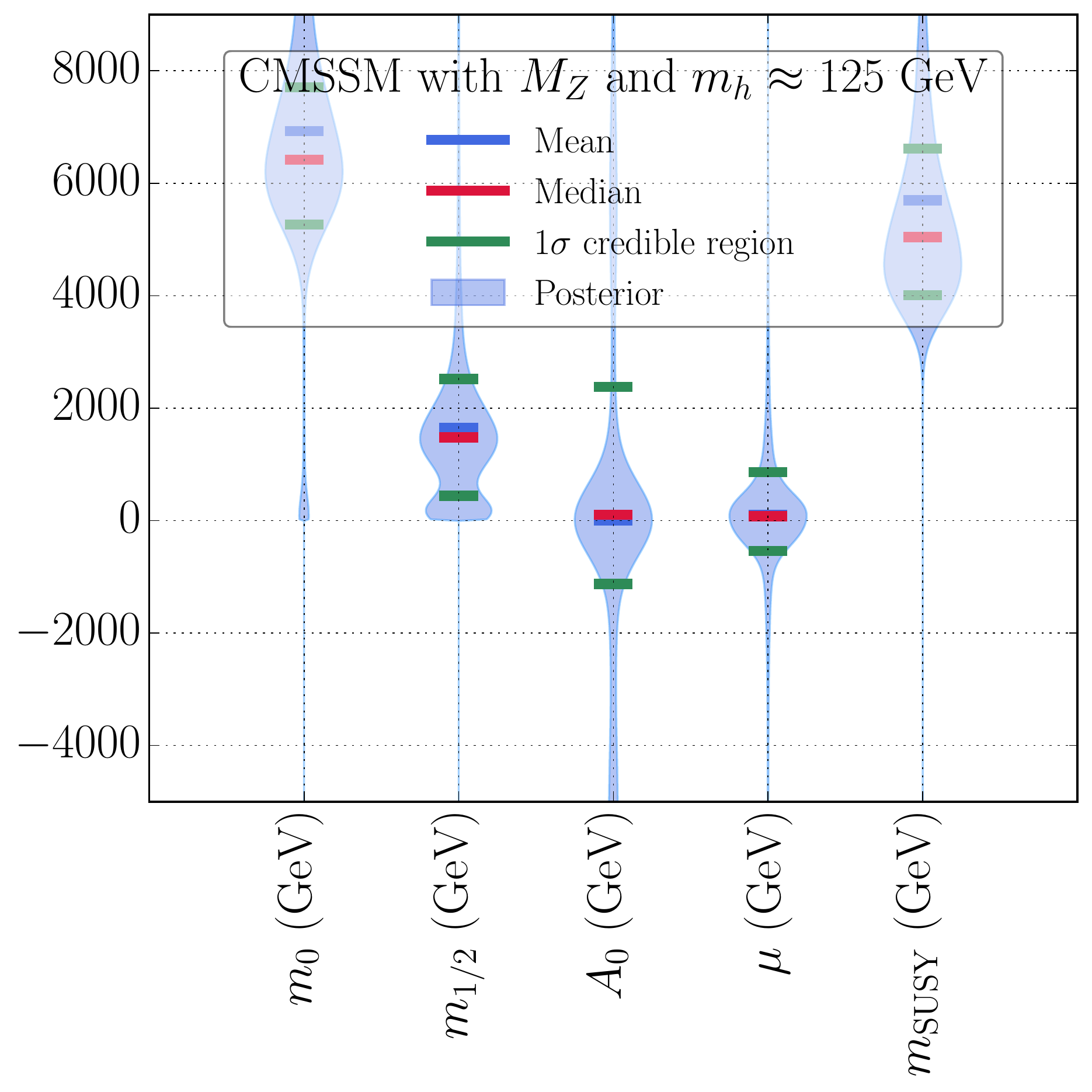}
        \caption{\CMSSM with \mz and \mh}
    \end{subfigure}
    
    \begin{subfigure}[t]{\figwidth}
        \includegraphics[height=\textwidth]{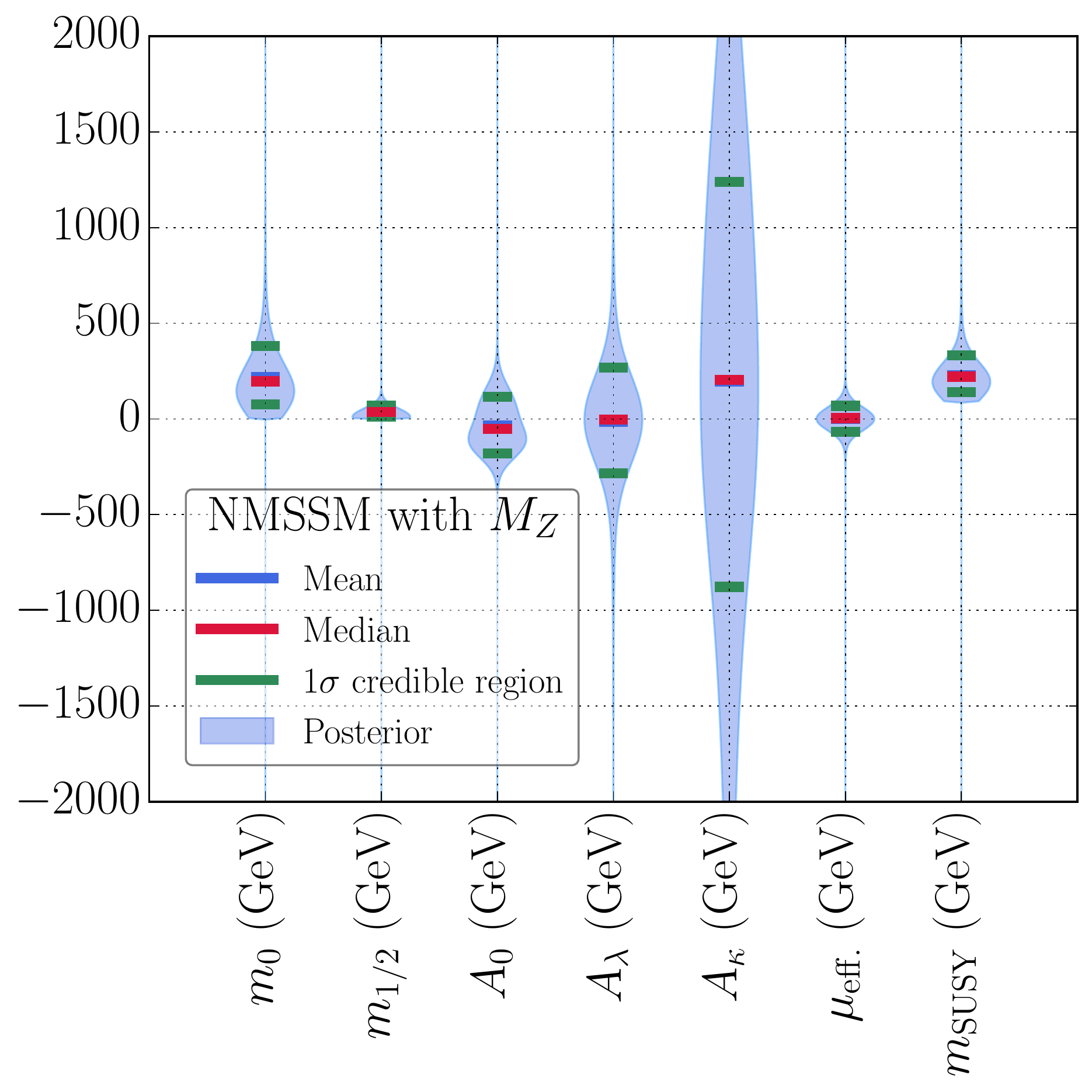}
        \caption{\CNMSSM with \mz}
    \end{subfigure}
    \begin{subfigure}[t]{\figwidth}
        \includegraphics[height=0.985\textwidth]{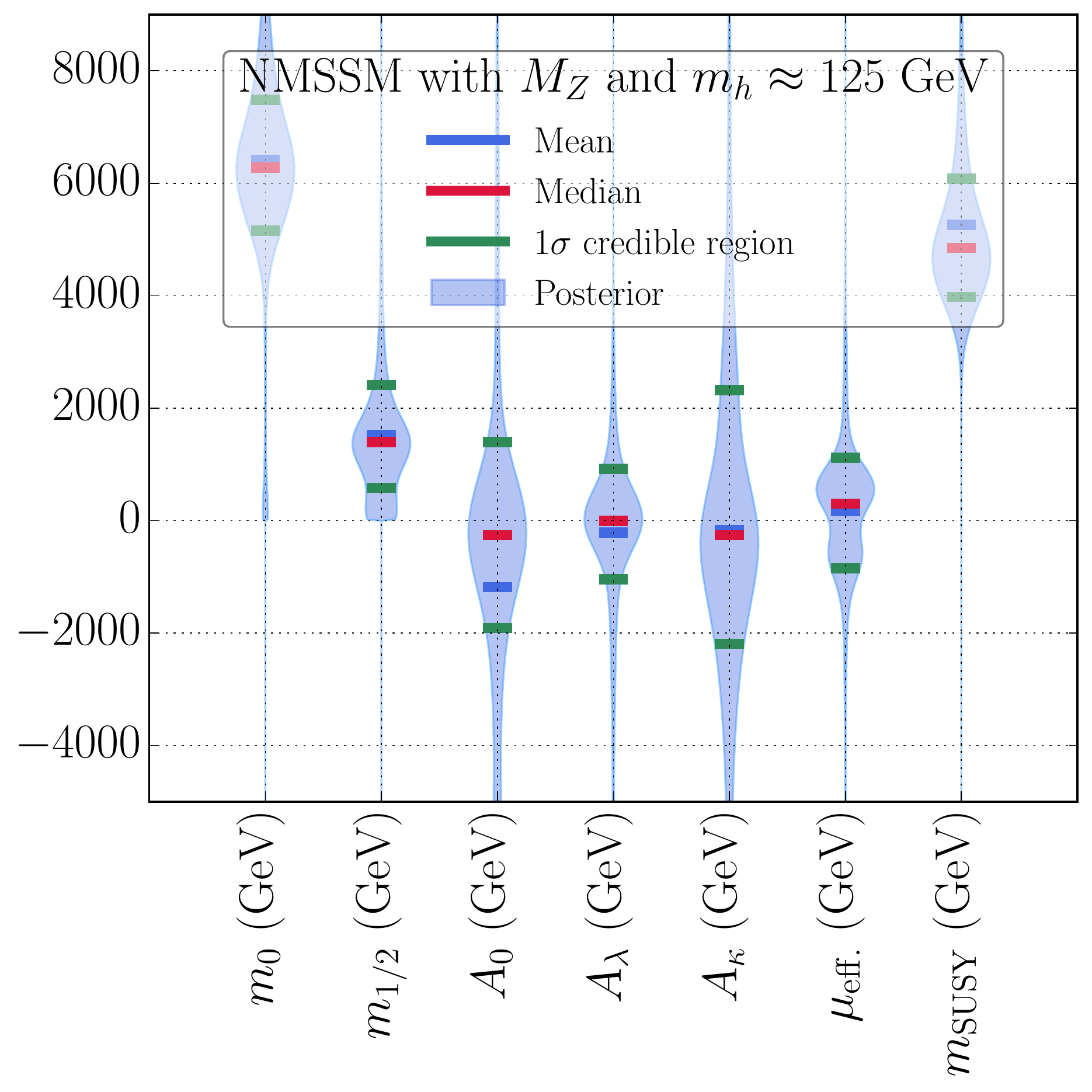}
        \caption{\CNMSSM with \mz and \mh}
    \end{subfigure} 
    \caption{Violin plots showing one-dimensional posterior and summary statistics for important dimensionful parameters in the \CMSSM (upper) and \CNMSSM (lower) with only \mz (left) and \mz and \mh (right) in the likelihood.}
    \label{fig:violin}
\end{figure}

We summarize the one-dimensional posterior for the dimensionful parameters in \reffig{fig:violin}. We see that in the \CMSSM and \CNMSSM with only \mz in the likelihood, the posterior favors $\msusy\lesssim 1\tev$. Once we consider \mz and \mh, however, we require $\msusy \gtrsim 4\tev$ and TeV-scale soft-breaking parameters. It is, therefore, not surprising to see no signature of supersymmetric particles until the current data set of the LHC in the regard that our Higgs mass is $125\gev$.

As a byproduct of our investigations, we calculated the Bayes factor between the semi-constrained \NMSSM and \CMSSM, though with appreciable uncertainty as in \refcite{Fowlie:2014faa}. The Bayes factor measures the change in relative plausibility of two models in light of data. With \mz only, our lower estimate of the Bayes factors favored the \CNMSSM by a factor of about $3$, whereas our upper estimate favored it by a factor of about $20$. With \mz and \mh, our lower estimate favored the \CMSSM by a factor of about $3$, whereas our upper estimate favored the \CNMSSM by a factor of about $6$. This agrees reasonably with a Bayes factor for different data calculated in \refcite{Fowlie:2014faa}. The lower estimates may be more accurate as they were found by importance sampling; however, since there were significant differences between estimates from importance sampling and ordinary summation, we present our results with caution, and do not make a definitive model selection statement. To improve the accuracy of our evidence estimates requires more computational resources, or, possibly, sampling techniques which are more specialised for exploring the very strong degeneracies that can be induced by the naturalness priors in scans constrained only by measurements of \mz and \mh. 















The minimum fine-tuning measures found in our scan are shown in \reftable{Table:FT}. For both the \CMSSM and \CNMSSM, we found minimum fine-tuning measures of about zero for our measures based upon the Jacobian; about $0.3$ for EW fine-tuning; and about $0.1$ for BGEN fine-tuning. If we require that $\mh \approx 125 \gev$, all fine-tuning measures increase, though in this case the Jacobian-based measures and $\Delta_\text{BG}$ in the \CNMSSM are substantially less than those in the \CMSSM. 
The EW measure, $\Delta_\text{EW}$, is very similar in each model. To avoid confusion, it should be stressed again that the numbers are to be compared or interpreted considering the dimensionality or the physical meaning of each measure.

\newcommand{\ten}[2]{\ensuremath{{#1}\times 10^{#2}}}

\begin{table}[htbp]
\centering
\begin{tabular}{lcccc}
\toprule
& \multicolumn{2}{c}{\mz} & \multicolumn{2}{c}{\mz and $\mh \approx 125\gev$}\\
\cmidrule(r){2-3}\cmidrule(r){4-5}
& \CMSSM & \NMSSM & \CMSSM & \NMSSM\\
\midrule
$\Delta_\jac\big|_{\mgut}$ & \ten{3}{-9} & \ten{2}{-10} & 0.004 & \ten{8}{-7}\\
$\Delta_\jac\big|_{\msusy}$ & \ten{6}{-7} & \ten{2}{-10} & 0.005 & \ten{8}{-7}\\
$\Delta_\text{EW}$ & 0.3 & 0.3 & 48.7 & 47.4\\
$\Delta_\text{BG}$ & 0.1 & 0.2 & 451.9 & 133.2\\
\bottomrule
\end{tabular}
\caption{Minimum fine-tuning measures (defined in \refsec{Subsec:Measures}) found in our scans with only \mz in the
likelihood and with the requirement that $\mh \approx 125\gev$.}
\label{Table:FT}
\end{table}

\section{Conclusions}\label{Sec:Conclusions}

After introducing fine-tuning in the context of Bayesian statistics with the Standard Model as an example, we presented a comprehensive analysis of fine-tuning in a minimal and next-to-minimal supersymmetric model. Results of a Bayesian analysis were contrasted with traditional fine-tuning measures, for parameter inference and, briefly, for model selection.

For parameter inference, conditioning upon only \mz we found qualitative agreement between regions favored by the posterior density and regions of low fine-tuning, as measured by, \eg the BGEN measure. Weak-scale soft-breaking masses, \ie $\msusy \sim M_Z$, were favored in a Bayesian analysis, in agreement with heuristic arguments from naturalness. This provided numerical support for our argument, made in the introduction, that naturalness arguments are underpinned by Bayesian statistics. Adding LHC measurements of the Higgs mass to our likelihood pushed the posterior for the soft-breaking masses into a multi-TeV region, as expected.
%
%
This study completes our preliminary work\cite{Kim:2013uxa} and our argument that Bayesian statistics is the correct framework for understanding fine-tuning and naturalness in supersymmetric models.




\section*{Acknowledgements}
 This work was supported by the National Research Foundation
 of Korea (NRF) grant funded by the Korea government (MSIT)
 (No. NRF-2015R1C1A1A02037830), and by IBS under the project
 code, IBS-R018-D1. The work of DH was supported by the
 University of Adelaide and through an Australian Government
 Research Training Program Scholarship, and is also
 supported by the Grant Agency of the Czech Republic (GACR),
 contract 17-04902S. This research, in part, was supported by
 the ARC Centre of Excellence for Particle Physics at the
 Tera-scale, grant CE110001004. The work of PA was also
 supported by Australian Research Council grant, FT160100274.
 The final step of sampling was supported by the National
 Institute of Supercomputing and Network Center/Korea
 Institute of Science and Technology Information
 (KSC-2017-S1-0024).
\appendix

\section{\CMSSM and \CNMSSM Jacobians}\label{App:Jacobian}
In this appendix we present analytic expressions for the Jacobians
that appear in the effective priors as discussed in
\refsec{Sec:EffectivePrior}.

\subsection{\CMSSM Jacobian}
In the \CMSSM, the relevant Jacobian arises from making the change
of variables 
\begin{equation}
\{|\mu_0|, B_0\mu_0\} \to \{\mzpole^2, \tanb\}.
\end{equation}
By performing this trade in two steps, namely, by first exchanging
the high-scale values of the Lagrangian parameters for their values
at the EW scale, and subsequently trading these for the parameters
$\mzpole^2$ and $\tanb$, the full Jacobian factorizes,
\begin{equation} \label{eq:cmssm-jacobian-defn}
\jac^{\text{\text{\CMSSM}}} \equiv \jac^{\text{\CMSSM}}_{\msusy} \jac^{\text{\CMSSM}}_{\mgut} ,
\end{equation}
where the Jacobian determinants on the right-hand side arise from
this series of variable changes, \ie $\{|\mu_0|,B_0\mu_0\} \to
\{|\mu|,B\mu\} \to \{\mzpole^2,\tanb\}$. The
various prior probability density functions are related by
\begin{align}
\pg{\mu_0, B_0\mu_0, \dots}{\text{\CMSSM}} &= \jac^{\text{\CMSSM}}_{\mgut}
\pg{\mu, B\mu, \dots}{\text{\CMSSM}} \label{eq:cmssm-high-scale-jac-defn}
\\
&= \jac^{\text{\CMSSM}}_{\msusy} \jac^{\text{\CMSSM}}_{\mgut}
\pg{\mzpole^2,\tanb,\dots}{\text{\CMSSM}} .
\label{eq:cmssm-susy-scale-jac-defn}
\end{align}
The elements of the two Jacobian matrices that are required read
\begin{equation} \label{eq:cmssm-jacobian-matrices}
J^{\text{\CMSSM}}_{\mgut} = \begin{pmatrix}
\frac{\partial \mu}{\partial \mu_0} &
\frac{\partial \mu}{\partial B_0 \mu_0} \\
\frac{\partial B\mu}{\partial \mu_0} &
\frac{\partial B\mu}{\partial B_0 \mu_0}
\end{pmatrix} , \quad
J^{\text{\CMSSM}}_{\msusy} = \begin{pmatrix}
\frac{\partial \mzpole^2}{\partial \mu} &
\frac{\partial \mzpole^2}{\partial B\mu} \\
\frac{\partial \tanb}{\partial \mu} &
\frac{\partial \tanb}{\partial B\mu}
\end{pmatrix} ,
\end{equation}
with $\jac^{\text{\CMSSM}}_{\mgut} = |\det
J^{\text{\CMSSM}}_{\mgut}|$ and
$\jac^{\text{\CMSSM}}_{\msusy} = |\det J^{\text{\CMSSM}}_{\msusy}|$.

The construction of the Jacobian matrices requires evaluating
derivatives of the functions that implement the changes of
variables from the initial high-scale parameters to the EW
parameters. The first of these trades, $\{|\mu_0|, B_0\mu_0\} \to
\{|\mu|, B\mu\}$, is achieved by integrating the two-loop RGEs
from the GUT scale to the SUSY scale. The dependence of $\mu$ and
$B\mu$ on the \CMSSM parameters defined at $\mgut$ can be explicitly
expressed using semi-analytic solutions to the RGEs, with the
result that
\begin{align}
 \mu(\msusy) &= c^{\mu}_{\mu_0}(\msusy) \mu_0,
 \label{eq:cmssm-semi-analytic-mu-soln} \\
 B\mu(\msusy) &= c^{B\mu}_{B_0\mu_0}(\msusy) B_0 \mu_0
 + c^{B\mu}_{\mu_0 \mhalf}(\msusy) \mu_0 \mhalf
 + c^{B\mu}_{\mu_0 \azero}(\msusy) \mu_0 \azero .
 \label{eq:cmssm-semi-analytic-Bmu-soln}
\end{align}
The elements of $J^{\text{\CMSSM}}_{\mgut}$ can immediately be read
from these expressions. The dimensionless coefficients $c^i_j$
depend only on the running of the gauge and Yukawa couplings;
however, in the absence of exact analytic solutions to the two-loop
RGEs they must be evaluated by numerical integration of the RGEs.

As noted in \refsec{Sec:EffectivePrior}, the subsequent change of
variables from $\{\mu, B\mu\}$ to $\{\mzpole^2, \tanb\}$ is done
by solving the EWSB conditions to write the former pair as functions
of $\mzpole^2$ and $\tanb$. In the \MSSM, the requirement that the 
neutral scalar Higgs fields acquire VEVs of the form given in
\refeq{eq:higgs-vev-convention} leads to the two EWSB conditions,
\begin{align}
(\mu^2 + \mhd^2) v_1 + \frac{\bar{g}^2}{8} (v_1^2 - v_2^2 ) v_1
- B\mu v_2 - t_1 = 0 , \label{eq:mssm-ewsb-condition-1} \\
(\mu^2 + \mhu^2) v_2 - \frac{\bar{g}^2}{8} (v_1^2 - v_2^2) v_2
- B\mu v_1 - t_2 = 0 , \label{eq:mssm-ewsb-condition-2}
\end{align}
where
\begin{equation} \label{eq:tadpole-notation-defn}
t_1 = -\frac{\partial \Delta V_{\text{eff.}}^{\text{\MSSM}}}
{\partial v_1} ,
\quad t_2 = -\frac{\partial \Delta V_{\text{eff.}}^{\text{\MSSM}}}
{\partial v_2}
\end{equation}
contain the one- and two-loop corrections to the Coleman-Weinberg
potential in the \MSSM. \refeq{eq:mssm-ewsb-condition-1} and
\refeq{eq:mssm-ewsb-condition-2} define the VEVs $v_1$ and $v_2$
implicitly in terms of $\mu$ and $B\mu$, allowing the required
derivatives to be written in the form\footnote{Although it is
possible to solve the EWSB conditions explicitly for the VEVs in
the \MSSM at tree-level, once higher-order corrections are also
included this is no longer the case. It is then more straightforward
to utilize the EWSB conditions in the form of
\refeq{eq:mssm-ewsb-condition-1} and
\refeq{eq:mssm-ewsb-condition-2} instead. This approach is also
more appropriate when we consider the \NMSSM, where it is not
possible to solve the EWSB conditions explicitly, even at
tree-level.}
\begin{equation} \label{eq:mssm-vev-derivatives}
\begin{pmatrix}
a_{11} & a_{12} \\
a_{21} & a_{22}
\end{pmatrix}
\begin{pmatrix}
\frac{\partial v_1}{\partial p} \\
\frac{\partial v_2}{\partial p}
\end{pmatrix} =
\begin{pmatrix}
b_1^{(p)} \\ b_2^{(p)}
\end{pmatrix}
\end{equation}
for $p = \mu, B\mu$. The coefficients appearing on the left-hand
side of \refeq{eq:mssm-vev-derivatives} are given by 
(assuming $\mu$ to be real)
\begin{align}
a_{11} &= \mhd^2 + \mu^2 + \frac{\bar{g}^2}{8}(3 v_1^2 - v_2^2)
- \frac{\partial t_1}{\partial v_1} , \label{eq:mssm-df1dv1} \\
a_{12} &= a_{21} = -B\mu - \frac{\bar{g}^2}{4} v_1 v_2 -
\frac{\partial t_1}{\partial v_2} , \label{eq:mssm-df1dv2} \\
a_{22} &= \mhu^2 + \mu^2 - \frac{\bar{g}^2}{8} ( v_1^2 - 3 v_2^2 )
- \frac{\partial t_2}{\partial v_2} , \label{eq:mssm-df2dv2} 
\end{align}
while the derivatives of the EWSB conditions with respect to the
Lagrangian parameters read
\begin{gather}
b_1^{(\mu)} = -2 \mu v_1 + \frac{\partial t_1}{\partial \mu} ,
\quad
b_2^{(\mu)} = -2 \mu v_2 + \frac{\partial t_2}{\partial \mu} ,
\label{eq:mssm-ewsb-mu-derivs} \\
b_1^{(B\mu)} = v_2 + \frac{\partial t_1}{\partial B\mu} ,
\quad
b_2^{(B\mu)} = v_1 + \frac{\partial t_2}{\partial B\mu}.
\label{eq:mssm-ewsb-Bmu-derivs}
\end{gather}
The elements of the Jacobian matrix $J^{\text{\CMSSM}}_{\msusy}$ are
then related to the solution of \refeq{eq:mssm-vev-derivatives}
through
\begin{align}
\frac{\partial \mzpole^2}{\partial p} &= \left ( 
\frac{\bar{g}^2 v_1}{2} - \frac{\partial \re \Pi_{ZZ}^T}
{\partial v_1} \right ) \frac{\partial v_1}{\partial p} + \left (
\frac{\bar{g}^2 v_2}{2} - \frac{\partial \re \Pi_{ZZ}^T}
{\partial v_2} \right ) \frac{\partial v_2}{\partial p} \nonumber \\
& \quad {} - \frac{\partial \re \Pi_{ZZ}^T}{\partial \mu}
\frac{\partial \mu}{\partial p} - \frac{\partial \re \Pi_{ZZ}^T}
{\partial B\mu} \frac{\partial B\mu}{\partial p} , 
\label{eq:mssm-mzsq-derivatives} \\
\frac{\partial \tanb}{\partial p} &= \frac{1}{v_1}
\frac{\partial v_2}{\partial p} - \frac{v_2}{v_1^2}
\frac{\partial v_1}{\partial p} , \label{eq:mssm-tanb-derivatives}
\end{align}
for each of $p = \mu, B\mu$. In arriving at
\refeq{eq:mssm-mzsq-derivatives}, we approximate the solution of
\begin{equation*}
\mzpole^2 = \mzrun^2 - \re \Pi_{ZZ}^T(p^2 = \mzpole^2)
\end{equation*}
for $\mzpole$ by evaluating the $Z$-boson self-energy at the
external momentum $p^2 = \mzrun^2 = \bar{g}^2 (v_1^2 + v_2^2) / 4$.
Although it is possible to evaluate the above derivatives completely
analytically, the resulting expressions are quite long and unwieldy.
As described in \refsec{Sec:NumericalMethods}, for the results 
presented here we have instead computed these derivatives
numerically using \softsusy.

\subsection{\CNMSSM Jacobian}
The calculation of the Jacobian in the \CNMSSM proceeds in a similar
fashion to the approach used in the \CMSSM. As mentioned in
\refsec{Sec:EffectivePrior}, in this model we trade the GUT
scale parameters $\lambda_0$, $\kappa_0$, and $m_{S_0}^2$ for the
low-energy parameters $\mzpole^2$, $\tanb$ and $\lambda$. An
initial exchange of parameters defined at the GUT scale for their
low-energy counterparts, \ie $\lambda$, $\kappa$, and $m_S^2$,
generates a factor of $\jac^{\text{\CNMSSM}}_{\mgut} \equiv | \det
J^{\text{\CNMSSM}}_{\mgut}|$, where the Jacobian matrix has the form
\begin{equation} \label{eq:nmssm-gut-scale-jacobian-matrix}
J^{\text{\CNMSSM}}_{\mgut} = \begin{pmatrix}
\frac{\partial \lambda}{\partial \lambda_0} &
\frac{\partial \lambda}{\partial \kappa_0} &
\frac{\partial \lambda}{\partial m_{S_0}^2} \\
\frac{\partial \kappa}{\partial \lambda_0} &
\frac{\partial \kappa}{\partial \kappa_0} &
\frac{\partial \kappa}{\partial m_{S_0}^2} \\
\frac{\partial m_S^2}{\partial \lambda_0} &
\frac{\partial m_S^2}{\partial \kappa_0} &
\frac{\partial m_S^2}{\partial m_{S_0}^2}
\end{pmatrix}.
\end{equation}
The elements in the last column of this matrix are easily seen to be
given by
\begin{equation*}
\frac{\partial \lambda_0}{\partial m_{S_0}^2} =
\frac{\partial \kappa_0}{\partial m_{S_0}^2} = 0 , \quad
\frac{\partial m_S^2}{\partial m_{S_0}^2} =
c^{m_S^2}_{m_{S_0}^2}(\msusy) ,
\end{equation*}
where the last expression contains the coefficient of $m_{S_0}^2$
in the  semi-analytic solution for $m_S^2$,
\refeq{eq:snmssm-semi-analytic-soft-mass}. Unlike in the case of
the \CMSSM, the dependence of the low-energy parameters on
$\lambda_0$ and $\kappa_0$ cannot be given explicitly, and these
derivatives, along with the coefficient $c^{m_S^2}_{m_{S_0}^2}$,
must be evaluated numerically.

The Jacobian matrix associated with the second change of
variables, $\{\lambda, \kappa, m_S^2\} \to \{\lambda, \mzpole^2,
\tanb \}$, reads
\begin{equation} \label{eq:nmssm-susy-scale-jacobian-matrix}
J^{\text{\CNMSSM}}_{\msusy} = \begin{pmatrix}
\frac{\partial \mzpole^2}{\partial \kappa} &
\frac{\partial \mzpole^2}{\partial m_S^2} & 0 \\
\frac{\partial \tanb}{\partial \kappa} &
\frac{\partial \tanb}{\partial m_S^2} & 0 \\
0 & 0 & 1
\end{pmatrix} ,
\end{equation}
where it should be noted that, since $\lambda$ remains an input
parameter, it is taken to be the case that $\mzpole^2$
and $\tanb$ are independent of $\lambda$. The determinant of this
matrix, $\jac^{\text{\CNMSSM}}_{\msusy} \equiv | \det
J^{\text{\CNMSSM}}_{\msusy}|$, when combined with
$\jac^{\text{\CNMSSM}}_{\mgut}$,
yields the full Jacobian appearing in the effective priors in the
\CNMSSM,
\begin{equation} \label{eq:nmssm-jacobian-defn}
\jac^{\text{\CNMSSM}} = \jac^{\text{\CNMSSM}}_{\msusy} \jac^{\text{\CNMSSM}}_{\mgut} .
\end{equation}
The derivatives of $\mzpole^2$ and $\tanb$ can once again be
expressed in terms of derivatives of the Higgs and singlet VEVs,
$v_1$, $v_2$ and $s$. \refeq{eq:mssm-tanb-derivatives} continues to
hold in the \CNMSSM, with $p = \kappa, m_S^2$, while the dependence
on the additional singlet VEV leads to an expression of the form
\begin{align}
\frac{\partial \mzpole^2}{\partial p} &= \left ( 
\frac{\bar{g}^2 v_1}{2} - \frac{\partial \re \Pi_{ZZ}^T}{\partial v_1} \right ) \frac{\partial v_1}{\partial p} + \left (
\frac{\bar{g}^2 v_2}{2} - \frac{\partial \re \Pi_{ZZ}^T}
{\partial v_2} \right ) \frac{\partial v_2}{\partial p} 
- \frac{\partial \re \Pi_{ZZ}^T}{\partial s} \frac{\partial s}
{\partial p} \nonumber \\
& \quad {} - \frac{\partial \re \Pi_{ZZ}^T}{\partial \kappa}
\frac{\partial \kappa}{\partial p} - \frac{\partial \re \Pi_{ZZ}^T}
{\partial m_S^2} \frac{\partial m_S^2}{\partial p}
\end{align}
for the required derivatives of $\mzpole^2$.

Analytic formulas for the derivatives of the VEVs are most 
conveniently obtained from the three EWSB conditions,
\begin{align}
&\left ( \mhd^2 + \frac{\lambda^2 s^2}{2} \right ) v_1
+ \frac{\lambda^2 v_2^2 v_1}{2} + \frac{\bar{g}^2}{8} (v_1^2
- v_2^2 ) v_1 - \frac{s v_2}{\sqrt{2}} \left ( a_\lambda
+ \frac{\lambda \kappa s}{\sqrt{2}} \right ) - t_1 = 0 ,
\label{eq:nmssm-ewsb-condition-1} \\
& \left ( \mhu^2 + \frac{\lambda^2 s^2}{2} \right ) v_2
+ \frac{\lambda^2 v_1^2 v_2}{2} - \frac{\bar{g}^2}{8} (v_1^2 
- v_2^2) v_2 - \frac{s v_1}{\sqrt{2}} \left ( a_\lambda
+ \frac{\lambda \kappa s}{\sqrt{2}} \right ) - t_2 = 0 ,
\label{eq:nmssm-ewsb-condition-2} \\
&\left [ m_S^2 + \frac{\lambda^2 (v_1^2 + v_2^2)}{2} \right ] s
+ \kappa^2 s^3 + \frac{a_\kappa s^2}{\sqrt{2}}
- v_1 v_2 \left ( \frac{a_\lambda}{\sqrt{2}} 
+ \lambda \kappa s \right ) - t_3 = 0 ,
\label{eq:nmssm-ewsb-condition-3}
\end{align}
where we take there to be no additional sources of CP-violation,
and write the one- and two-loop corrections to the effective 
potential as
\begin{equation} \label{eq:nmssm-tadpole-contributions}
t_1 = -\frac{\partial \Delta V^{\text{\NMSSM}}_{\text{eff.}}}{\partial v_1}
, \quad
t_2 = -\frac{\partial \Delta V^{\text{\NMSSM}}_{\text{eff.}}}{\partial v_2}
, \quad
t_3 = -\frac{\partial \Delta V^{\text{\NMSSM}}_{\text{eff.}}}{\partial s}.
\end{equation}
The quantities $\partial v_1 / \partial p$, $\partial v_2 /
\partial p$ and $\partial s / \partial p$ are then once again
obtained by solving a linear system of the form
\begin{equation} \label{eq:nmssm-vev-derivatives}
X \begin{pmatrix}
\frac{\partial v_1}{\partial p} \\
\frac{\partial v_2}{\partial p} \\
\frac{\partial s}{\partial p}
\end{pmatrix} =
\begin{pmatrix}
y_1^{(p)} \\ y_2^{(p)} \\ y_3^{(p)}
\end{pmatrix} .
\end{equation}
The elements of the $3 \times 3$ matrix $X$ are easily found to
be given by
\begin{align}
x_{11} &= \mhd^2 + \frac{\lambda^2}{2} (s^2 + v_2^2)
+ \frac{\bar{g}^2}{8} (3 v_1^2 - v_2^2) 
- \frac{\partial t_1}{\partial v_1} , \label{eq:nmssm-df1dv1} \\
x_{12} &= x_{21} = \left ( \lambda^2 - \frac{\bar{g}^2}{4} \right )
v_1 v_2 - \frac{s}{\sqrt{2}} \left ( a_\lambda
+ \frac{\lambda \kappa s}{\sqrt{2}} \right ) 
- \frac{\partial t_1}{\partial v_2} , \label{eq:nmssm-df1dv2} \\
x_{13} &= x_{31} = \lambda^2 s v_1 - \frac{v_2}{\sqrt{2}}
( a_\lambda + \sqrt{2} \lambda \kappa s ) 
- \frac{\partial t_1}{\partial s} , \label{eq:nmssm-df1ds} \\
x_{22} &= \mhu^2 + \frac{\lambda^2}{2} (s^2 + v_1^2)
- \frac{\bar{g}^2}{8}(v_1^2 - 3 v_2^2)
- \frac{\partial t_2}{\partial v_2} , \label{eq:nmssm-df2dv2} \\
x_{23} &= x_{32} = \lambda^2 s v_2 - \frac{v_1}{\sqrt{2}}
( a_\lambda + \sqrt{2} \lambda \kappa s )
- \frac{\partial t_2}{\partial s} , \label{eq:nmssm-df2ds} \\
x_{33} &= m_S^2 + \frac{\lambda^2}{2} (v_1^2 + v_2^2)
+ \kappa^2 s^2 + \sqrt{2} a_\kappa s - \lambda \kappa v_1 v_2
- \frac{\partial t_3}{\partial s} . \label{eq:nmssm-df3ds}
\end{align}
Similarly, the derivatives of the EWSB conditions with respect
to $\kappa$ and $m_S^2$ appearing on the right-hand side of
\refeq{eq:nmssm-vev-derivatives} are simply\footnote{Note that
we allow $a_\kappa$ to vary independently of $\kappa$.}
\begin{gather}
y_1^{(\kappa)} = \frac{\lambda s^2 v_2}{2}
+ \frac{\partial t_1}{\partial \kappa} , \quad 
y_2^{(\kappa)} = \frac{\lambda s^2 v_1}{2}
+ \frac{\partial t_2}{\partial \kappa} , \quad
y_3^{(\kappa)} = -2 \kappa s^3 + \lambda v_1 v_2 s
+ \frac{\partial t_3}{\partial \kappa} ,
\label{eq:nmssm-ewsb-kappa-derivs} \\
y_1^{(m_S^2)} = \frac{\partial t_1}{\partial m_S^2}, \quad
y_2^{(m_S^2)} = \frac{\partial t_2}{\partial m_S^2}, \quad
y_3^{(m_S^2)} = -s + \frac{\partial t_3}{\partial m_S^2} .
\label{eq:nmssm-ewsb-ms2-derivs}
\end{gather}

\section{EW Fine-Tuning Contributions} \label{App:ew-tuning}
The tuning measure $\Delta_{\text{EW}}$ defined in
\refeq{eq:ew-tuning-defn} in \refsec{Subsec:Measures} quantifies
the competition between the terms contributing to the EWSB
condition determining $\mzrun$. The $C_i$ are given by the 
absolute values of the terms entering into the prediction of
$\mzrun$ in the model, \ie the terms on the right-hand side of
\refeq{eq:mssm-mz-prediction} or
\refeq{eq:nmssm-mz-prediction}, excluding the self-energy
correction. In the \MSSM we consider the coefficients
\begin{equation}
\begin{gathered}
    C_\mu = -\mu^2 , \quad 
    C_{H_d} = \frac{\mhd^2}{\tan^2\beta - 1} , \quad
    C_{H_u} = -\frac{\mhu^2 \tan^2\beta}{\tan^2\beta - 1} , \\
    C_{t_1} = -\frac{t_1}{v_1(\tan^2\beta - 1)} , \quad
    C_{t_2} = \frac{t_2 \tan^2\beta}{v_2 (\tan^2\beta - 1)} .
\end{gathered} \label{eq:mssm-delta-ew-coefficients}
\end{equation}
Here the quantities $t_1$ and $t_2$ are the Coleman-Weinberg
contributions defined in \refeq{eq:tadpole-notation-defn} and
previously absorbed into $\bar{m}_{H_{u,d}}^2$ in
\refsec{Subsec:Measures}. The coefficients considered in the \NMSSM
are similar, with the only differences being that $\mu \to
\mueff$ and $t_1$, $t_2$ are instead given by
\refeq{eq:nmssm-tadpole-contributions}.

Separating the Coleman-Weinberg pieces allows to see how the loop
corrections in the Higgs potential cancel the tree level parameters
delicately.  The ideal case would be $|C_i| \sim
\mathcal{O}(\mzrun^2)$, while reality pushes them to much larger
values. In the case of large \tanb, the prediction for $\mzrun$
is well approximated by
\begin{equation} \label{eq:large-tanb-mz-limit}
    \frac{1}{2}\mzrun^2 \approx -\mu_{(\text{eff.})}^2
    - \bar{m}^2_{H_u},
\end{equation}
so that $C_{\mu}$, $C_{H_u}$ and $C_{t_2}$ play the most important roles in the
determining $\Delta_{\text{EW}}$.

\bibliography{NMSSM,NON_INSPIRE}

\providecommand{\href}[2]{#2}\begingroup\raggedright\begin{thebibliography}{100}

\bibitem{Agashe:2014kda}
{\bfseries Particle Data Group} Collaboration, K.~A. Olive {\em et~al.},
  ``{Review of Particle Physics},''
\href{http://dx.doi.org/10.1088/1674-1137/38/9/090001}{{\em Chin. Phys.}
  {\bfseries C38} (2014) 090001}.

\bibitem{Chatrchyan:2012xdj}
{\bfseries CMS} Collaboration, S.~Chatrchyan {\em et~al.}, ``{Observation of a
  new boson at a mass of 125 GeV with the CMS experiment at the LHC},''
  \href{http://dx.doi.org/10.1016/j.physletb.2012.08.021}{{\em Phys. Lett.}
  {\bfseries B716} (2012) 30--61},
\href{http://arxiv.org/abs/1207.7235}{{\ttfamily arXiv:1207.7235 [hep-ex]}}.

\bibitem{Aad:2012tfa}
{\bfseries ATLAS} Collaboration, G.~Aad {\em et~al.}, ``{Observation of a new
  particle in the search for the Standard Model Higgs boson with the ATLAS
  detector at the LHC},''
  \href{http://dx.doi.org/10.1016/j.physletb.2012.08.020}{{\em Phys. Lett.}
  {\bfseries B716} (2013) 1--29},
\href{http://arxiv.org/abs/1207.7214}{{\ttfamily arXiv:1207.7214 [hep-ex]}}.

\bibitem{Barbieri:1987fn}
R.~Barbieri and G.~F. Giudice, ``{Upper Bounds on Supersymmetric Particle
  Masses},''
\href{http://dx.doi.org/10.1016/0550-3213(88)90171-X}{{\em Nucl. Phys.}
  {\bfseries B306} (1988) 63}.

\bibitem{Ellis:1986yg}
J.~R. Ellis, K.~Enqvist, D.~V. Nanopoulos, and F.~Zwirner, ``{Observables in
  Low-Energy Superstring Models},''
\href{http://dx.doi.org/10.1142/S0217732386000105}{{\em Mod. Phys. Lett.}
  {\bfseries A1} (1986) 57}.

\bibitem{Witten:1981nf}
E.~Witten, ``{Dynamical Breaking of Supersymmetry},''
\href{http://dx.doi.org/10.1016/0550-3213(81)90006-7}{{\em Nucl. Phys.}
  {\bfseries B188} (1981) 513}.

\bibitem{Weinberg:1975gm}
S.~Weinberg, ``{Implications of Dynamical Symmetry Breaking},''
\href{http://dx.doi.org/10.1103/PhysRevD.13.974}{{\em Phys. Rev.} {\bfseries
  D13} (1976) 974--996}.

\bibitem{Weinberg:1979bn}
S.~Weinberg, ``{Implications of Dynamical Symmetry Breaking: An Addendum},''
\href{http://dx.doi.org/10.1103/PhysRevD.19.1277}{{\em Phys. Rev.} {\bfseries
  D19} (1979) 1277--1280}.

\bibitem{Susskind:1978ms}
L.~Susskind, ``{Dynamics of Spontaneous Symmetry Breaking in the Weinberg-Salam
  Theory},''
\href{http://dx.doi.org/10.1103/PhysRevD.20.2619}{{\em Phys. Rev.} {\bfseries
  D20} (1979) 2619--2625}.

\bibitem{Gildener:1976ai}
E.~Gildener, ``{Gauge Symmetry Hierarchies},''
\href{http://dx.doi.org/10.1103/PhysRevD.14.1667}{{\em Phys. Rev.} {\bfseries
  D14} (1976) 1667}.

\bibitem{Draper:2011aa}
P.~Draper, P.~Meade, M.~Reece, and D.~Shih, ``{Implications of a 125 GeV Higgs
  for the MSSM and Low-Scale SUSY Breaking},''
  \href{http://dx.doi.org/10.1103/PhysRevD.85.095007}{{\em Phys. Rev.}
  {\bfseries D85} (2012) 095007},
\href{http://arxiv.org/abs/1112.3068}{{\ttfamily arXiv:1112.3068 [hep-ph]}}.

\bibitem{Akula:2011aa}
S.~Akula, B.~Altunkaynak, D.~Feldman, P.~Nath, and G.~Peim, ``{Higgs Boson Mass
  Predictions in SUGRA Unification, Recent LHC-7 Results, and Dark Matter},''
  \href{http://dx.doi.org/10.1103/PhysRevD.85.075001}{{\em Phys. Rev.}
  {\bfseries D85} (2012) 075001},
\href{http://arxiv.org/abs/1112.3645}{{\ttfamily arXiv:1112.3645 [hep-ph]}}.

\bibitem{Fowlie:2012im}
A.~Fowlie, M.~Kazana, K.~Kowalska, S.~Munir, L.~Roszkowski, E.~M. Sessolo,
  S.~Trojanowski, and Y.-L.~S. Tsai, ``{The CMSSM Favoring New Territories: The
  Impact of New LHC Limits and a 125 GeV Higgs},''
  \href{http://dx.doi.org/10.1103/PhysRevD.86.075010}{{\em Phys. Rev.}
  {\bfseries D86} (2012) 075010},
\href{http://arxiv.org/abs/1206.0264}{{\ttfamily arXiv:1206.0264 [hep-ph]}}.

\bibitem{Kadastik:2011aa}
M.~Kadastik, K.~Kannike, A.~Racioppi, and M.~Raidal, ``{Implications of the 125
  GeV Higgs boson for scalar dark matter and for the CMSSM phenomenology},''
  \href{http://dx.doi.org/10.1007/JHEP05(2012)061}{{\em JHEP} {\bfseries 05}
  (2012) 061},
\href{http://arxiv.org/abs/1112.3647}{{\ttfamily arXiv:1112.3647 [hep-ph]}}.

\bibitem{Buchmueller:2011ab}
O.~Buchmueller {\em et~al.}, ``{Higgs and Supersymmetry},''
  \href{http://dx.doi.org/10.1140/epjc/s10052-012-2020-3}{{\em Eur. Phys. J.}
  {\bfseries C72} (2012) 2020},
\href{http://arxiv.org/abs/1112.3564}{{\ttfamily arXiv:1112.3564 [hep-ph]}}.

\bibitem{Cao:2011sn}
J.~Cao, Z.~Heng, D.~Li, and J.~M. Yang, ``{Current experimental constraints on
  the lightest Higgs boson mass in the constrained MSSM},''
  \href{http://dx.doi.org/10.1016/j.physletb.2012.03.052}{{\em Phys. Lett.}
  {\bfseries B710} (2012) 665--670},
\href{http://arxiv.org/abs/1112.4391}{{\ttfamily arXiv:1112.4391 [hep-ph]}}.

\bibitem{Draper:2013oza}
P.~Draper, G.~Lee, and C.~E.~M. Wagner, ``{Precise estimates of the Higgs mass
  in heavy supersymmetry},''
  \href{http://dx.doi.org/10.1103/PhysRevD.89.055023}{{\em Phys. Rev.}
  {\bfseries D89} no.~5, (2014) 055023},
\href{http://arxiv.org/abs/1312.5743}{{\ttfamily arXiv:1312.5743 [hep-ph]}}.

\bibitem{Bagnaschi:2014rsa}
E.~Bagnaschi, G.~F. Giudice, P.~Slavich, and A.~Strumia, ``{Higgs Mass and
  Unnatural Supersymmetry},''
  \href{http://dx.doi.org/10.1007/JHEP09(2014)092}{{\em JHEP} {\bfseries 09}
  (2014) 092},
\href{http://arxiv.org/abs/1407.4081}{{\ttfamily arXiv:1407.4081 [hep-ph]}}.

\bibitem{Vega:2015fna}
J.~P. Vega and G.~Villadoro, ``{SusyHD: Higgs mass Determination in
  Supersymmetry},'' \href{http://dx.doi.org/10.1007/JHEP07(2015)159}{{\em JHEP}
  {\bfseries 07} (2015) 159},
\href{http://arxiv.org/abs/1504.05200}{{\ttfamily arXiv:1504.05200 [hep-ph]}}.

\bibitem{Bahl:2016brp}
H.~Bahl and W.~Hollik, ``{Precise prediction for the light MSSM Higgs boson
  mass combining effective field theory and fixed-order calculations},''
  \href{http://dx.doi.org/10.1140/epjc/s10052-016-4354-8}{{\em Eur. Phys. J.}
  {\bfseries C76} no.~9, (2016) 499},
\href{http://arxiv.org/abs/1608.01880}{{\ttfamily arXiv:1608.01880 [hep-ph]}}.

\bibitem{Athron:2016fuq}
P.~Athron, J.-h. Park, T.~Steudtner, D.~Stöckinger, and A.~Voigt, ``{Precise
  Higgs mass calculations in (non-)minimal supersymmetry at both high and low
  scales},'' \href{http://dx.doi.org/10.1007/JHEP01(2017)079}{{\em JHEP}
  {\bfseries 01} (2017) 079},
\href{http://arxiv.org/abs/1609.00371}{{\ttfamily arXiv:1609.00371 [hep-ph]}}.

\bibitem{Bahl:2017aev}
H.~Bahl, S.~Heinemeyer, W.~Hollik, and G.~Weiglein, ``{Reconciling EFT and
  hybrid calculations of the light MSSM Higgs-boson mass},''
\href{http://arxiv.org/abs/1706.00346}{{\ttfamily arXiv:1706.00346 [hep-ph]}}.

\bibitem{Harlander:2017kuc}
R.~V. Harlander, J.~Klappert, and A.~Voigt, ``{Higgs mass prediction in the
  MSSM at three-loop level in a pure $\overline{\text{DR}}$ context},''
\href{http://arxiv.org/abs/1708.05720}{{\ttfamily arXiv:1708.05720 [hep-ph]}}.

\bibitem{Fayet:1974pd}
P.~Fayet, ``{Supergauge Invariant Extension of the Higgs Mechanism and a Model
  for the electron and Its Neutrino},''
\href{http://dx.doi.org/10.1016/0550-3213(75)90636-7}{{\em Nucl. Phys.}
  {\bfseries B90} (1975) 104--124}.

\bibitem{Fayet:1976et}
P.~Fayet, ``{Supersymmetry and Weak, Electromagnetic and Strong
  Interactions},''
\href{http://dx.doi.org/10.1016/0370-2693(76)90319-1}{{\em Phys. Lett.}
  {\bfseries 64B} (1976) 159}.

\bibitem{Fayet:1977yc}
P.~Fayet, ``{Spontaneously Broken Supersymmetric Theories of Weak,
  Electromagnetic and Strong Interactions},''
\href{http://dx.doi.org/10.1016/0370-2693(77)90852-8}{{\em Phys. Lett.}
  {\bfseries 69B} (1977) 489}.

\bibitem{Nilles:1982dy}
H.~P. Nilles, M.~Srednicki, and D.~Wyler, ``{Weak Interaction Breakdown Induced
  by Supergravity},''
\href{http://dx.doi.org/10.1016/0370-2693(83)90460-4}{{\em Phys. Lett.}
  {\bfseries 120B} (1983) 346}.

\bibitem{Frere:1983ag}
J.~M. Frere, D.~R.~T. Jones, and S.~Raby, ``{Fermion Masses and Induction of
  the Weak Scale by Supergravity},''
\href{http://dx.doi.org/10.1016/0550-3213(83)90606-5}{{\em Nucl. Phys.}
  {\bfseries B222} (1983) 11--19}.

\bibitem{Derendinger:1983bz}
J.~P. Derendinger and C.~A. Savoy, ``{Quantum Effects and SU(2) x U(1) Breaking
  in Supergravity Gauge Theories},''
\href{http://dx.doi.org/10.1016/0550-3213(84)90162-7}{{\em Nucl. Phys.}
  {\bfseries B237} (1984) 307--328}.

\bibitem{Veselov:1985gd}
A.~I. Veselov, M.~I. Vysotsky, and K.~A. Ter-Martirosian, ``{LOW-ENERGY
  SUPERGRAVITY AND THE LIGHT t QUARK},'' {\em Sov. Phys. JETP} {\bfseries 63}
  (1986) 489.
[Zh. Eksp. Teor. Fiz.90,838(1986)].

\bibitem{Ellis:1988er}
J.~R. Ellis, J.~F. Gunion, H.~E. Haber, L.~Roszkowski, and F.~Zwirner, ``{Higgs
  Bosons in a Nonminimal Supersymmetric Model},''
\href{http://dx.doi.org/10.1103/PhysRevD.39.844}{{\em Phys. Rev.} {\bfseries
  D39} (1989) 844}.

\bibitem{Drees:1988fc}
M.~Drees, ``{Supersymmetric Models with Extended Higgs Sector},''
\href{http://dx.doi.org/10.1142/S0217751X89001448}{{\em Int. J. Mod. Phys.}
  {\bfseries A4} (1989) 3635}.

\bibitem{Maniatis:2009re}
M.~Maniatis, ``{The Next-to-Minimal Supersymmetric extension of the Standard
  Model reviewed},'' \href{http://dx.doi.org/10.1142/S0217751X10049827}{{\em
  Int. J. Mod. Phys.} {\bfseries A25} (2010) 3505--3602},
\href{http://arxiv.org/abs/0906.0777}{{\ttfamily arXiv:0906.0777 [hep-ph]}}.

\bibitem{Ellwanger:2009dp}
U.~Ellwanger, C.~Hugonie, and A.~M. Teixeira, ``{The Next-to-Minimal
  Supersymmetric Standard Model},''
  \href{http://dx.doi.org/10.1016/j.physrep.2010.07.001}{{\em Phys. Rept.}
  {\bfseries 496} (2010) 1--77},
\href{http://arxiv.org/abs/0910.1785}{{\ttfamily arXiv:0910.1785 [hep-ph]}}.

\bibitem{BasteroGil:2000bw}
M.~Bastero-Gil, C.~Hugonie, S.~F. King, D.~P. Roy, and S.~Vempati, ``{Does LEP
  prefer the NMSSM?},''
  \href{http://dx.doi.org/10.1016/S0370-2693(00)00930-8}{{\em Phys. Lett.}
  {\bfseries B489} (2000) 359--366},
\href{http://arxiv.org/abs/hep-ph/0006198}{{\ttfamily arXiv:hep-ph/0006198
  [hep-ph]}}.

\bibitem{Dermisek:2005ar}
R.~Dermisek and J.~F. Gunion, ``{Escaping the large fine tuning and little
  hierarchy problems in the next to minimal supersymmetric model and $h \to aa$
  decays},'' \href{http://dx.doi.org/10.1103/PhysRevLett.95.041801}{{\em Phys.
  Rev. Lett.} {\bfseries 95} (2005) 041801},
\href{http://arxiv.org/abs/hep-ph/0502105}{{\ttfamily arXiv:hep-ph/0502105
  [hep-ph]}}.

\bibitem{Ellwanger:2011mu}
U.~Ellwanger, G.~Espitalier-Noel, and C.~Hugonie, ``{Naturalness and Fine
  Tuning in the NMSSM: Implicatifons of Early LHC Results},''
  \href{http://dx.doi.org/10.1007/JHEP09(2011)105}{{\em JHEP} {\bfseries 09}
  (2011) 105},
\href{http://arxiv.org/abs/1107.2472}{{\ttfamily arXiv:1107.2472 [hep-ph]}}.

\bibitem{King:2012is}
S.~F. King, M.~Muhlleitner, and R.~Nevzorov, ``{NMSSM Higgs Benchmarks Near 125
  GeV},'' \href{http://dx.doi.org/10.1016/j.nuclphysb.2012.02.010}{{\em Nucl.
  Phys.} {\bfseries B860} (2012) 207--244},
\href{http://arxiv.org/abs/1201.2671}{{\ttfamily arXiv:1201.2671 [hep-ph]}}.

\bibitem{Kang:2012sy}
Z.~Kang, J.~Li, and T.~Li, ``{On Naturalness of the MSSM and NMSSM},''
  \href{http://dx.doi.org/10.1007/JHEP11(2012)024}{{\em JHEP} {\bfseries 11}
  (2012) 024},
\href{http://arxiv.org/abs/1201.5305}{{\ttfamily arXiv:1201.5305 [hep-ph]}}.

\bibitem{Gunion:2012zd}
J.~F. Gunion, Y.~Jiang, and S.~Kraml, ``{The Constrained NMSSM and Higgs near
  125 GeV},'' \href{http://dx.doi.org/10.1016/j.physletb.2012.03.027}{{\em
  Phys. Lett.} {\bfseries B710} (2012) 454--459},
\href{http://arxiv.org/abs/1201.0982}{{\ttfamily arXiv:1201.0982 [hep-ph]}}.

\bibitem{Cao:2012fz}
J.-J. Cao, Z.-X. Heng, J.~M. Yang, Y.-M. Zhang, and J.-Y. Zhu, ``{A SM-like
  Higgs near 125 GeV in low energy SUSY: a comparative study for MSSM and
  NMSSM},'' \href{http://dx.doi.org/10.1007/JHEP03(2012)086}{{\em JHEP}
  {\bfseries 03} (2012) 086},
\href{http://arxiv.org/abs/1202.5821}{{\ttfamily arXiv:1202.5821 [hep-ph]}}.

\bibitem{Ellwanger:2012ke}
U.~Ellwanger and C.~Hugonie, ``{Higgs bosons near 125 GeV in the NMSSM with
  constraints at the GUT scale},''
  \href{http://dx.doi.org/10.1155/2012/625389}{{\em Adv. High Energy Phys.}
  {\bfseries 2012} (2012) 625389},
\href{http://arxiv.org/abs/1203.5048}{{\ttfamily arXiv:1203.5048 [hep-ph]}}.

\bibitem{King:2012tr}
S.~F. King, M.~Mühlleitner, R.~Nevzorov, and K.~Walz, ``{Natural NMSSM Higgs
  Bosons},'' \href{http://dx.doi.org/10.1016/j.nuclphysb.2013.01.020}{{\em
  Nucl. Phys.} {\bfseries B870} (2013) 323--352},
\href{http://arxiv.org/abs/1211.5074}{{\ttfamily arXiv:1211.5074 [hep-ph]}}.

\bibitem{Gherghetta:2012gb}
T.~Gherghetta, B.~von Harling, A.~D. Medina, and M.~A. Schmidt, ``{The
  Scale-Invariant NMSSM and the 126 GeV Higgs Boson},''
  \href{http://dx.doi.org/10.1007/JHEP02(2013)032}{{\em JHEP} {\bfseries 02}
  (2013) 032},
\href{http://arxiv.org/abs/1212.5243}{{\ttfamily arXiv:1212.5243 [hep-ph]}}.

\bibitem{Li:2011ab}
T.~Li, J.~A. Maxin, D.~V. Nanopoulos, and J.~W. Walker, ``{A Higgs Mass Shift
  to 125 GeV and A Multi-Jet Supersymmetry Signal: Miracle of the Flippons at
  the $\sqrt{s} = 7$ TeV LHC},''
  \href{http://dx.doi.org/10.1016/j.physletb.2012.02.086}{{\em Phys. Lett.}
  {\bfseries B710} (2012) 207--214},
\href{http://arxiv.org/abs/1112.3024}{{\ttfamily arXiv:1112.3024 [hep-ph]}}.

\bibitem{Moroi:2011aa}
T.~Moroi, R.~Sato, and T.~T. Yanagida, ``{Extra Matters Decree the Relatively
  Heavy Higgs of Mass about 125 GeV in the Supersymmetric Model},''
  \href{http://dx.doi.org/10.1016/j.physletb.2012.02.012}{{\em Phys. Lett.}
  {\bfseries B709} (2012) 218--221},
\href{http://arxiv.org/abs/1112.3142}{{\ttfamily arXiv:1112.3142 [hep-ph]}}.

\bibitem{Kyae:2012ea}
B.~Kyae and J.-C. Park, ``{Hidden Sector Assisted 125 GeV Higgs},''
  \href{http://dx.doi.org/10.1103/PhysRevD.86.031701}{{\em Phys. Rev.}
  {\bfseries D86} (2012) 031701},
\href{http://arxiv.org/abs/1203.1656}{{\ttfamily arXiv:1203.1656 [hep-ph]}}.

\bibitem{Boudjema:2012cq}
F.~Boudjema and G.~Drieu La~Rochelle, ``{Beyond the MSSM Higgs bosons at 125
  GeV},'' \href{http://dx.doi.org/10.1103/PhysRevD.86.015018}{{\em Phys. Rev.}
  {\bfseries D86} (2012) 015018},
\href{http://arxiv.org/abs/1203.3141}{{\ttfamily arXiv:1203.3141 [hep-ph]}}.

\bibitem{Basak:2012bd}
T.~Basak and S.~Mohanty, ``{Triplet-Singlet Extension of the MSSM with a 125
  GeV Higgs and Dark Matter},''
  \href{http://dx.doi.org/10.1103/PhysRevD.86.075031}{{\em Phys. Rev.}
  {\bfseries D86} (2012) 075031},
\href{http://arxiv.org/abs/1204.6592}{{\ttfamily arXiv:1204.6592 [hep-ph]}}.

\bibitem{Athron:2012sq}
P.~Athron, S.~F. King, D.~J. Miller, S.~Moretti, and R.~Nevzorov,
  ``{Constrained Exceptional Supersymmetric Standard Model with a Higgs Near
  125 GeV},'' \href{http://dx.doi.org/10.1103/PhysRevD.86.095003}{{\em Phys.
  Rev.} {\bfseries D86} (2012) 095003},
\href{http://arxiv.org/abs/1206.5028}{{\ttfamily arXiv:1206.5028 [hep-ph]}}.

\bibitem{Benakli:2012cy}
K.~Benakli, M.~D. Goodsell, and F.~Staub, ``{Dirac Gauginos and the 125 GeV
  Higgs},'' \href{http://dx.doi.org/10.1007/JHEP06(2013)073}{{\em JHEP}
  {\bfseries 06} (2013) 073},
\href{http://arxiv.org/abs/1211.0552}{{\ttfamily arXiv:1211.0552 [hep-ph]}}.

\bibitem{An:2012vp}
H.~An, T.~Liu, and L.-T. Wang, ``{125 GeV Higgs Boson, Enhanced Di-photon Rate,
  and Gauged $U(1)_{PQ}$-Extended MSSM},''
  \href{http://dx.doi.org/10.1103/PhysRevD.86.075030}{{\em Phys. Rev.}
  {\bfseries D86} (2012) 075030},
\href{http://arxiv.org/abs/1207.2473}{{\ttfamily arXiv:1207.2473 [hep-ph]}}.

\bibitem{Kyae:2012rv}
B.~Kyae and J.-C. Park, ``{A Singlet-Extension of the MSSM for 125 GeV Higgs
  with the Least Tuning},''
  \href{http://dx.doi.org/10.1103/PhysRevD.87.075021}{{\em Phys. Rev.}
  {\bfseries D87} (2013) 075021},
\href{http://arxiv.org/abs/1207.3126}{{\ttfamily arXiv:1207.3126 [hep-ph]}}.

\bibitem{Bae:2012ir}
K.~J. Bae, T.~H. Jung, and H.~D. Kim, ``{125 GeV Higgs boson as a
  pseudo-Goldstone boson in supersymmetry with vectorlike matters},''
  \href{http://dx.doi.org/10.1103/PhysRevD.87.015014}{{\em Phys. Rev.}
  {\bfseries D87} no.~1, (2013) 015014},
\href{http://arxiv.org/abs/1208.3748}{{\ttfamily arXiv:1208.3748 [hep-ph]}}.

\bibitem{Bhattacharyya:2012qj}
G.~Bhattacharyya and T.~S. Ray, ``{Pushing the SUSY Higgs mass towards 125 GeV
  with a color adjoint},''
  \href{http://dx.doi.org/10.1103/PhysRevD.87.015017}{{\em Phys. Rev.}
  {\bfseries D87} no.~1, (2013) 015017},
\href{http://arxiv.org/abs/1210.0594}{{\ttfamily arXiv:1210.0594 [hep-ph]}}.

\bibitem{Craig:2012bs}
N.~Craig and A.~Katz, ``{A Supersymmetric Higgs Sector with Chiral D-terms},''
  \href{http://dx.doi.org/10.1007/JHEP05(2013)015}{{\em JHEP} {\bfseries 05}
  (2013) 015},
\href{http://arxiv.org/abs/1212.2635}{{\ttfamily arXiv:1212.2635 [hep-ph]}}.

\bibitem{Basso:2013vla}
L.~Basso, ``{Minimal Z' models and the 125 GeV Higgs boson},''
  \href{http://dx.doi.org/10.1016/j.physletb.2013.07.025}{{\em Phys. Lett.}
  {\bfseries B725} (2013) 322--326},
\href{http://arxiv.org/abs/1303.1084}{{\ttfamily arXiv:1303.1084 [hep-ph]}}.

\bibitem{Maru:2013ooa}
N.~Maru and N.~Okada, ``{Diphoton decay excess and 125 GeV Higgs boson in
  gauge-Higgs unification},''
  \href{http://dx.doi.org/10.1103/PhysRevD.87.095019}{{\em Phys. Rev.}
  {\bfseries D87} no.~9, (2013) 095019},
\href{http://arxiv.org/abs/1303.5810}{{\ttfamily arXiv:1303.5810 [hep-ph]}}.

\bibitem{Galloway:2013dma}
J.~Galloway, M.~A. Luty, Y.~Tsai, and Y.~Zhao, ``{Induced Electroweak Symmetry
  Breaking and Supersymmetric Naturalness},''
  \href{http://dx.doi.org/10.1103/PhysRevD.89.075003}{{\em Phys. Rev.}
  {\bfseries D89} no.~7, (2014) 075003},
\href{http://arxiv.org/abs/1306.6354}{{\ttfamily arXiv:1306.6354 [hep-ph]}}.

\bibitem{Bandyopadhyay:2013lca}
P.~Bandyopadhyay, K.~Huitu, and A.~Sabanci, ``{Status of $Y=0$ Triplet Higgs
  with supersymmetry in the light of $\sim 125$ GeV Higgs discovery},''
  \href{http://dx.doi.org/10.1007/JHEP10(2013)091}{{\em JHEP} {\bfseries 10}
  (2013) 091},
\href{http://arxiv.org/abs/1306.4530}{{\ttfamily arXiv:1306.4530 [hep-ph]}}.

\bibitem{Bharucha:2013ela}
A.~Bharucha, A.~Goudelis, and M.~McGarrie, ``{En-gauging Naturalness},''
  \href{http://dx.doi.org/10.1140/epjc/s10052-014-2858-7}{{\em Eur. Phys. J.}
  {\bfseries C74} (2014) 2858},
\href{http://arxiv.org/abs/1310.4500}{{\ttfamily arXiv:1310.4500 [hep-ph]}}.

\bibitem{Chang:2014zva}
C.-H. Chang, T.-F. Feng, Y.-L. Yan, H.-B. Zhang, and S.-M. Zhao, ``{Spontaneous
  R -parity violation in the minimal gauged (B-L) supersymmetry with a 125 GeV
  Higgs boson},'' \href{http://dx.doi.org/10.1103/PhysRevD.90.035013}{{\em
  Phys. Rev.} {\bfseries D90} no.~3, (2014) 035013},
\href{http://arxiv.org/abs/1401.4586}{{\ttfamily arXiv:1401.4586 [hep-ph]}}.

\bibitem{Bertuzzo:2014bwa}
E.~Bertuzzo, C.~Frugiuele, T.~Gregoire, and E.~Ponton, ``{Dirac gauginos, R
  symmetry and the 125 GeV Higgs},''
  \href{http://dx.doi.org/10.1007/JHEP04(2015)089}{{\em JHEP} {\bfseries 04}
  (2015) 089},
\href{http://arxiv.org/abs/1402.5432}{{\ttfamily arXiv:1402.5432 [hep-ph]}}.

\bibitem{Dimopoulos:2014aua}
S.~Dimopoulos, K.~Howe, and J.~March-Russell, ``{Maximally Natural
  Supersymmetry},''
  \href{http://dx.doi.org/10.1103/PhysRevLett.113.111802}{{\em Phys. Rev.
  Lett.} {\bfseries 113} (2014) 111802},
\href{http://arxiv.org/abs/1404.7554}{{\ttfamily arXiv:1404.7554 [hep-ph]}}.

\bibitem{Bertuzzo:2014sma}
E.~Bertuzzo and C.~Frugiuele, ``{Natural SM-like 126 GeV Higgs boson via
  nondecoupling D terms},''
  \href{http://dx.doi.org/10.1103/PhysRevD.93.035019}{{\em Phys. Rev.}
  {\bfseries D93} no.~3, (2016) 035019},
\href{http://arxiv.org/abs/1412.2765}{{\ttfamily arXiv:1412.2765 [hep-ph]}}.

\bibitem{Ding:2015wma}
R.~Ding, T.~Li, F.~Staub, C.~Tian, and B.~Zhu, ``{Supersymmetric standard
  models with a pseudo-Dirac gluino from hybrid F - and D -term supersymmetry
  breaking},'' \href{http://dx.doi.org/10.1103/PhysRevD.92.015008}{{\em Phys.
  Rev.} {\bfseries D92} no.~1, (2015) 015008},
\href{http://arxiv.org/abs/1502.03614}{{\ttfamily arXiv:1502.03614 [hep-ph]}}.

\bibitem{Belanger:2015cra}
G.~Bélanger, J.~Da~Silva, U.~Laa, and A.~Pukhov, ``{Probing U(1) extensions of
  the MSSM at the LHC Run I and in dark matter searches},''
  \href{http://dx.doi.org/10.1007/JHEP09(2015)151}{{\em JHEP} {\bfseries 09}
  (2015) 151},
\href{http://arxiv.org/abs/1505.06243}{{\ttfamily arXiv:1505.06243 [hep-ph]}}.

\bibitem{Kim:2015yha}
D.~Kim and B.~Kyae, ``{Naturalness-guided Gluino Mass Bound from the Minimal
  Mixed Mediation of SUSY Breaking},''
  \href{http://dx.doi.org/10.1103/PhysRevD.92.075025}{{\em Phys. Rev.}
  {\bfseries D92} no.~7, (2015) 075025},
\href{http://arxiv.org/abs/1507.07611}{{\ttfamily arXiv:1507.07611 [hep-ph]}}.

\bibitem{Capdevilla:2015qwa}
R.~M. Capdevilla, A.~Delgado, and A.~Martin, ``{Light Stops in a minimal U(1)x
  extension of the MSSM},''
  \href{http://dx.doi.org/10.1103/PhysRevD.92.115020}{{\em Phys. Rev.}
  {\bfseries D92} no.~11, (2015) 115020},
\href{http://arxiv.org/abs/1509.02472}{{\ttfamily arXiv:1509.02472 [hep-ph]}}.

\bibitem{Nakai:2015swg}
Y.~Nakai, M.~Reece, and R.~Sato, ``{SUSY Higgs Mass and Collider Signals with a
  Hidden Valley},'' \href{http://dx.doi.org/10.1007/JHEP03(2016)143}{{\em JHEP}
  {\bfseries 03} (2016) 143},
\href{http://arxiv.org/abs/1511.00691}{{\ttfamily arXiv:1511.00691 [hep-ph]}}.

\bibitem{Okada:2016wlm}
N.~Okada and H.~M. Tran, ``{125 GeV Higgs boson mass and muon $g-2$ in 5D
  MSSM},'' \href{http://dx.doi.org/10.1103/PhysRevD.94.075016}{{\em Phys. Rev.}
  {\bfseries D94} no.~7, (2016) 075016},
\href{http://arxiv.org/abs/1606.05329}{{\ttfamily arXiv:1606.05329 [hep-ph]}}.

\bibitem{Hebbar:2017fit}
A.~Hebbar, G.~Lazarides, and Q.~Shafi, ``{$\psi'$MSSM: Light Sterile Neutrinos,
  Dark Matter, and New Resonances},''
\href{http://arxiv.org/abs/1706.09630}{{\ttfamily arXiv:1706.09630 [hep-ph]}}.

\bibitem{Badziak:2017kjk}
M.~Badziak and K.~Harigaya, ``{Minimal Non-Abelian Supersymmetric Twin
  Higgs},''
\href{http://arxiv.org/abs/1707.09071}{{\ttfamily arXiv:1707.09071 [hep-ph]}}.

\bibitem{Gregory}
P.~Gregory, {\em {Bayesian Logical Data Analysis for the Physical Sciences}}.
\newblock Cambridge University Press, 2005.

\bibitem{Jaynes:2003}
E.~T. Jaynes, {\em {Probability theory: the logic of science}}.
\newblock Cambridge University Press, 2003.

\bibitem{Earman}
J.~Earman, {\em {Bayes or bust?: a critical examination of Bayesian
  confirmation theory}}.
\newblock MIT Press, 1992.

\bibitem{Allanach:2007qk}
B.~C. Allanach, K.~Cranmer, C.~G. Lester, and A.~M. Weber, ``{Natural priors,
  CMSSM fits and LHC weather forecasts},''
  \href{http://dx.doi.org/10.1088/1126-6708/2007/08/023}{{\em JHEP} {\bfseries
  08} (2007) 023},
\href{http://arxiv.org/abs/0705.0487}{{\ttfamily arXiv:0705.0487 [hep-ph]}}.

\bibitem{Cabrera:2008tj}
M.~E. Cabrera, J.~A. Casas, and R.~Ruiz~de Austri, ``{Bayesian approach and
  Naturalness in MSSM analyses for the LHC},''
  \href{http://dx.doi.org/10.1088/1126-6708/2009/03/075}{{\em JHEP} {\bfseries
  03} (2009) 075},
\href{http://arxiv.org/abs/0812.0536}{{\ttfamily arXiv:0812.0536 [hep-ph]}}.

\bibitem{Fowlie:2014xha}
A.~Fowlie, ``{CMSSM, naturalness and the ``fine-tuning price'' of the Very
  Large Hadron Collider},''
  \href{http://dx.doi.org/10.1103/PhysRevD.90.015010}{{\em Phys. Rev.}
  {\bfseries D90} (2014) 015010},
\href{http://arxiv.org/abs/1403.3407}{{\ttfamily arXiv:1403.3407 [hep-ph]}}.

\bibitem{Fichet:2012sn}
S.~Fichet, ``{Quantified naturalness from Bayesian statistics},''
  \href{http://dx.doi.org/10.1103/PhysRevD.86.125029}{{\em Phys. Rev.}
  {\bfseries D86} (2012) 125029},
\href{http://arxiv.org/abs/1204.4940}{{\ttfamily arXiv:1204.4940 [hep-ph]}}.

\bibitem{Cabrera:2010dh}
M.~E. Cabrera, ``{Bayesian Study and Naturalness in MSSM Forecast for the
  LHC},'' in {\em {Proceedings, 45th Rencontres de Moriond on Electroweak
  Interactions and Unified Theories}}.
\newblock 2010.
\newblock
\href{http://arxiv.org/abs/1005.2525}{{\ttfamily arXiv:1005.2525 [hep-ph]}}.
\newblock

\bibitem{Fowlie:2014faa}
A.~Fowlie, ``{Is the CNMSSM more credible than the CMSSM?},''
  \href{http://dx.doi.org/10.1140/epjc/s10052-014-3105-y}{{\em Eur. Phys. J.}
  {\bfseries C74} no.~10, (2014) 3105},
\href{http://arxiv.org/abs/1407.7534}{{\ttfamily arXiv:1407.7534 [hep-ph]}}.

\bibitem{Fowlie:2015uga}
A.~Fowlie, ``{The little-hierarchy problem is a little problem: understanding
  the difference between the big- and little-hierarchy problems with Bayesian
  probability},''
\href{http://arxiv.org/abs/1506.03786}{{\ttfamily arXiv:1506.03786 [hep-ph]}}.

\bibitem{Clarke:2016jzm}
J.~D. Clarke and P.~Cox, ``{Naturalness made easy: two-loop naturalness bounds
  on minimal SM extensions},''
  \href{http://dx.doi.org/10.1007/JHEP02(2017)129}{{\em JHEP} {\bfseries 02}
  (2017) 129},
\href{http://arxiv.org/abs/1607.07446}{{\ttfamily arXiv:1607.07446 [hep-ph]}}.

\bibitem{Fundira:2017vip}
P.~Fundira and A.~Purves, ``{Bayesian naturalness, simplicity, and testability
  applied to the $B-L$ MSSM GUT using GPU Monte Carlo},''
\href{http://arxiv.org/abs/1708.07835}{{\ttfamily arXiv:1708.07835 [hep-ph]}}.

\bibitem{Kim:2013uxa}
D.~Kim, P.~Athron, C.~Balázs, B.~Farmer, and E.~Hutchison, ``{Bayesian
  naturalness of the CMSSM and CNMSSM},''
  \href{http://dx.doi.org/10.1103/PhysRevD.90.055008}{{\em Phys. Rev.}
  {\bfseries D90} no.~5, (2014) 055008},
\href{http://arxiv.org/abs/1312.4150}{{\ttfamily arXiv:1312.4150 [hep-ph]}}.

\bibitem{LopezFogliani:2009np}
D.~E. Lopez-Fogliani, L.~Roszkowski, R.~Ruiz~de Austri, and T.~A. Varley, ``{A
  Bayesian Analysis of the Constrained NMSSM},''
  \href{http://dx.doi.org/10.1103/PhysRevD.80.095013}{{\em Phys. Rev.}
  {\bfseries D80} (2009) 095013},
\href{http://arxiv.org/abs/0906.4911}{{\ttfamily arXiv:0906.4911 [hep-ph]}}.

\bibitem{Kowalska:2012gs}
K.~Kowalska, S.~Munir, L.~Roszkowski, E.~M. Sessolo, S.~Trojanowski, and
  Y.-L.~S. Tsai, ``{Constrained next-to-minimal supersymmetric standard model
  with a 126 GeV Higgs boson: A global analysis},''
  \href{http://dx.doi.org/10.1103/PhysRevD.87.115010}{{\em Phys. Rev.}
  {\bfseries D87} (2013) 115010},
\href{http://arxiv.org/abs/1211.1693}{{\ttfamily arXiv:1211.1693 [hep-ph]}}.

\bibitem{Akula:2012kk}
S.~Akula, P.~Nath, and G.~Peim, ``{Implications of the Higgs Boson Discovery
  for mSUGRA},'' \href{http://dx.doi.org/10.1016/j.physletb.2012.09.007}{{\em
  Phys. Lett.} {\bfseries B717} (2012) 188--192},
\href{http://arxiv.org/abs/1207.1839}{{\ttfamily arXiv:1207.1839 [hep-ph]}}.

\bibitem{Williams:2015bfa}
A.~J. Williams, ``{Explaining the Fermi Galactic Centre Excess in the CMSSM},''
\href{http://arxiv.org/abs/1510.00714}{{\ttfamily arXiv:1510.00714 [hep-ph]}}.

\bibitem{Diamanti:2015kma}
R.~Diamanti, M.~E.~C. Catalan, and S.~Ando, ``{Dark matter protohalos in a nine
  parameter MSSM and implications for direct and indirect detection},''
  \href{http://dx.doi.org/10.1103/PhysRevD.92.065029}{{\em Phys. Rev.}
  {\bfseries D92} no.~6, (2015) 065029},
\href{http://arxiv.org/abs/1506.01529}{{\ttfamily arXiv:1506.01529 [hep-ph]}}.

\bibitem{Catalan:2015cna}
M.~E. Cabrera-Catalan, S.~Ando, C.~Weniger, and F.~Zandanel, ``{Indirect and
  direct detection prospect for TeV dark matter in the nine parameter MSSM},''
  \href{http://dx.doi.org/10.1103/PhysRevD.92.035018}{{\em Phys. Rev.}
  {\bfseries D92} no.~3, (2015) 035018},
\href{http://arxiv.org/abs/1503.00599}{{\ttfamily arXiv:1503.00599 [hep-ph]}}.

\bibitem{Casas:2014eca}
J.~A. Casas, J.~M. Moreno, S.~Robles, K.~Rolbiecki, and B.~Zaldívar, ``{What
  is a Natural SUSY scenario?},''
  \href{http://dx.doi.org/10.1007/JHEP06(2015)070}{{\em JHEP} {\bfseries 06}
  (2015) 070},
\href{http://arxiv.org/abs/1407.6966}{{\ttfamily arXiv:1407.6966 [hep-ph]}}.

\bibitem{Roszkowski:2014wqa}
L.~Roszkowski, E.~M. Sessolo, and A.~J. Williams, ``{What next for the CMSSM
  and the NUHM: Improved prospects for superpartner and dark matter
  detection},'' \href{http://dx.doi.org/10.1007/JHEP08(2014)067}{{\em JHEP}
  {\bfseries 08} (2014) 067},
\href{http://arxiv.org/abs/1405.4289}{{\ttfamily arXiv:1405.4289 [hep-ph]}}.

\bibitem{Strege:2014ija}
C.~Strege, G.~Bertone, G.~J. Besjes, S.~Caron, R.~Ruiz~de Austri, A.~Strubig,
  and R.~Trotta, ``{Profile likelihood maps of a 15-dimensional MSSM},''
  \href{http://dx.doi.org/10.1007/JHEP09(2014)081}{{\em JHEP} {\bfseries 09}
  (2014) 081},
\href{http://arxiv.org/abs/1405.0622}{{\ttfamily arXiv:1405.0622 [hep-ph]}}.

\bibitem{Fowlie:2014awa}
A.~Fowlie and M.~Raidal, ``{Prospects for constrained supersymmetry at
  $\sqrt{s}={33}\,\text {TeV} $ and $\sqrt{s}={100}\,\text {TeV} $
  proton-proton super-colliders},''
  \href{http://dx.doi.org/10.1140/epjc/s10052-014-2948-6}{{\em Eur. Phys. J.}
  {\bfseries C74} (2014) 2948},
\href{http://arxiv.org/abs/1402.5419}{{\ttfamily arXiv:1402.5419 [hep-ph]}}.

\bibitem{AbdusSalam:2013qba}
S.~S. AbdusSalam, ``{Stop-mass prediction in naturalness scenarios within
  MSSM-25},'' \href{http://dx.doi.org/10.1142/S0217751X14501607}{{\em Int. J.
  Mod. Phys.} {\bfseries A29} no.~27, (2014) 1450160},
\href{http://arxiv.org/abs/1312.7830}{{\ttfamily arXiv:1312.7830 [hep-ph]}}.

\bibitem{Cabrera:2013jya}
M.~E. Cabrera, A.~Casas, R.~Ruiz~de Austri, and G.~Bertone, ``{LHC and dark
  matter phenomenology of the NUGHM},''
  \href{http://dx.doi.org/10.1007/JHEP12(2014)114}{{\em JHEP} {\bfseries 12}
  (2014) 114},
\href{http://arxiv.org/abs/1311.7152}{{\ttfamily arXiv:1311.7152 [hep-ph]}}.

\bibitem{Arina:2013zca}
C.~Arina and M.~E. Cabrera, ``{Multi-lepton signatures at LHC from sneutrino
  dark matter},'' \href{http://dx.doi.org/10.1007/JHEP04(2014)100}{{\em JHEP}
  {\bfseries 04} (2014) 100},
\href{http://arxiv.org/abs/1311.6549}{{\ttfamily arXiv:1311.6549 [hep-ph]}}.

\bibitem{deAustri:2013saa}
R.~Ruiz~de Austri and C.~Pérez de~los Heros, ``{Impact of nucleon matrix
  element uncertainties on the interpretation of direct and indirect dark
  matter search results},''
  \href{http://dx.doi.org/10.1088/1475-7516/2013/11/049}{{\em JCAP} {\bfseries
  1311} (2013) 049},
\href{http://arxiv.org/abs/1307.6668}{{\ttfamily arXiv:1307.6668 [hep-ph]}}.

\bibitem{Fowlie:2013oua}
A.~Fowlie, K.~Kowalska, L.~Roszkowski, E.~M. Sessolo, and Y.-L.~S. Tsai,
  ``{Dark matter and collider signatures of the MSSM},''
  \href{http://dx.doi.org/10.1103/PhysRevD.88.055012}{{\em Phys. Rev.}
  {\bfseries D88} (2013) 055012},
\href{http://arxiv.org/abs/1306.1567}{{\ttfamily arXiv:1306.1567 [hep-ph]}}.

\bibitem{Antusch:2013kna}
S.~Antusch, C.~Gross, V.~Maurer, and C.~Sluka, ``{A flavour GUT model with
  $\theta_{13}^{PMNS} \simeq \theta_C/\sqrt2$},''
  \href{http://dx.doi.org/10.1016/j.nuclphysb.2013.11.003}{{\em Nucl. Phys.}
  {\bfseries B877} (2013) 772--791},
\href{http://arxiv.org/abs/1305.6612}{{\ttfamily arXiv:1305.6612 [hep-ph]}}.

\bibitem{Cabrera:2012vu}
M.~E. Cabrera, J.~A. Casas, and R.~Ruiz~de Austri, ``{The health of SUSY after
  the Higgs discovery and the XENON100 data},''
  \href{http://dx.doi.org/10.1007/JHEP07(2013)182}{{\em JHEP} {\bfseries 07}
  (2013) 182},
\href{http://arxiv.org/abs/1212.4821}{{\ttfamily arXiv:1212.4821 [hep-ph]}}.

\bibitem{Strege:2012bt}
C.~Strege, G.~Bertone, F.~Feroz, M.~Fornasa, R.~Ruiz~de Austri, and R.~Trotta,
  ``{Global Fits of the cMSSM and NUHM including the LHC Higgs discovery and
  new XENON100 constraints},''
  \href{http://dx.doi.org/10.1088/1475-7516/2013/04/013}{{\em JCAP} {\bfseries
  1304} (2013) 013},
\href{http://arxiv.org/abs/1212.2636}{{\ttfamily arXiv:1212.2636 [hep-ph]}}.

\bibitem{Balazs:2012bx}
C.~Balázs and S.~K. Gupta, ``{Peccei-Quinn violating minimal supergravity and
  a 126 GeV Higgs boson},''
  \href{http://dx.doi.org/10.1103/PhysRevD.87.035023}{{\em Phys. Rev.}
  {\bfseries D87} no.~3, (2013) 035023},
\href{http://arxiv.org/abs/1212.1708}{{\ttfamily arXiv:1212.1708 [hep-ph]}}.

\bibitem{Balazs:2013qva}
C.~Balazs, A.~Buckley, D.~Carter, B.~Farmer, and M.~White, ``{Should we still
  believe in constrained supersymmetry?},''
  \href{http://dx.doi.org/10.1140/epjc/s10052-013-2563-y}{{\em Eur. Phys. J.}
  {\bfseries C73} (2013) 2563},
\href{http://arxiv.org/abs/1205.1568}{{\ttfamily arXiv:1205.1568 [hep-ph]}}.

\bibitem{Roszkowski:2012uf}
L.~Roszkowski, E.~M. Sessolo, and Y.-L.~S. Tsai, ``{Bayesian Implications of
  Current LHC Supersymmetry and Dark Matter Detection Searches for the
  Constrained MSSM},'' \href{http://dx.doi.org/10.1103/PhysRevD.86.095005}{{\em
  Phys. Rev.} {\bfseries D86} (2012) 095005},
\href{http://arxiv.org/abs/1202.1503}{{\ttfamily arXiv:1202.1503 [hep-ph]}}.

\bibitem{Strege:2011pk}
C.~Strege, G.~Bertone, D.~G. Cerdeno, M.~Fornasa, R.~Ruiz~de Austri, and
  R.~Trotta, ``{Updated global fits of the cMSSM including the latest LHC SUSY
  and Higgs searches and XENON100 data},''
  \href{http://dx.doi.org/10.1088/1475-7516/2012/03/030}{{\em JCAP} {\bfseries
  1203} (2012) 030},
\href{http://arxiv.org/abs/1112.4192}{{\ttfamily arXiv:1112.4192 [hep-ph]}}.

\bibitem{Fowlie:2011mb}
A.~Fowlie, A.~Kalinowski, M.~Kazana, L.~Roszkowski, and Y.~L.~S. Tsai,
  ``{Bayesian Implications of Current LHC and XENON100 Search Limits for the
  Constrained MSSM},'' \href{http://dx.doi.org/10.1103/PhysRevD.85.075012}{{\em
  Phys. Rev.} {\bfseries D85} (2012) 075012},
\href{http://arxiv.org/abs/1111.6098}{{\ttfamily arXiv:1111.6098 [hep-ph]}}.

\bibitem{Bertone:2011pq}
G.~Bertone, D.~G. Cerdeno, M.~Fornasa, L.~Pieri, R.~Ruiz~de Austri, and
  R.~Trotta, ``{Complementarity of Indirect and Accelerator Dark Matter
  Searches},'' \href{http://dx.doi.org/10.1103/PhysRevD.85.055014}{{\em Phys.
  Rev.} {\bfseries D85} (2012) 055014},
\href{http://arxiv.org/abs/1111.2607}{{\ttfamily arXiv:1111.2607
  [astro-ph.HE]}}.

\bibitem{Cabrera:2011ds}
M.~E. Cabrera, J.~A. Casas, V.~A. Mitsou, R.~Ruiz~de Austri, and J.~Terron,
  ``{Histogram comparison as a powerful tool for the search of new physics at
  LHC. Application to CMSSM},''
  \href{http://dx.doi.org/10.1007/JHEP04(2012)133}{{\em JHEP} {\bfseries 04}
  (2012) 133},
\href{http://arxiv.org/abs/1109.3759}{{\ttfamily arXiv:1109.3759 [hep-ph]}}.

\bibitem{Allanach:2011ya}
B.~C. Allanach and M.~J. Dolan, ``{Supersymmetry With Prejudice: Fitting the
  Wrong Model to LHC Data},''
  \href{http://dx.doi.org/10.1103/PhysRevD.86.055022}{{\em Phys. Rev.}
  {\bfseries D86} (2012) 055022},
\href{http://arxiv.org/abs/1107.2856}{{\ttfamily arXiv:1107.2856 [hep-ph]}}.

\bibitem{Bertone:2011nj}
G.~Bertone, D.~G. Cerdeno, M.~Fornasa, R.~Ruiz~de Austri, C.~Strege, and
  R.~Trotta, ``{Global fits of the cMSSM including the first LHC and XENON100
  data},'' \href{http://dx.doi.org/10.1088/1475-7516/2012/01/015}{{\em JCAP}
  {\bfseries 1201} (2012) 015},
\href{http://arxiv.org/abs/1107.1715}{{\ttfamily arXiv:1107.1715 [hep-ph]}}.

\bibitem{Fowlie:2011vf}
A.~Fowlie and L.~Roszkowski, ``{Reconstructing ATLAS SU3 in the CMSSM and
  relaxed phenomenological supersymmetry models},''
\href{http://arxiv.org/abs/1106.5117}{{\ttfamily arXiv:1106.5117 [hep-ph]}}.

\bibitem{Allanach:2011ut}
B.~C. Allanach, ``{Impact of CMS Multi-jets and Missing Energy Search on CMSSM
  Fits},'' \href{http://dx.doi.org/10.1103/PhysRevD.83.095019}{{\em Phys. Rev.}
  {\bfseries D83} (2011) 095019},
\href{http://arxiv.org/abs/1102.3149}{{\ttfamily arXiv:1102.3149 [hep-ph]}}.

\bibitem{Feroz:2011bj}
F.~Feroz, K.~Cranmer, M.~Hobson, R.~Ruiz~de Austri, and R.~Trotta,
  ``{Challenges of Profile Likelihood Evaluation in Multi-Dimensional SUSY
  Scans},'' \href{http://dx.doi.org/10.1007/JHEP06(2011)042}{{\em JHEP}
  {\bfseries 06} (2011) 042},
\href{http://arxiv.org/abs/1101.3296}{{\ttfamily arXiv:1101.3296 [hep-ph]}}.

\bibitem{Ripken:2010ja}
J.~Ripken, J.~Conrad, and P.~Scott, ``{Implications for constrained
  supersymmetry of combined H.E.S.S. observations of dwarf galaxies, the
  Galactic halo and the Galactic Centre},''
  \href{http://dx.doi.org/10.1088/1475-7516/2011/11/004}{{\em JCAP} {\bfseries
  1111} (2011) 004},
\href{http://arxiv.org/abs/1012.3939}{{\ttfamily arXiv:1012.3939
  [astro-ph.HE]}}.

\bibitem{Cabrera:2010xx}
M.~E. Cabrera, J.~A. Casas, R.~Ruiz~de Austri, and R.~Trotta, ``{Quantifying
  the tension between the Higgs mass and $(g-2)_\mu$ in the CMSSM},''
  \href{http://dx.doi.org/10.1103/PhysRevD.84.015006}{{\em Phys. Rev.}
  {\bfseries D84} (2011) 015006},
\href{http://arxiv.org/abs/1011.5935}{{\ttfamily arXiv:1011.5935 [hep-ph]}}.

\bibitem{Akrami:2010cz}
Y.~Akrami, C.~Savage, P.~Scott, J.~Conrad, and J.~Edsjo, ``{Statistical
  coverage for supersymmetric parameter estimation: a case study with direct
  detection of dark matter},''
  \href{http://dx.doi.org/10.1088/1475-7516/2011/07/002}{{\em JCAP} {\bfseries
  1107} (2011) 002},
\href{http://arxiv.org/abs/1011.4297}{{\ttfamily arXiv:1011.4297 [hep-ph]}}.

\bibitem{Cabrera:2009dm}
M.~E. Cabrera, J.~A. Casas, and R.~Ruiz~de Austri, ``{MSSM Forecast for the
  LHC},'' \href{http://dx.doi.org/10.1007/JHEP05(2010)043}{{\em JHEP}
  {\bfseries 05} (2010) 043},
\href{http://arxiv.org/abs/0911.4686}{{\ttfamily arXiv:0911.4686 [hep-ph]}}.

\bibitem{Akrami:2009hp}
Y.~Akrami, P.~Scott, J.~Edsjo, J.~Conrad, and L.~Bergstrom, ``{A Profile
  Likelihood Analysis of the Constrained MSSM with Genetic Algorithms},''
  \href{http://dx.doi.org/10.1007/JHEP04(2010)057}{{\em JHEP} {\bfseries 04}
  (2010) 057},
\href{http://arxiv.org/abs/0910.3950}{{\ttfamily arXiv:0910.3950 [hep-ph]}}.

\bibitem{Roszkowski:2009ye}
L.~Roszkowski, R.~Ruiz~de Austri, and R.~Trotta, ``{Efficient reconstruction of
  CMSSM parameters from LHC data: A Case study},''
  \href{http://dx.doi.org/10.1103/PhysRevD.82.055003}{{\em Phys. Rev.}
  {\bfseries D82} (2010) 055003},
\href{http://arxiv.org/abs/0907.0594}{{\ttfamily arXiv:0907.0594 [hep-ph]}}.

\bibitem{Roszkowski:2009sm}
L.~Roszkowski, R.~Ruiz~de Austri, R.~Trotta, Y.-L.~S. Tsai, and T.~A. Varley,
  ``{Global fits of the Non-Universal Higgs Model},''
  \href{http://dx.doi.org/10.1103/PhysRevD.83.015014,
  10.1103/PhysRevD.83.039901}{{\em Phys. Rev.} {\bfseries D83} no.~1, (2011)
  015014}, \href{http://arxiv.org/abs/0903.1279}{{\ttfamily arXiv:0903.1279
  [hep-ph]}}.
[Erratum: \textit{Phys. Rev.} \textbf{D83} no. 3, (2011) 039901].

\bibitem{Trotta:2008bp}
R.~Trotta, F.~Feroz, M.~P. Hobson, L.~Roszkowski, and R.~Ruiz~de Austri, ``{The
  Impact of priors and observables on parameter inferences in the Constrained
  MSSM},'' \href{http://dx.doi.org/10.1088/1126-6708/2008/12/024}{{\em JHEP}
  {\bfseries 12} (2008) 024},
\href{http://arxiv.org/abs/0809.3792}{{\ttfamily arXiv:0809.3792 [hep-ph]}}.

\bibitem{Feroz:2008wr}
F.~Feroz, B.~C. Allanach, M.~Hobson, S.~S. AbdusSalam, R.~Trotta, and A.~M.
  Weber, ``{Bayesian Selection of sign(mu) within mSUGRA in Global Fits
  Including WMAP5 Results},''
  \href{http://dx.doi.org/10.1088/1126-6708/2008/10/064}{{\em JHEP} {\bfseries
  10} (2008) 064},
\href{http://arxiv.org/abs/0807.4512}{{\ttfamily arXiv:0807.4512 [hep-ph]}}.

\bibitem{Allanach:2008iq}
B.~C. Allanach and D.~Hooper, ``{Panglossian Prospects for Detecting Neutralino
  Dark Matter in Light of Natural Priors},''
  \href{http://dx.doi.org/10.1088/1126-6708/2008/10/071}{{\em JHEP} {\bfseries
  10} (2008) 071},
\href{http://arxiv.org/abs/0806.1923}{{\ttfamily arXiv:0806.1923 [hep-ph]}}.

\bibitem{Allanach:2008tu}
B.~C. Allanach, M.~J. Dolan, and A.~M. Weber, ``{Global Fits of the Large
  Volume String Scenario to WMAP5 and Other Indirect Constraints Using Markov
  Chain Monte Carlo},''
  \href{http://dx.doi.org/10.1088/1126-6708/2008/08/105}{{\em JHEP} {\bfseries
  08} (2008) 105},
\href{http://arxiv.org/abs/0806.1184}{{\ttfamily arXiv:0806.1184 [hep-ph]}}.

\bibitem{Allanach:2008zn}
B.~C. Allanach, ``{SUSY Predictions and SUSY Tools at the LHC},''
  \href{http://dx.doi.org/10.1140/epjc/s10052-008-0695-2}{{\em Eur. Phys. J.}
  {\bfseries C59} (2009) 427--443},
\href{http://arxiv.org/abs/0805.2088}{{\ttfamily arXiv:0805.2088 [hep-ph]}}.

\bibitem{Roszkowski:2007fd}
L.~Roszkowski, R.~Ruiz~de Austri, and R.~Trotta, ``{Implications for the
  Constrained MSSM from a new prediction for $b \to s \gamma$},''
  \href{http://dx.doi.org/10.1088/1126-6708/2007/07/075}{{\em JHEP} {\bfseries
  07} (2007) 075},
\href{http://arxiv.org/abs/0705.2012}{{\ttfamily arXiv:0705.2012 [hep-ph]}}.

\bibitem{Roszkowski:2006mi}
L.~Roszkowski, R.~Ruiz~de Austri, and R.~Trotta, ``{On the detectability of the
  CMSSM light Higgs boson at the Tevatron},''
  \href{http://dx.doi.org/10.1088/1126-6708/2007/04/084}{{\em JHEP} {\bfseries
  04} (2007) 084},
\href{http://arxiv.org/abs/hep-ph/0611173}{{\ttfamily arXiv:hep-ph/0611173
  [hep-ph]}}.

\bibitem{Olive:2016xmw}
{\bfseries Particle Data Group} Collaboration, C.~Patrignani {\em et~al.},
  ``{Review of Particle Physics},''
\href{http://dx.doi.org/10.1088/1674-1137/40/10/100001}{{\em Chin. Phys.}
  {\bfseries C40} no.~10, (2016) 100001}.

\bibitem{Kim:1983dt}
J.~E. Kim and H.~P. Nilles, ``{The mu Problem and the Strong CP Problem},''
\href{http://dx.doi.org/10.1016/0370-2693(84)91890-2}{{\em Phys. Lett.}
  {\bfseries B138} (1984) 150}.

\bibitem{Chamseddine:1982jx}
A.~H. Chamseddine, R.~L. Arnowitt, and P.~Nath, ``{Locally Supersymmetric Grand
  Unification},''
\href{http://dx.doi.org/10.1103/PhysRevLett.49.970}{{\em Phys. Rev. Lett.}
  {\bfseries 49} (1982) 970}.

\bibitem{Arnowitt:1992aq}
R.~L. Arnowitt and P.~Nath, ``{SUSY mass spectrum in SU(5) supergravity grand
  unification},''
\href{http://dx.doi.org/10.1103/PhysRevLett.69.725}{{\em Phys. Rev. Lett.}
  {\bfseries 69} (1992) 725--728}.

\bibitem{AbdusSalam:2011fc}
S.~S. AbdusSalam {\em et~al.}, ``{Benchmark Models, Planes, Lines and Points
  for Future SUSY Searches at the LHC},''
  \href{http://dx.doi.org/10.1140/epjc/s10052-011-1835-7}{{\em Eur. Phys. J.}
  {\bfseries C71} (2011) 1835},
\href{http://arxiv.org/abs/1109.3859}{{\ttfamily arXiv:1109.3859 [hep-ph]}}.

\bibitem{Kane:1993td}
G.~L. Kane, C.~F. Kolda, L.~Roszkowski, and J.~D. Wells, ``{Study of
  constrained minimal supersymmetry},''
  \href{http://dx.doi.org/10.1103/PhysRevD.49.6173}{{\em Phys. Rev.} {\bfseries
  D49} (1994) 6173--6210},
\href{http://arxiv.org/abs/hep-ph/9312272}{{\ttfamily arXiv:hep-ph/9312272
  [hep-ph]}}.

\bibitem{Allanach:2001kg}
B.~C. Allanach, ``{SOFTSUSY: a program for calculating supersymmetric
  spectra},'' \href{http://dx.doi.org/10.1016/S0010-4655(01)00460-X}{{\em
  Comput. Phys. Commun.} {\bfseries 143} (2002) 305--331},
\href{http://arxiv.org/abs/hep-ph/0104145}{{\ttfamily arXiv:hep-ph/0104145
  [hep-ph]}}.

\bibitem{Allanach:2013kza}
B.~C. Allanach, P.~Athron, L.~C. Tunstall, A.~Voigt, and A.~G. Williams,
  ``{Next-to-Minimal SOFTSUSY},''
  \href{http://dx.doi.org/10.1016/j.cpc.2014.04.015}{{\em Comput. Phys.
  Commun.} {\bfseries 185} (2014) 2322--2339},
\href{http://arxiv.org/abs/1311.7659}{{\ttfamily arXiv:1311.7659 [hep-ph]}}.

\bibitem{Coleman:1973jx}
S.~R. Coleman and E.~J. Weinberg, ``{Radiative Corrections as the Origin of
  Spontaneous Symmetry Breaking},''
\href{http://dx.doi.org/10.1103/PhysRevD.7.1888}{{\em Phys. Rev.} {\bfseries
  D7} (1973) 1888--1910}.

\bibitem{Chan:1997bi}
K.~L. Chan, U.~Chattopadhyay, and P.~Nath, ``{Naturalness, weak scale
  supersymmetry and the prospect for the observation of supersymmetry at the
  Tevatron and at the CERN LHC},''
  \href{http://dx.doi.org/10.1103/PhysRevD.58.096004}{{\em Phys. Rev.}
  {\bfseries D58} (1998) 096004},
\href{http://arxiv.org/abs/hep-ph/9710473}{{\ttfamily arXiv:hep-ph/9710473
  [hep-ph]}}.

\bibitem{Feng:1999hg}
J.~L. Feng and T.~Moroi, ``{Supernatural supersymmetry: Phenomenological
  implications of anomaly mediated supersymmetry breaking},''
  \href{http://dx.doi.org/10.1103/PhysRevD.61.095004}{{\em Phys. Rev.}
  {\bfseries D61} (2000) 095004},
\href{http://arxiv.org/abs/hep-ph/9907319}{{\ttfamily arXiv:hep-ph/9907319
  [hep-ph]}}.

\bibitem{Feng:1999mn}
J.~L. Feng, K.~T. Matchev, and T.~Moroi, ``{Multi-TeV scalars are natural in
  minimal supergravity},''
  \href{http://dx.doi.org/10.1103/PhysRevLett.84.2322}{{\em Phys. Rev. Lett.}
  {\bfseries 84} (2000) 2322--2325},
\href{http://arxiv.org/abs/hep-ph/9908309}{{\ttfamily arXiv:hep-ph/9908309
  [hep-ph]}}.

\bibitem{Baer:2012up}
H.~Baer, V.~Barger, P.~Huang, A.~Mustafayev, and X.~Tata, ``{Radiative natural
  SUSY with a 125 GeV Higgs boson},''
  \href{http://dx.doi.org/10.1103/PhysRevLett.109.161802}{{\em Phys. Rev.
  Lett.} {\bfseries 109} (2012) 161802},
\href{http://arxiv.org/abs/1207.3343}{{\ttfamily arXiv:1207.3343 [hep-ph]}}.

\bibitem{deCarlos:1993rbr}
B.~de~Carlos and J.~A. Casas, ``{One loop analysis of the electroweak breaking
  in supersymmetric models and the fine tuning problem},''
  \href{http://dx.doi.org/10.1016/0370-2693(93)90940-J}{{\em Phys. Lett.}
  {\bfseries B309} (1993) 320--328},
\href{http://arxiv.org/abs/hep-ph/9303291}{{\ttfamily arXiv:hep-ph/9303291
  [hep-ph]}}.

\bibitem{deCarlos:1993ca}
B.~de~Carlos and J.~A. Casas, ``{The Fine tuning problem of the electroweak
  symmetry breaking mechanism in minimal SUSY models},'' in {\em {16th
  International Warsaw Meeting on Elementary Particle Physics: New Physics at
  New Experiments Kazimierz, Poland, May 24-28, 1993}}.
\newblock 1993.
\newblock
\href{http://arxiv.org/abs/hep-ph/9310232}{{\ttfamily arXiv:hep-ph/9310232
  [hep-ph]}}.
\newblock

\bibitem{Chankowski:1997zh}
P.~H. Chankowski, J.~R. Ellis, and S.~Pokorski, ``{The Fine tuning price of
  LEP},'' \href{http://dx.doi.org/10.1016/S0370-2693(98)00060-4}{{\em Phys.
  Lett.} {\bfseries B423} (1998) 327--336},
\href{http://arxiv.org/abs/hep-ph/9712234}{{\ttfamily arXiv:hep-ph/9712234
  [hep-ph]}}.

\bibitem{Agashe:1997kn}
K.~Agashe and M.~Graesser, ``{Improving the fine tuning in models of low-energy
  gauge mediated supersymmetry breaking},''
  \href{http://dx.doi.org/10.1016/S0550-3213(97)00569-5}{{\em Nucl. Phys.}
  {\bfseries B507} (1997) 3--34},
\href{http://arxiv.org/abs/hep-ph/9704206}{{\ttfamily arXiv:hep-ph/9704206
  [hep-ph]}}.

\bibitem{Wright:1998mk}
D.~Wright, ``{Naturally nonminimal supersymmetry},''
\href{http://arxiv.org/abs/hep-ph/9801449}{{\ttfamily arXiv:hep-ph/9801449
  [hep-ph]}}.

\bibitem{Kane:1998im}
G.~L. Kane and S.~F. King, ``{Naturalness implications of LEP results},''
  \href{http://dx.doi.org/10.1016/S0370-2693(99)00190-2}{{\em Phys. Lett.}
  {\bfseries B451} (1999) 113--122},
\href{http://arxiv.org/abs/hep-ph/9810374}{{\ttfamily arXiv:hep-ph/9810374
  [hep-ph]}}.

\bibitem{BasteroGil:1999gu}
M.~Bastero-Gil, G.~L. Kane, and S.~F. King, ``{Fine tuning constraints on
  supergravity models},''
  \href{http://dx.doi.org/10.1016/S0370-2693(00)00002-2}{{\em Phys. Lett.}
  {\bfseries B474} (2000) 103--112},
\href{http://arxiv.org/abs/hep-ph/9910506}{{\ttfamily arXiv:hep-ph/9910506
  [hep-ph]}}.

\bibitem{Feng:1999zg}
J.~L. Feng, K.~T. Matchev, and T.~Moroi, ``{Focus points and naturalness in
  supersymmetry},'' \href{http://dx.doi.org/10.1103/PhysRevD.61.075005}{{\em
  Phys. Rev.} {\bfseries D61} (2000) 075005},
\href{http://arxiv.org/abs/hep-ph/9909334}{{\ttfamily arXiv:hep-ph/9909334
  [hep-ph]}}.

\bibitem{Allanach:2000ii}
B.~C. Allanach, J.~P.~J. Hetherington, M.~A. Parker, and B.~R. Webber,
  ``{Naturalness reach of the large hadron collider in minimal supergravity},''
  {\em JHEP} {\bfseries 08} (2000) 017,
\href{http://arxiv.org/abs/hep-ph/0005186}{{\ttfamily arXiv:hep-ph/0005186
  [hep-ph]}}.

\bibitem{Barbieri:2005kf}
R.~Barbieri and L.~J. Hall, ``{Improved naturalness and the two Higgs doublet
  model},''
\href{http://arxiv.org/abs/hep-ph/0510243}{{\ttfamily arXiv:hep-ph/0510243
  [hep-ph]}}.

\bibitem{Allanach:2006jc}
B.~C. Allanach, ``{Naturalness priors and fits to the constrained minimal
  supersymmetric standard model},''
  \href{http://dx.doi.org/10.1016/j.physletb.2006.02.052}{{\em Phys. Lett.}
  {\bfseries B635} (2006) 123--130},
\href{http://arxiv.org/abs/hep-ph/0601089}{{\ttfamily arXiv:hep-ph/0601089
  [hep-ph]}}.

\bibitem{Gripaios:2006nn}
B.~Gripaios and S.~M. West, ``{Improved Higgs naturalness with or without
  supersymmetry},'' \href{http://dx.doi.org/10.1103/PhysRevD.74.075002}{{\em
  Phys. Rev.} {\bfseries D74} (2006) 075002},
\href{http://arxiv.org/abs/hep-ph/0603229}{{\ttfamily arXiv:hep-ph/0603229
  [hep-ph]}}.

\bibitem{Dermisek:2006py}
R.~Dermisek, J.~F. Gunion, and B.~McElrath, ``{Probing NMSSM Scenarios with
  Minimal Fine-Tuning by Searching for Decays of the Upsilon to a Light CP-Odd
  Higgs Boson},'' \href{http://dx.doi.org/10.1103/PhysRevD.76.051105}{{\em
  Phys. Rev.} {\bfseries D76} (2007) 051105},
\href{http://arxiv.org/abs/hep-ph/0612031}{{\ttfamily arXiv:hep-ph/0612031
  [hep-ph]}}.

\bibitem{Barbieri:2006dq}
R.~Barbieri, L.~J. Hall, and V.~S. Rychkov, ``{Improved naturalness with a
  heavy Higgs: An Alternative road to LHC physics},''
  \href{http://dx.doi.org/10.1103/PhysRevD.74.015007}{{\em Phys. Rev.}
  {\bfseries D74} (2006) 015007},
\href{http://arxiv.org/abs/hep-ph/0603188}{{\ttfamily arXiv:hep-ph/0603188
  [hep-ph]}}.

\bibitem{Kobayashi:2006fh}
T.~Kobayashi, H.~Terao, and A.~Tsuchiya, ``{Fine-tuning in gauge mediated
  supersymmetry breaking models and induced top Yukawa coupling},''
  \href{http://dx.doi.org/10.1103/PhysRevD.74.015002}{{\em Phys. Rev.}
  {\bfseries D74} (2006) 015002},
\href{http://arxiv.org/abs/hep-ph/0604091}{{\ttfamily arXiv:hep-ph/0604091
  [hep-ph]}}.

\bibitem{Perelstein:2012qg}
M.~Perelstein and B.~Shakya, ``{XENON100 implications for naturalness in the
  MSSM, NMSSM, and $\lambda$-supersymmetry model},''
  \href{http://dx.doi.org/10.1103/PhysRevD.88.075003}{{\em Phys. Rev.}
  {\bfseries D88} no.~7, (2013) 075003},
\href{http://arxiv.org/abs/1208.0833}{{\ttfamily arXiv:1208.0833 [hep-ph]}}.

\bibitem{Antusch:2012gv}
S.~Antusch, L.~Calibbi, V.~Maurer, M.~Monaco, and M.~Spinrath, ``{Naturalness
  of the Non-Universal MSSM in the Light of the Recent Higgs Results},''
  \href{http://dx.doi.org/10.1007/JHEP01(2013)187}{{\em JHEP} {\bfseries 01}
  (2013) 187},
\href{http://arxiv.org/abs/1207.7236}{{\ttfamily arXiv:1207.7236}}.

\bibitem{Cheng:2012pe}
T.~Cheng, J.~Li, T.~Li, X.~Wan, Y.~k. Wang, and S.-h. Zhu, ``{Toward the
  Natural and Realistic NMSSM with and without $R$-Parity},''
\href{http://arxiv.org/abs/1207.6392}{{\ttfamily arXiv:1207.6392 [hep-ph]}}.

\bibitem{CahillRowley:2012rv}
M.~W. Cahill-Rowley, J.~L. Hewett, A.~Ismail, and T.~G. Rizzo, ``{The Higgs
  Sector and Fine-Tuning in the pMSSM},''
  \href{http://dx.doi.org/10.1103/PhysRevD.86.075015}{{\em Phys. Rev.}
  {\bfseries D86} (2012) 075015},
\href{http://arxiv.org/abs/1206.5800}{{\ttfamily arXiv:1206.5800 [hep-ph]}}.

\bibitem{Ross:2012nr}
G.~G. Ross, K.~Schmidt-Hoberg, and F.~Staub, ``{The Generalised NMSSM at One
  Loop: Fine Tuning and Phenomenology},''
  \href{http://dx.doi.org/10.1007/JHEP08(2012)074}{{\em JHEP} {\bfseries 08}
  (2012) 074},
\href{http://arxiv.org/abs/1205.1509}{{\ttfamily arXiv:1205.1509 [hep-ph]}}.

\bibitem{Boehm:2013gst}
C.~Boehm, P.~S.~B. Dev, A.~Mazumdar, and E.~Pukartas, ``{Naturalness of Light
  Neutralino Dark Matter in pMSSM after LHC, XENON100 and Planck Data},''
  \href{http://dx.doi.org/10.1007/JHEP06(2013)113}{{\em JHEP} {\bfseries 06}
  (2013) 113},
\href{http://arxiv.org/abs/1303.5386}{{\ttfamily arXiv:1303.5386 [hep-ph]}}.

\bibitem{Anderson:1994dz}
G.~W. Anderson and D.~J. Castano, ``{Measures of fine tuning},''
  \href{http://dx.doi.org/10.1016/0370-2693(95)00051-L}{{\em Phys. Lett.}
  {\bfseries B347} (1995) 300--308},
\href{http://arxiv.org/abs/hep-ph/9409419}{{\ttfamily arXiv:hep-ph/9409419
  [hep-ph]}}.

\bibitem{Anderson:1994tr}
G.~W. Anderson and D.~J. Castano, ``{Naturalness and superpartner masses or
  when to give up on weak scale supersymmetry},''
  \href{http://dx.doi.org/10.1103/PhysRevD.52.1693}{{\em Phys. Rev.} {\bfseries
  D52} (1995) 1693--1700},
\href{http://arxiv.org/abs/hep-ph/9412322}{{\ttfamily arXiv:hep-ph/9412322
  [hep-ph]}}.

\bibitem{Anderson:1995cp}
G.~W. Anderson and D.~J. Castano, ``{Challenging weak scale supersymmetry at
  colliders},'' \href{http://dx.doi.org/10.1103/PhysRevD.53.2403}{{\em Phys.
  Rev.} {\bfseries D53} (1996) 2403--2410},
\href{http://arxiv.org/abs/hep-ph/9509212}{{\ttfamily arXiv:hep-ph/9509212
  [hep-ph]}}.

\bibitem{Anderson:1996ew}
G.~W. Anderson, D.~J. Castano, and A.~Riotto, ``{Naturalness lowers the upper
  bound on the lightest Higgs boson mass in supersymmetry},''
  \href{http://dx.doi.org/10.1103/PhysRevD.55.2950}{{\em Phys. Rev.} {\bfseries
  D55} (1997) 2950--2954},
\href{http://arxiv.org/abs/hep-ph/9609463}{{\ttfamily arXiv:hep-ph/9609463
  [hep-ph]}}.

\bibitem{Ciafaloni:1996zh}
P.~Ciafaloni and A.~Strumia, ``{Naturalness upper bounds on gauge mediated soft
  terms},'' \href{http://dx.doi.org/10.1016/S0550-3213(97)00138-7}{{\em Nucl.
  Phys.} {\bfseries B494} (1997) 41--53},
\href{http://arxiv.org/abs/hep-ph/9611204}{{\ttfamily arXiv:hep-ph/9611204
  [hep-ph]}}.

\bibitem{Barbieri:1998uv}
R.~Barbieri and A.~Strumia, ``{About the fine tuning price of LEP},''
  \href{http://dx.doi.org/10.1016/S0370-2693(98)00577-2}{{\em Phys. Lett.}
  {\bfseries B433} (1998) 63--66},
\href{http://arxiv.org/abs/hep-ph/9801353}{{\ttfamily arXiv:hep-ph/9801353
  [hep-ph]}}.

\bibitem{Giusti:1998gz}
L.~Giusti, A.~Romanino, and A.~Strumia, ``{Natural ranges of supersymmetric
  signals},'' \href{http://dx.doi.org/10.1016/S0550-3213(99)00153-4}{{\em Nucl.
  Phys.} {\bfseries B550} (1999) 3--31},
\href{http://arxiv.org/abs/hep-ph/9811386}{{\ttfamily arXiv:hep-ph/9811386
  [hep-ph]}}.

\bibitem{Casas:2003jx}
J.~A. Casas, J.~R. Espinosa, and I.~Hidalgo, ``{The MSSM fine tuning problem: A
  Way out},'' \href{http://dx.doi.org/10.1088/1126-6708/2004/01/008}{{\em JHEP}
  {\bfseries 01} (2004) 008},
\href{http://arxiv.org/abs/hep-ph/0310137}{{\ttfamily arXiv:hep-ph/0310137
  [hep-ph]}}.

\bibitem{Casas:2004uu}
J.~A. Casas, J.~R. Espinosa, and I.~Hidalgo, ``{A Relief to the supersymmetric
  fine tuning problem},'' in {\em {String phenomenology. Proceedings, 2nd
  International Conference, Durham, UK, July 29-August 4, 2003}}, pp.~76--85.
\newblock 2004.
\newblock
\href{http://arxiv.org/abs/hep-ph/0402017}{{\ttfamily arXiv:hep-ph/0402017
  [hep-ph]}}.
\newblock

\bibitem{Casas:2004gh}
J.~A. Casas, J.~R. Espinosa, and I.~Hidalgo, ``{Implications for new physics
  from fine-tuning arguments. 1. Application to SUSY and seesaw cases},''
  \href{http://dx.doi.org/10.1088/1126-6708/2004/11/057}{{\em JHEP} {\bfseries
  11} (2004) 057},
\href{http://arxiv.org/abs/hep-ph/0410298}{{\ttfamily arXiv:hep-ph/0410298
  [hep-ph]}}.

\bibitem{Casas:2006bd}
J.~A. Casas, J.~R. Espinosa, and I.~Hidalgo, ``{Expectations for LHC from
  naturalness: modified versus SM Higgs sector},''
  \href{http://dx.doi.org/10.1016/j.nuclphysb.2007.04.024}{{\em Nucl. Phys.}
  {\bfseries B777} (2007) 226--252},
\href{http://arxiv.org/abs/hep-ph/0607279}{{\ttfamily arXiv:hep-ph/0607279
  [hep-ph]}}.

\bibitem{Kitano:2005wc}
R.~Kitano and Y.~Nomura, ``{A Solution to the supersymmetric fine-tuning
  problem within the MSSM},''
  \href{http://dx.doi.org/10.1016/j.physletb.2005.10.003}{{\em Phys. Lett.}
  {\bfseries B631} (2005) 58--67},
\href{http://arxiv.org/abs/hep-ph/0509039}{{\ttfamily arXiv:hep-ph/0509039
  [hep-ph]}}.

\bibitem{Athron:2007ry}
P.~Athron and D.~J. Miller, ``{A New Measure of Fine Tuning},''
  \href{http://dx.doi.org/10.1103/PhysRevD.76.075010}{{\em Phys. Rev.}
  {\bfseries D76} (2007) 075010},
\href{http://arxiv.org/abs/0705.2241}{{\ttfamily arXiv:0705.2241 [hep-ph]}}.

\bibitem{Feroz:2013hea}
F.~Feroz, M.~P. Hobson, E.~Cameron, and A.~N. Pettitt, ``{Importance Nested
  Sampling and the MultiNest Algorithm},''
\href{http://arxiv.org/abs/1306.2144}{{\ttfamily arXiv:1306.2144
  [astro-ph.IM]}}.

\bibitem{Feroz:2007kg}
F.~Feroz and M.~P. Hobson, ``{Multimodal nested sampling: an efficient and
  robust alternative to MCMC methods for astronomical data analysis},''
  \href{http://dx.doi.org/10.1111/j.1365-2966.2007.12353.x}{{\em Mon. Not. Roy.
  Astron. Soc.} {\bfseries 384} (2008) 449},
\href{http://arxiv.org/abs/0704.3704}{{\ttfamily arXiv:0704.3704 [astro-ph]}}.

\bibitem{Feroz:2008xx}
F.~Feroz, M.~P. Hobson, and M.~Bridges, ``{MultiNest: an efficient and robust
  Bayesian inference tool for cosmology and particle physics},''
  \href{http://dx.doi.org/10.1111/j.1365-2966.2009.14548.x}{{\em Mon. Not. Roy.
  Astron. Soc.} {\bfseries 398} (2009) 1601--1614},
\href{http://arxiv.org/abs/0809.3437}{{\ttfamily arXiv:0809.3437 [astro-ph]}}.

\bibitem{Fowlie:2016hew}
A.~Fowlie and M.~H. Bardsley, ``{Superplot: a graphical interface for plotting
  and analysing MultiNest output},''
  \href{http://dx.doi.org/10.1140/epjp/i2016-16391-0}{{\em Eur. Phys. J. Plus}
  {\bfseries 131} no.~11, (2016) 391},
\href{http://arxiv.org/abs/1603.00555}{{\ttfamily arXiv:1603.00555
  [physics.data-an]}}.

\end{thebibliography}\endgroup
\bibliographystyle{utphys}
\end{document}